\documentclass[aps, pra, superscriptaddress, a4paper, showpacs, twocolumn, english, 10pt]{revtex4-2}
\usepackage{bbm, amsthm, bm,textcomp, nicefrac,geometry,ragged2e}
\usepackage[dvipsnames]{xcolor}
\usepackage{float}
\usepackage[bbgreekl]{mathbbol}
\usepackage{graphicx,epstopdf,color,verbatim,enumitem}
\usepackage{pbox,array}
\usepackage{mathrsfs}
\usepackage{amssymb}
\usepackage{textgreek}
\usepackage{mathtools}
\usepackage{stackrel}
\usepackage[thinlines]{easytable}
\usepackage{amsmath}
\usepackage[utf8]{inputenc}
\usepackage{enumitem}
\usepackage{xr}
\makeatletter
\usepackage{soul}
\usepackage[normalem]{ulem}
\usepackage[caption=false]{subfig}
\usepackage{hyperref}
\usepackage{blkarray}
\usepackage[T1]{fontenc}
\usepackage[caption=false]{subfig}
\usepackage{babel}
\usepackage[linesnumbered,ruled]{algorithm2e}

\hypersetup{
    colorlinks = true,
    linkcolor = purple,
    citecolor = purple,
    filecolor = purple,      
    urlcolor = purple,
}

\newcommand{\ket}[1]{\left|#1\right\rangle}

\newcommand{\bra}[1]{\left\langle #1\right|}
\newcommand{\bracket}[2]{\left\langle #1|#2\right\rangle}
\newcommand{\proj}[1]{\ket{#1}\!\bra{#1}}

\newcommand{\px}{\sigma_x}

\newcommand{\pz}{\sigma_z}
\newcommand{\id}{\mathbb{1}}

\newcommand{\luca}{\color{magenta}}
\newcommand{\cit}{{\luca [CITE]}}

\newcommand{\albie}{\color{brown}}

\newcommand{\done}{\color{black}}

\begin{document}

\title{Self-Configuring Quantum Networks with Superposition of Trajectories}

\author{Albie Chan*}
    \affiliation{Institute for Quantum Computing, University of Waterloo, Waterloo, ON N2L 3G1, Canada}
    \thanks{This author contributed equally.}
    \affiliation{Department of Physics \& Astronomy, University of Waterloo, Waterloo, ON N2L 3G1, Canada}
\author{Zheng Shi*}
    \affiliation{Institute for Quantum Computing, University of Waterloo, Waterloo, ON N2L 3G1, Canada}
    \thanks{This author contributed equally.}
    \affiliation{Department of Physics \& Astronomy, University of Waterloo, Waterloo, ON N2L 3G1, Canada}
\author{Jorge Miguel-Ramiro}
    \affiliation{Universit\"at Innsbruck, Institut f\"ur Theoretische Physik, Technikerstra{\ss}e 21a, 6020 Innsbruck, Austria}
\author{Luca Dellantonio}
    \affiliation{Institute for Quantum Computing, University of Waterloo, Waterloo, ON N2L 3G1, Canada}
    \affiliation{Department of Physics \& Astronomy, University of Waterloo, Waterloo, ON N2L 3G1, Canada}
    \affiliation{Department of Physics and Astronomy, University of Exeter, Stocker Road, Exeter EX4 4QL, United Kingdom}
\author{Christine A. Muschik}
    \affiliation{Institute for Quantum Computing, University of Waterloo, Waterloo, ON N2L 3G1, Canada}
    \affiliation{Department of Physics \& Astronomy, University of Waterloo, Waterloo, ON N2L 3G1, Canada}
    \affiliation{Perimeter Institute for Theoretical Physics, Waterloo, Ontario N2L 2Y5, Canada}
\author{Wolfgang D\"ur}
    \affiliation{Universit\"at Innsbruck, Institut f\"ur Theoretische Physik, Technikerstra{\ss}e 21a, 6020 Innsbruck, Austria}

\begin{abstract}
Quantum networks are a backbone of future quantum technologies thanks to their role in communication and scalable quantum computing. However, their performance is challenged by noise and decoherence. We propose a self-configuring approach that integrates superposed quantum paths with variational quantum optimization techniques. This allows networks to dynamically optimize the superposition of noisy paths across multiple nodes to establish high-fidelity connections between different parties. Our framework {\done is in principle capable of adapting to unknown noise without requiring detailed characterization or benchmarking of the corresponding quantum channels}. We also discuss the role of vacuum coherence, a quantum effect central to path superposition that impacts protocol performance. Additionally, we demonstrate that our approach remains beneficial even in the presence of imperfections in the generation of path superposition. 

\end{abstract}

\maketitle

\section{Introduction}
Quantum networks can become a foundational technology within societal infrastructures~\cite{Kimble2008, Wehner2018, Acin_2018, Kozlowski2019,Azuma2023}, offering unique opportunities and applications including cryptography~\cite{Gisin2002, Pirandola_2020}, distributed quantum computing~\cite{Cirac_distributed, Caleffi_2024}, quantum interfaces between processor modules as a path towards scalability~\cite{Howe2024Scalable} and enhanced distributed sensing~\cite{Kessler2014, Sekatski2020, Kim_2024}. Central to the functionality of these networks is the transmission of quantum information, i.e., the transfer of quantum states encoded in information carriers through quantum channels and entanglement distribution between parties~\cite{Gisin2007, Northup2014, Chen2021}. 

Nevertheless, the reliability of such tasks is challenged by the presence of decoherence and noise within the quantum channels~\cite{Preskill_2018, Bharti2022, Coutinho_2022,morruiz24}. Extensive efforts have been made to address these challenges in quantum communication scenarios, spanning various approaches including quantum error correction~\cite{knill97, Duur_2007}, entanglement purification~\cite{Duur_2007, Dur99, Pan_2001}, network tomography and benchmarking~\cite{Tomography2010, Eisert_2020, Helsen_2023}, and optimized network routing for entanglement distribution~\cite{Das2018, Pant_2019, Bugalho_2023}. More recently, works have exploited the use of \textit{superposed trajectories} where the information carrier is sent through multiple noisy paths in coherent superposition~\cite{Chiribella2019, Kristjnsson_2020, Abbott2020, MiguelRamiro2021, Rubino2021, sqem1, sqem2, kristjansson24} or various superposed orders~\cite{Procopio2015,Ebler2018,Guerin2019,Caleffi2020,Guo2020,Chiribella2021}. Such approaches have revealed advantageous quantum enhancements in contexts such as quantum computation~\cite{Araujo2014,  Procopio2015, Rubino2021, Lee2023, simonov23,Ali2025}, quantum communications~\cite{chandra21,koudia22,koudia23,caleffi23,Kechrimparis24,tang24demonstration,wu25general,mondal25path}, quantum metrology~\cite{Mothe24,chen25nonlinear,Huang2024}, general noise mitigation~\cite{ Chiribella2021, sqem1, sqem2}, entanglement purification~\cite{Miguelramiro24}, randomized benchmarking~\cite{Miguel2021}, and charging quantum batteries~\cite{lai24}. 

In this paper, we develop a general framework for designing \textit{self-configuring} quantum networks with superposition of trajectories. By employing Variational Quantum Optimization (VQO) techniques~\cite{Moll_2018, Cerezo_2021}, we aim to determine the most effective way to coherently combine available paths between two parties in a network. Our protocol generally achieves enhanced connections with markedly higher fidelity than any classical mixture or single-path strategy, even in the absence of detailed noise characterization for the network; this underscores the intrinsic advantage of quantum path superposition as a noise-robust resource. Moreover, our approach is applicable to quantum networks with multiple intermediate nodes and nested superpositions of paths, thus being compatible with both long-distance quantum communication and short-distance chip-like connections for quantum computation. These features are relevant in the development of robust and adaptable quantum networks suitable for real-world applications.


In addition, our work directly addresses the assumption commonly applied to superposed trajectories and quantum orders (e.g., quantum switch) that the degree of freedom (DOF) controlling the paths superposed remains noiseless~\cite{Chiribella2019, Abbott2020,  Rubino2021,  Caleffi2020, Guo2020}. In contrast to this idealized scenario, we analyze cases where noise affects this path DOF via fluctuations of superposition amplitudes. Notably, we demonstrate that path superposition can still yield advantages even if the path DOF experiences noise, and especially when the noise strength is no stronger than that affecting the information carriers. This yields key insights into parameter regimes where our protocol is operationally effective.

The paper is structured as follows. In Sec.~\ref{sec:important_summary}, we give a high-level overview of the key ideas, motivations and findings in this work, then proceed to review basic concepts and notation in Sec.~\ref{sec:back}. In Sec.~\ref{sec:tools} we introduce the essential tools needed to implement self-configuring networks: the general network setting, the relevant mathematical formalism underlying path superposition, and VQO. We then showcase the capability of these networks using {\done two versions of the protocol}, applicable to direct two-node communication scenarios (Sec.~\ref{subsec:two_node_protocol_desc}) and the general multi-node scenarios (Sec.~\ref{sec:multi_node_protocol_desc}), respectively. We numerically test the performance of each protocol (Secs.~\ref{sec:numerics_2_node} and~\ref{sec:numerics_multinode}, respectively) in a variety of relevant scenarios, and provide detailed analyses to support and interpret the observations. Finally, we draw conclusions about our results and provide an outlook for future work in Sec.~\ref{sec:conclusionss}.

\section{Summary of key ideas and findings}
\label{sec:important_summary}
In this section, we present a high-level overview of the general setting, key concepts, findings, and principal contributions of our approach.

\begin{figure}
{\includegraphics[width=\columnwidth]{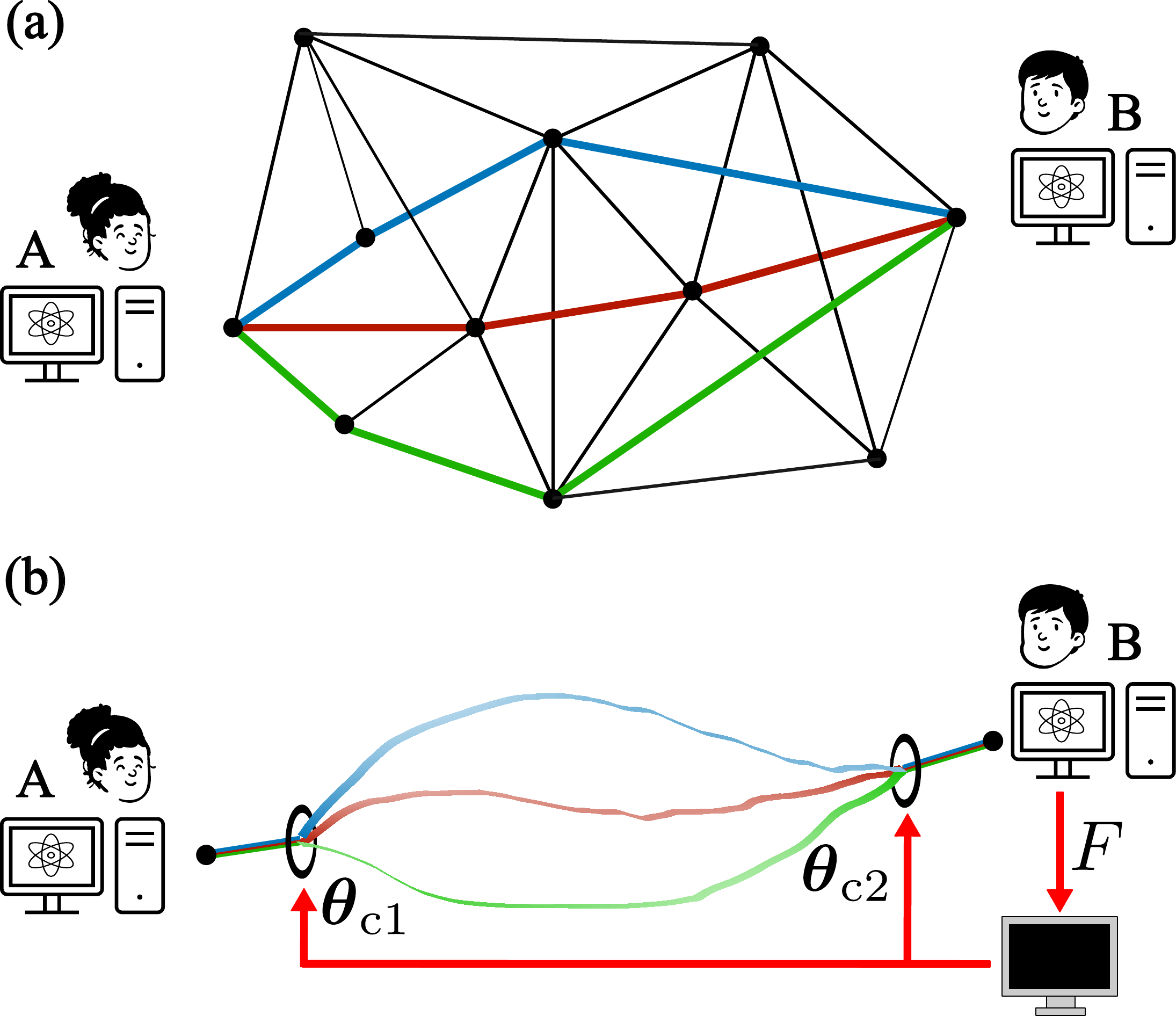}}
    \caption{General setting. Two arbitrary parties in a quantum network seek to communicate. (a) Different paths can be used to connect parties Alice (A) and Bob (B). (b) These paths can also be utilized in coherent superposition to establish the connection. Our self-configuring framework employs a quantum-classical feedback loop (solid red arrows), in which a classical processor optimizes the fidelity $F$ of a quantum state transmitted from Alice to Bob by iteratively updating the parameters that control the path superposition, $\boldsymbol{\theta}_{\mathrm{c}1}$ and $\boldsymbol{\theta}_{\mathrm{c}2}$. Here we show the direct two-node configuration, with the full protocol illustrated in Fig.~\ref{fig:vqo_2_node}. For the more general multi-node configuration, see Fig.~\ref{fig:vqo_multi_node}.}
    \label{fig:setting}
\end{figure}

We propose a framework that enables the constituents of a given quantum network to autonomously determine an optimized way to communicate via superposed paths, see Fig.~\ref{fig:setting}. Our approach generally yields improvements in transmission fidelity compared with any single path across the network. The core innovation of the work is a self-configuring element, which dynamically optimizes the amplitudes and phases of the path superposition; importantly, it adapts automatically to the network topology, and requires no prior knowledge of noise properties. This is accomplished through a feedback loop, where a quantum state is repeatedly transmitted and classical feedback is used to improve the fidelity of the transmitted state. We demonstrate the essential role of vacuum coherence in our approach, and systematically characterize its robustness against noise that directly affects the path DOF (which controls the path superposition). Our approach can be readily applied to arbitrary multi-node networks where each node can generate new superpositions in a nested fashion. These contributions potentially advance the development of robust quantum communication in realistic networks.


We consider general quantum networks where two quantum devices (Alice and Bob) can communicate through a complicated series of nodes and path channels (Fig.~\ref{fig:setting}(a)). Rather than attempting to identify and utilize a single ``best'' path between Alice and Bob, information is sent into a coherent superposition of available paths, so that multiple paths are traversed at once. Our approach adjusts the amplitudes and phases of the path superposition in a feedback loop, such that the transmission fidelity $F$ between Alice and Bob is maximized (Fig.~\ref{fig:setting}(b) and Sec.~\ref{sec:tools}). Via analytical calculations and numerical simulations, we provide a comprehensive performance analysis of the protocol tailored to two different network configurations: two nodes connected by multiple paths with identical and non-identical noise types (Sec.~\ref{sec:two_node_protocol}), and multi-node network topologies featuring nested superpositions and randomized noise parameters (Sec.~\ref{sec:multi_node_protocol}).


\textit{Two nodes} -- In this basic configuration, the quantum network consists of Alice and Bob connected by multiple noisy paths whose actions are represented by quantum channels (Fig.~\ref{fig:vqo_2_node}, Protocol~\ref{table:two_node_protocol_summary} and Sec.~\ref{subsec:two_node_protocol_desc}). This configuration provides not only a minimal testbed for protocol performance in a variety of noise strength regimes, but also a gateway to direct analytical insights regarding the fidelity advantage achievable through coherent path superposition (App.~\ref{app:analytical_ex}). {\done We examine two protocol variants: a deterministic version, which averages the fidelities from all possible path DOF measurement outcomes, and a probabilistic version, which postselects outcomes with more favorable fidelities. We note that the probabilistic version can be used for the heralded distribution of Bell pairs, a resource for quantum teleportation, despite having a success probability less than unity. Generally, the deterministic variant maximizes the information throughput at the cost of additional local unitary corrections, while the probabilistic variant yields higher fidelity advantages without requiring a large measurement budget}.

The degree of noise mitigation achievable in our protocol depends on what we refer to as the ``vacuum coherence'' of each available channel -- the extent to which the environment of the channel remains unaffected by the presence of an information carrier (Fig.~\ref{fig:identical_detprob_vacamp} and Sec.~\ref{sec:identical_chs_vac_coh}). The vacuum coherence of a channel is quantified by its vacuum extension and therefore fully determined by its microscopic details (Sec.~\ref{sec:tools}, Apps.~\ref{app:Proof} and \ref{app:microscopic}). For concreteness, in most of the demonstrations of our protocol, we specify the vacuum coherence by adopting a simple repeated-interaction microscopic model (App.~\ref{app:microscopic}).

For common types of noise, including dephasing, depolarizing, and amplitude damping, substantial fidelity improvements are attained compared to the best single-path strategy. This is demonstrated for up to four identical channels (Fig.~\ref{fig:identical_detprob_micro} and Sec.~\ref{sec:identical_chs_micro}), two nonidentical channels (Fig.~\ref{fig:two_ineq_channels} and Sec.~\ref{subsec:non_identical_chs_2}), and three nonidentical channels (Fig.~\ref{fig:three_ineq_channels} and Sec.~\ref{subsec:non_identical_chs_3}). {\done We find the deterministic variant of our protocol is advantageous only above a vacuum coherence threshold that depends on the noise type, while the probabilistic variant offers an advantage as long as vacuum coherence exists.} Even when the path DOF is itself noisy (App.~\ref{app:noise_path_dof}), meaningful mitigation can still be achieved: the probabilistic variant of the protocol is more robust to the path DOF noise, and the deterministic variant is advantageous when the path DOF noise does not exceed the noise acting on individual paths. 

\textit{Multiple nodes} -- In the more general situation, the network includes additional nodes that can serve as intermediate hops or relays between Alice and Bob (Fig.~\ref{fig:vqo_multi_node}, Protocol~\ref{table:multi_node_protocol_summary} and Sec.~\ref{sec:multi_node_protocol_desc}). Paths are split and recombined at stages delimited by Alice, Bob, and intermediate nodes, producing nested superpositions across the network. Each node in the network may prepare and manipulate the path superposition, allowing us to dynamically select and coherently combine multiple paths at different stages.

Numerical analyses on 5- and 12- stage networks (containing 6 and 29 nodes respectively), with randomly sampled noise strengths for the latter, demonstrate that our self-configuring approach is effective in situations with varying complexity (Figs.~\ref{fig:multinode_simple}, \ref{fig:multinode_complex} and Sec.~\ref{sec:numerics_multinode}). {\done Specifically, it is capable of achieving significant fidelity improvements within reasonable iteration numbers and in the presence of statistical noise}.

The above findings establish that self-configuring quantum networks built using superposed trajectories offer a powerful and flexible tool for enhancing quantum communication. Our approach is broadly applicable to networks with unknown, fluctuating, or heterogeneous noise, and can adapt on-the-fly to changing network conditions. Our approach’s reliance on self-configuration, rather than detailed knowledge of the network, holds promise for both near-term quantum hardware and future large-scale infrastructures.






\section{Background} \label{sec:back}
In this section, we review relevant concepts and notations used throughout this work.

\subsection{Quantum channels}
\label{subsec:chs}
In an ideal scenario, channels connecting parties in a quantum network enable perfect quantum information transmission and entanglement distribution. However, in practice, noise and decoherence effects are unavoidable. Mathematically, these are modeled using completely positive trace preserving (CPTP) quantum maps, described through the Kraus operator (operator-sum) representation \cite{nielsen_chuang_2010}. Here, a channel $\mathcal{C}$ mapping $\rho_{\mathrm{t}}^{\mathrm{in}}$ to  $\rho^{\mathrm{out}}_{\mathrm{t}}$ is defined as
\begin{equation}
\mathcal{C}(\rho^{\mathrm{in}}_{\mathrm{t}}) = \sum_{s}{K}_{s}\rho^{\mathrm{in}}_{\mathrm{t}}{K}_{s}^{\dag} = \rho^{\mathrm{out}}_{\mathrm{t}},
\label{eq:channel}
\end{equation}
where $K_{s}$ are Kraus operators satisfying $\sum_{s}K_{s}^{\dag}K_{s} = \id$. Some relevant instances of quantum channels (which we employ in this work), are dephasing, depolarizing, and amplitude damping maps~\cite{nielsen_chuang_2010}. Their corresponding Kraus operators are
\begin{align}
    & \text{Dephasing: } K_0=\sqrt{1-p_0} \id,  K_1=\sqrt{p_0} {\sigma_{3}}
    ,   \label{eq:dephasingnoise}\\
   & \text{Depolarizing: }  K_0=\sqrt{1-p_0} \id,  K_i=\sqrt{\frac{p_0}{3}} {\sigma_i}, 
   \label{eq:depolarizingnoise}\\
   \begin{split}
    & \text{Amplitude damping: } \\ & \hspace{1cm} K_0=\ket{0}\bra{0} + \sqrt{1-p_{0}}\ket{1}\bra{1}, \\ & \hspace{1cm} K_1=\sqrt{p_{0}}\ket{0}\bra{1},  \label{eq:amplitudedampingnoise}
    \end{split}
\end{align}
where $p_{0}$ is the noise probability associated with the channel and the operators $\sigma_{i}$ correspond to the Pauli matrices ${X}$, ${Y}$, and ${Z}$ for $i=1,2,3$, respectively. The first two channels are classified as unital, as they drive any input state towards the maximally mixed state $\id/2$. In contrast, amplitude damping is non-unital, corresponding to an irreversible energy loss process where an arbitrary state relaxes toward the ground state $\ket{0}$. While our approach is intended to work with arbitrary (including random) noises, we employ these channels for their inherent simplicity in demonstrating the underlying features.

\subsection{Stinespring dilation: Environmental formalism}
\label{subsec:stinespring}
The Stinespring theorem states that any quantum channel can be conceived as a unitary operation acting on a larger Hilbert space~\cite{nielsen_chuang_2010}. More concretely, the effect of a channel $\mathcal{E}$ acting on a system $S$ can be understood as a unitary operation $\Lambda$ that describes the joint evolution of $S$ and an auxiliary environment ``$\mathrm{e}$''. The action of the channel on an arbitrary state of the system $\rho$ can be described by tracing out the Hilbert space of the environment, i.e., 
\begin{equation}
    \hat{\mathcal{E}}(\rho) = \operatorname{tr}_{\mathrm{e}} \!\left[ \Lambda \left(  \rho \otimes\ket{\varepsilon}_{\mathrm{e}}\bra{\varepsilon}_{\mathrm{e}} \right) \Lambda^\dagger \right].
\label{eq:steinspring}
\end{equation}
where $\ket{\varepsilon}_{\mathrm{e}}$ is the initial state of the environment.

The Stinespring dilation provides an explicit unitary extension on a larger Hilbert space, while the Kraus decomposition expresses the evolution in terms of a sum over operators acting on the system. These two representations of CPTP maps are related: the latter can be derived from the former by expanding $\Lambda$ in the environment basis (spanned by states $\ket{s}$). If we consider the case of a system initially in a pure state $\ket{\psi}$, the expansion reads
\begin{equation}
\label{eq:ss_uni_expand}
 \Lambda \left( \ket{\psi}\otimes\ket{\varepsilon}_{\mathrm{e}} \right) = \sum_{s}K_{s}\ket{\psi}\otimes \ket{s}_{\mathrm{e}},
\end{equation}
where the Kraus operators are identified as
\begin{equation}
\label{eq:ss_kraus_expand}
K_{s} = \bra{s}_{\mathrm{e}}\Lambda\ket{\varepsilon}_{\mathrm{e}}.
\end{equation}
By tracing out ``$\mathrm{e}$'' and denoting $\rho = \ket{\psi}\bra{\psi}$, we obtain the standard Kraus decomposition of the CPTP map, 
\begin{equation}
 \hat{\mathcal{E}}(\rho) = \operatorname{tr}_{\mathrm{e}} \!\left[ \sum_{s, s'} K_{s}\rho K_{s'}^{\dagger} \otimes \ket{s}_{\mathrm{e}}\bra{s'}_{\mathrm{e}} \right]= \sum_{s}K_{s}\rho K_{s}^{\dagger}.
\end{equation}

\subsection{Figure of merit: Choi–Jamiołkowski (CJ) fidelty}
\label{subsec:cj}
To evaluate the performance of our approach, we need to quantify the ``closeness'' between ideal and noisy realizations of our network setup. A useful quantity is the state fidelity $F$, defined between an input state $\rho^{\mathrm{in}}$ and noisy output state $\rho^{\mathrm{out}}$ as
\begin{equation}
F(\rho^{\mathrm{in}}, \rho^{\mathrm{out}}) = \big\{\operatorname{tr}\big(\sqrt{\rho^{\mathrm{in}}}\rho^{\mathrm{out}}\sqrt{\rho^{\mathrm{in}}}\big)^{\frac{1}{2}}\big\}^{2},
\label{eq:fid}
\end{equation}
where $F = 1$ implies that $\rho^{\rm in} = \rho^{\rm out}$. For quantum channels, $F=1$ is achieved when the noise is absent.

Our approach proceeds by sending entangled input states across quantum channels. As a result of the Choi–Jamiołkowski (CJ) isomorphism~\cite{jozsa1994fidelity, Glischrist2005}, maximally entangled input states  allow us to completely characterize a channel and capture the ``worst-case'' effect of noise on an arbitrary input state. The Bell states are defined as two-qubit maximally entangled quantum states
\begin{align}
\ket{\mathrm{\Phi}_{ij}} = (\id \otimes \px^j \pz^i) \ket{\mathrm{\Phi}_{00}}, \label{eq:bellstate}
\end{align}
where $i,j \in\{0,1\}$, and $\ket{{\Phi}_{00}} \equiv |\Phi^{+}\rangle =  (\ket{00} + \ket{11})/\sqrt{2}$. In this work, we choose $\rho^{\mathrm{in}}_{\mathrm{ab}} = \ket{\Phi^{+}}_{\mathrm{ab}}\bra{\Phi^{+}}_{\mathrm{ab}}$ where we designate `b' (Bob) as the half of the Bell pair being transmitted and subjected to noise, while the other half `a' (Alice) is retained by the sender. For a channel with Kraus operators $K_s$, the output state is therefore
\begin{equation}
\label{eq:rho_out_fcj}
    \rho^{\mathrm{out}}_{\mathrm{ab}} = \sum_{s}(\id_{\mathrm{a}} \otimes [K_{s}]_{\mathrm{b}})\ket{\Phi^{+}}_{\mathrm{ab}}\bra{\Phi^{+}}_{\mathrm{ab}}(\mathbb{1}_{\mathrm{a}} \otimes [K_{s}^{\dagger}]_{\mathrm{b}}).
\end{equation}
Since $\ket{\Phi^{+}}$ is pure, Eq.~\eqref{eq:fid} simplifies to  
\begin{equation}
\label{eq:fid_simple}
F_{\mathrm{CJ}} = \bra{\Phi^{+}}\rho^{\mathrm{out}}_{\mathrm{ab}}\ket{\Phi^{+}}. 
\end{equation}
The quantity $F_{\mathrm{CJ}}$ provides a theoretical \textit{lower} bound on the fidelity of an arbitrary single-qubit state transmitted through the network~\cite{Dur2005}. 
For the dephasing noise Eq.~\eqref{eq:dephasingnoise} and the depolarizing noise Eq.~\eqref{eq:depolarizingnoise}, $F_{\mathrm{CJ}}=1-p_{0}$; for the amplitude damping noise Eq.~\eqref{eq:amplitudedampingnoise},  $F_{\mathrm{CJ}} = (1+\sqrt{1-p_0})^2/4$.

{\done We note that estimating $F_{\mathrm{CJ}}$ is less costly than performing either a state tomography of $\rho^{\mathrm{out}}_{\mathrm{ab}}$ or a process tomography of the channel $\mathcal{E}$~\cite{flammia11}. Only three sets of Pauli operators $[\sigma_{i}]_{\mathrm{a}} \otimes [\sigma_{i}]_{\mathrm{b}}$ for $i=1,2,3$ need to be measured at both Alice and Bob, and $F_{\mathrm{CJ}}$ can be estimated by using
\begin{equation}
\label{eq:fid_estimate}
    \ket{\Phi^{+}}\bra{\Phi^{+}}=\frac{1}{4}\big(\id \otimes\id + \sigma_1 \otimes \sigma_1 - \sigma_2 \otimes \sigma_2 + \sigma_3 \otimes \sigma_3 \big).
\end{equation}}

\section{Tools for self-configuring quantum networks\label{sec:tools}}
In this section, we introduce the framework and tools for designing self-configuring quantum networks. First, we outline the general setting considered in this work. Next, we provide a brief overview of the formalisms governing superposition of trajectories, stressing the equivalence between different mathematical descriptions. Finally, we give a brief overview of the basic idea of the protocol employed in our work; the step-by-step details of implementation will be given in Secs.~\ref{subsec:two_node_protocol_desc} and \ref{subsec:multi_node}.

\subsection{General setting} \label{sec:setting}
Consider a general quantum network, as shown in Fig.~\ref{fig:setting}, where adjacent nodes or parties are connected by noisy quantum channels. To transmit quantum information or distribute entanglement, a link between two (or more) parties is established via a path comprising a sequence of channels [blue, red and green segments in Fig.~\ref{fig:setting}(a)]. Here, we further allow such links to be coherent superpositions of multiple paths [Fig.~\ref{fig:setting}(b)]. The task is to design variational approaches that, without requiring explicit knowledge of channel properties, determine the best way to connect two (or more) parties, including the use of superposed paths.

\subsection{Superposition of trajectories\label{sec:superpos_traj}}
The advantages of superposing quantum channels have recently been explored~\cite{Chiribella2019, Kristjnsson_2020, Abbott2020, MiguelRamiro2021, Rubino2021, sqem1, sqem2}, revealing broad applications across various areas of quantum information~\cite{Araujo2014,  Procopio2015, Chiribella2021, sqem1, sqem2, Miguelramiro24, Miguel2021}. Here, we briefly review the underlying mechanisms of this phenomenon, stressing the equivalence between existing mathematical descriptions~\cite{Chiribella2019, Abbott2020}.

We explain how path superposition affects the action of noisy quantum channels using a direct communication (two-node) setup between Alice and Bob. In this setup, we assume all paths have similar traversal times, so that information leaving Alice at a given time along different paths can arrive at Bob within a short time window {\done (this is not a strict requirement; see
Sec.~\ref{sec:exp_impl}). Furthermore, each path corresponds to a single quantum channel; for simplicity, we further assume that channels act independently.}

To keep track of the path superposition, one designates a path DOF (also known as the ``control'' DOF in literature) in a Hilbert space spanned by $\ket{i}_{\mathrm{c}}$, with $i=0,1,\dots,d-1$ denoting the path index (see Fig.~\ref{fig:setting}(b)). 
We assume that the system is in the coherent superposition of all $d$ paths
\begin{subequations}
    \label{eq:input}
    \begin{align}
        \rho^{\mathrm{in}} 
        &
        = 
        \ket{\psi}^\mathrm{in}\bra{\psi}^\mathrm{in}
        ,\label{eq:input_dm}
        \\
        \ket{\psi}^\mathrm{in}
        & 
        = 
        \sum_{i=0}^{d-1}a_{i}\ket{i}_{\mathrm{c}}\otimes\ket{\phi}^{\mathrm{in}}
        ,\label{eq:input_pure}
    \end{align}
\end{subequations}
where $a_{i}$ are the superposition amplitudes (we assume no phases for simplicity), $\sum_i \lvert a_i \rvert^2 =1$, and $\ket{\phi}^{\mathrm{in}}$ is an input state of the information carrier. The path superposition can be generated, for instance, by applying a path DOF unitary $U_{\mathrm{c}1} \in \mathrm{SU}(d)$ to the system in the fiducial state $\ket{0}_{\mathrm{c}}$ on the $0$th path immediately before it leaves Alice. Such a unitary is realized by, e.g, beamsplitters in an optical system. 

To determine the effect of network noise on the input Eqs.~\eqref{eq:input}, we purify all $d$ channels via the Stinespring dilation theorem. As discussed in Sec.~\ref{subsec:stinespring}, once the Hilbert space is expanded to include the environment DOF, the overall effect of a quantum channel is a joint unitary evolution of the system and its environment. If the system takes the $i$-th path and interacts with the environment of that channel, the environment state $|\varepsilon_i\rangle_{{\mathrm{e}_{i}}}$ evolves into one of the basis states $|s_i\rangle_{\mathrm{e}_i}$, while the corresponding Kraus operator $K^{(i)}_{s_i}$ is applied to the system, see Eq.~\eqref{eq:ss_uni_expand}. 
Therefore, we can express the purified output state as
\begin{align}
\ket{\psi}^{\mathrm{out}} =& \sum_{i=0}^{d-1}a_{i}\ket{i}_{\mathrm{c}}\bigotimes_{k \neq i} \ket{\varepsilon_k}_{\mathrm{e}_{k}}  \nonumber\\
& \otimes\bigg(\sum_{s_i} K^{(i)}_{s_i}\ket{\phi}^{\mathrm{in}}\otimes \ket{s_i}_{\mathrm{e}_{i}}\bigg),
\label{eq:superpos_env}
\end{align}
where we now keep track of the environment `e' comprising components $\mathrm{e}_{0}, \mathrm{e}_{1},\dots,\mathrm{e}_{d-1}$.

Equation~\eqref{eq:superpos_env} represents an extended quantum channel acting on both the input and path DOF; this is known as the vacuum extension formalism~\cite{Kristjnsson_2020}. To see this more explicitly, we expand the environment states
\begin{equation}
\ket{\varepsilon_i}_{\mathrm{e}{_{i}}} = \sum_{s_i}\alpha_{s_i}^{(i)}\ket{s_i}_{\mathrm{e}{_{i}}},\, \alpha_{s_i}^{(i)} = \bracket{s_i}{\varepsilon_i}_{\mathrm{e}{_{i}}},  \label{eq:vacuumamplitudes}
\end{equation}
where the \textit{vacuum amplitudes} $\alpha_{s_i}^{(i)}$ are inherent properties of the channels determined by their specific physical implementation~\cite{Abbott2020}, and satisfy the normalization condition $\sum_{s_i}|\alpha_{s_i}^{(i)}|^{2} = 1$. However, we mention for completeness that the normalization becomes an upper bound
\begin{equation}
    \sum_{s_i}|\alpha_{s_i}^{(i)}|^{2} \leq 1,
    \label{eq:vacamp_not_normalized}
\end{equation}
when $\alpha_{s_i}^{(i)}$ are understood as \textit{effective} vacuum amplitudes corresponding to a minimal set of \textit{orthogonal} Kraus operators. A proof of Eq.~\eqref{eq:vacamp_not_normalized} is given in App.~\ref{app:Proof}.

Substituting Eq.~\eqref{eq:vacuumamplitudes} into Eq.~\eqref{eq:superpos_env}, we find
\begin{equation}
\ket{\psi}^{{\mathrm{out}}} = \sum_{\boldsymbol{s}}K_{\boldsymbol{s}}\ket{\psi}^\mathrm{in}\otimes \ket{\boldsymbol{s}}_{\mathrm{e}},
\label{eq:extended_wf_kraus}
\end{equation}
where $\boldsymbol{s} = (s_{0},\dots,s_{d-1})$, $\ket{\boldsymbol{s}}_{\mathrm{e}} = \bigotimes_{i}\ket{s_i}_{\mathrm{e}_i}$, and we have defined the extended Kraus operators that act in the Hilbert space including both the path and system DOFs,
\begin{equation}
K_{\boldsymbol{s}} = \sum_{i=0}^{d-1}\bigg(\ket{i}_{\mathrm{c}}\bra{i}_{\mathrm{c}}\otimes K^{(i)}_{s_{i}} \prod_{k\neq i} \alpha^{(k)}_{s_{k}}\bigg).
\label{eq:extended_kraus}
\end{equation}
The component $K^{(i)}_{s_{i}}$ is a Kraus operator acting on the system on the $i$-th path, while each remaining vacuum amplitude $\alpha^{(k)}_{s_k}$ corresponds to an untraversed path $k \neq i$.
Tracing out all environment DOFs from Eq.~\eqref{eq:extended_wf_kraus}, we clearly see that the set of Kraus operators $K_{\boldsymbol{s}}$ defines a vacuum-extended quantum channel, with the operator-sum representation
\begin{equation}
\rho^{\mathrm{out}}
=
\sum_{\boldsymbol{s}} 
K_{\boldsymbol{s}}
\rho^{\rm in}
K^{\dagger}_{\boldsymbol{s}}.
\label{eq:kraus_global}
\end{equation}
Here, $K_{\boldsymbol{s}}$ is block-diagonal in the path Hilbert space, because the noises in all paths are assumed to be independent and do not directly act on the path DOF.


While Eq.~\eqref{eq:kraus_global} provides a compact description of paths in coherent superposition subject to independent noise, in practice it is more convenient to express the output state $\rho^{\mathrm{out}}$ explicitly in terms of its blocks in the path Hilbert space. Starting from Eq.~\eqref{eq:kraus_global} and making use of the normalization condition of $\alpha^{(i)}_{s_i}$, we find
\begin{equation}
\rho^{\mathrm{out}}
=\sum_{i=0}^{d-1}\rho^{\mathrm{out}(ii)}
 + \sum_{i\neq j}\rho^{\mathrm{out}(ij)},
\label{eq:supp_incoh_p_coh}
\end{equation}
with the diagonal $\rho^{\mathrm{out}(ii)}$ and off-diagonal $\rho^{\mathrm{out}(ij)}$ blocks in the path Hilbert space being
\begin{subequations}
    \label{eq:blocks_ci}
    \begin{align}
        \rho^{\mathrm{out}(ii)}
        &
        = 
        \sum_{s_i }K^{(i)}_{s_i} \rho^{\mathrm{in}(ii)}K_{s_i}^{(i)\dagger}
        ,\label{eq:incoh_part}
        \\
        \rho^{\mathrm{out}(ij)}
        & 
        = 
        F^{(i)}_{\mathrm{vio}}\rho^{\mathrm{in}(ij)}F^{(j)\dagger}_{\mathrm{vio}} \ \ (i\neq j)
        .\label{eq:coh_part}
    \end{align}
\end{subequations}
Here, the blocks of the input density matrix are given by $\rho^{\mathrm{in}(ij)}=a_{i}a^{*}_{j} \ket{\phi}^\mathrm{in}\bra{\phi}^\mathrm{in}$ (see Eqs.~\eqref{eq:input}),
and the \textit{vacuum interference operators} $F^{(i)}_{\mathrm{vio}}$ are linear combinations of the Kraus operators with the vacuum amplitudes as coefficients~\cite{Kristjnsson_2020}:
\begin{equation}
F^{(i)}_{\mathrm{vio}}=\sum_{s_i}\alpha^{(i)*}_{s_i} K^{(i)}_{s_i}.   \label{eq:vio} 
\end{equation}
The diagonal blocks Eq.~\eqref{eq:incoh_part} describe the action of each channel on the $i$-th path, and their contribution to Eq.~\eqref{eq:supp_incoh_p_coh} is a purely classical mixture. In contrast, the off-diagonal blocks Eq.~\eqref{eq:coh_part} capture the coherence between pairs of paths $i$ and $j$ through each $\alpha_{s}$ and $F_{\mathrm{vio}}$.



Both Eq.~\eqref{eq:extended_kraus} and Eqs.~\eqref{eq:supp_incoh_p_coh}--\eqref{eq:vio} imply that, in the presence of path superposition, a well-defined quantum map depends not only on the Kraus operators, but also on the vacuum amplitudes and in particular the vacuum interference operator $F_{\mathrm{vio}}$. 
In App.~\ref{app:microscopic_special}, we calculate $F_{\mathrm{vio}}$ for the common quantum channels in Eqs.~\eqref{eq:dephasingnoise}--\eqref{eq:amplitudedampingnoise} using a simple repeated-interaction microscopic model~\cite{attal06,grimmer16}, which we incorporate in most of our protocol simulations (Secs.~\ref{sec:two_node_protocol} and \ref{sec:multi_node_protocol}). {\done We further justify this choice in App.~\ref{app:microscopic_general} by analyzing a class of generic microscopic models, which we adopt in the simulations of Sec.~\ref{sec:multinode_complex}.}

Once the system is subject to the noise and arrives at Bob, we can apply a second path DOF unitary  $U_{\mathrm{c}2} \in \mathrm{SU}(d)$ to Eq.~\eqref{eq:supp_incoh_p_coh}. This induces either constructive or destructive interference (i.e., mixing) among the diagonal and off-diagonal blocks; both potentially enabling partial cancellation of noise in different paths. By measuring in the path DOF basis~\cite{sqem2}, we can achieve noise mitigation by selecting favorable measurement outcomes and/or applying unitary corrections to improve individual outcomes.

\subsection{Variational quantum optimization}
In realistic quantum networks, one must consider the possibilities of arbitrary noisy channels that may drift and fluctuate, as well as varying connectivity in intermediate modes between Alice and Bob. It then becomes necessary to introduce a ``self-configuring'' aspect that optimizes the properties of the network \emph{in situ}, maximizes noise mitigation and enhances communication between nodes -- all without prior knowledge of the noise.

We achieve this aspect by integrating an overall VQO routine in the path superposition protocol. Here, network components controlling the superposition of paths depend on tunable parameters. By variationally adjusting these parameters in a feedback loop with the help of a classical processor, we can optimize a cost function that characterizes the effectiveness of noise mitigation in a given communication task within the quantum network. Executing the path superposition protocol with the optimized parameters leads to a potentially enhanced transmission fidelity from Alice to Bob. In this sense, the network ``self-configures'' by adjusting parameters based on the average or the best-case fidelity, making it broadly applicable even when noise sources are unknown and/or dynamic.

When performing the routine, we assume that properties of each path in the network, namely the Kraus operators $K_s$ and the vacuum interference operator $F_{\mathrm{vio}}$, remain constant. In practice, the protocol will likely be effective as long as $K_s$ and $F_{\mathrm{vio}}$ drift slowly, such that the resulting optimal parameters are valid for a reasonably long amount of time after the optimization itself.

\section{Two-node protocol}
\label{sec:two_node_protocol}

In this section, we demonstrate how our protocol can optimize direct transmission of a qubit from Alice to Bob via $d$ superposed paths. These paths may be selected from a larger network based on resource constraints and overall noise levels. After describing the relevant protocol, we evaluate its performance via numerical simulation involving a variety of different communication scenarios. These scenarios permit direct investigation of different channel combinations, vacuum amplitudes, and noise affecting the path DOF.

\subsection{Protocol description}
\label{subsec:two_node_protocol_desc}

\begin{figure}
{\includegraphics[width=\columnwidth]{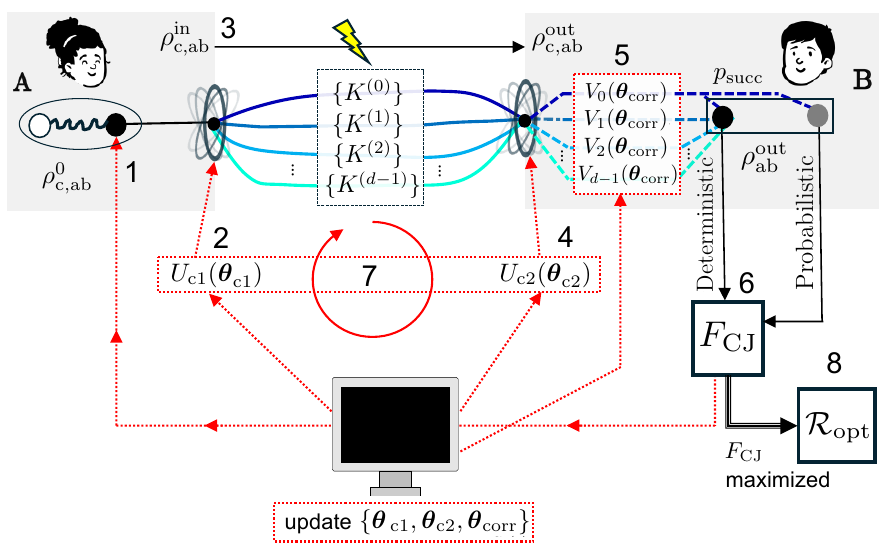}}
\caption{Schematic of the two-node protocol (Protocol~\ref{table:two_node_protocol_summary}). The directed lines indicate processes involving the classical hardware. All red lines/boxes involve the VQO algorithm. The gray boxes enclosing Alice and Bob denote all protocol aspects that are accessible to them (preparation and post-processing steps respectively). Alice and Bob's nodes are denoted as solid black circles residing on the edge of their respective boxes.}
\label{fig:vqo_2_node}
\end{figure}
\label{subsec:two_node}
The complete protocol is shown schematically in Fig.~\ref{fig:vqo_2_node}, summarized as Protocol~\ref{table:two_node_protocol_summary} and is described below:

\textit{Step 0 --} Choose $d$ non-interacting transmission paths from Alice to Bob [Fig.~\ref{fig:setting}(a)]. 

\textit{Step 1 --} Prepare a two-qubit state $\rho^{\mathrm{in}}_{\mathrm{ab}}=\ket{\Phi^{+}}_{\mathrm{ab}}\bra{\Phi^{+}}_{\mathrm{ab}}$ (see Sec.~\ref{subsec:cj}) as  input, with the system initially residing on one path, e.g., the 0th path $\ket{0}_{\mathrm{c}}$. This corresponds to the full system state $\rho^{0}_{\mathrm{c, ab}} = \ket{0}_{\mathrm{c}}\bra{0}_{\mathrm{c}} \otimes \rho^{\mathrm{in}}_{\mathrm{ab}}$.

\textit{Step 2 --} 
Apply the path DOF unitary $U_{\mathrm{c}1}$ (parameterized by $\boldsymbol{\theta}_{\mathrm{c}1}$) to generate the coherent path superposition, corresponding to the state $\ket{\psi}_{\mathrm{c}} = U_{\mathrm{c}1} \ket{0}_{\mathrm{c}}$. 
This requires $2d-2$ parameters ($d-1$ magnitudes and $d-1$ relative phases) and produces the state
\begin{equation}
\rho^{\mathrm{in}}_{\mathrm{c, ab}} = \ket{\psi}_{\mathrm{c}}\bra{\psi}_{\mathrm{c}} \otimes \rho_{\mathrm{ab}}^{\mathrm{in}}.
\end{equation}
where $\rho^{\mathrm{in}}_{\mathrm{c, ab}}$ takes the form of Eq.~\eqref{eq:input}. 

\textit{Step 3 --} Transmit the latter qubit in $\rho^{\mathrm{in}}_{\mathrm{ab}}$ through the noisy paths to obtain the output $\rho^{\mathrm{out}}_{\mathrm{c, ab}}$, keeping the other half at Alice. $\rho^{\mathrm{out}}_{\mathrm{c, ab}}$ now relates to $\rho^{\mathrm{in}}_{\mathrm{c, ab}}$ via Eqs.~\eqref{eq:kraus_global}--\eqref{eq:blocks_ci}.

\textit{Step 4 --} Apply the second path DOF unitary $U_{\mathrm{c}2}$ (parameterized by $\boldsymbol{\theta}_{\mathrm{c}2}$)  to $\rho^{\mathrm{out}}_{\mathrm{c, ab}}$, then measure the path DOF in the $\ket{j}_{\mathrm{c}}$ basis ($j = 0, 1, \dots, d-1$). For outcome $j$, the corresponding post-measured state is
\begin{equation}
\rho^{\mathrm{out}}_{\mathrm{ab}, j} = \frac{1}{P_j}\operatorname{tr}_{\mathrm{c}}\big(\ket{j}_{\mathrm{c}}\bra{j}_{\mathrm{c}}U_{\mathrm{c}2}\rho_{\mathrm{c, ab}}^{\mathrm{out}}U_{\mathrm{c}2}^\dagger\big),
\label{eq:postmeas_st}
\end{equation} 
where the corresponding probability $P_{j} = \bra{j}_{\mathrm{c}}U_{\mathrm{c}2}\rho_{\mathrm{c, ab}}^{\mathrm{out}}U_{\mathrm{c}2}^\dagger\ket{j}_{\mathrm{c}} = \operatorname{tr}\big(\ket{j}_{\mathrm{c}}\bra{j}_{\mathrm{c}}U_{\mathrm{c}2}\rho_{\mathrm{c, ab}}^{\mathrm{out}}U_{\mathrm{c}2}^\dagger\big)$. 
Generally, $U_{\mathrm{c}2} $ requires $d^{2}-1$ parameters since there are $d^{2}-1$ generators of $\mathrm{SU}(d)$. Nevertheless, as discussed in Sec.~\ref{sec:superpos_traj}, one can implement smaller families of $d$-dimensional unitaries with fewer parameters.

\textit{Step 5 --} Apply single-qubit local unitaries $V^{(j)}_{\mathrm{b}}$ (parameterized by $\boldsymbol\theta_{\mathrm{corr}}$) on qubit `b', now at Bob's node~\footnote{The local unitaries are applied to maximize the output CJ fidelity $F_{\mathrm{CJ}}$, see Step 7. $F_{\mathrm{CJ}}$ is upper bounded by the largest eigenvalue of the output density matrix, i.e., $\lambda_{\mathrm{max}} (\rho_{\mathrm{ab}}^{\mathrm{out}})$, and a unitary rotation can maximize the overlap of $\ket{\Phi^{+}}$ with an eigenvector corresponding to $\lambda_{\mathrm{max}}$.}. These are applied to each of the $d$ post-measurement states $\rho^{\mathrm{out}}_{\mathrm{ab}, j}$ in Eq.~\eqref{eq:postmeas_st}:
\begin{equation}
 \rho^{\mathrm{out,corr}}_{\mathrm{ab}, j} = (\mathrm{\mathbb{1}_{\mathrm{a}} }\otimes V^{(j)}_{\mathrm{b}})\rho^{\mathrm{out}}_{\mathrm{ab}, j}(\mathrm{\mathbb{1}_{\mathrm{a}} }\otimes V^{(j)}_{\mathrm{b}})^{\dagger},
 \label{eq:u3_corr}
\end{equation}
where $j = 0, 1, \dots, d-1$, $V_{\mathrm{b}}^{(j)} \in \mathrm{SU}(2)$ has the form
\begin{equation}
V_\mathrm{b}^{(j)}(\boldsymbol{\theta}_{\mathrm{corr}}) = 
\begin{pmatrix}
\mathrm{cos}{(\beta^{(j)}/2)} & -e^{\mathrm{i}\delta^{(j)}}\mathrm{sin}{(\beta^{(j)}/2)} \\
e^{\mathrm{i}\gamma^{(j)}}\mathrm{sin}{(\beta^{(j)}/2)} & e^{\mathrm{i}(\gamma^{(j)} + \delta^{(j)})}\mathrm{cos}{(\beta^{(j)}/2)}
\end{pmatrix}.
\label{eq:u3}
\end{equation} 
and $\boldsymbol\theta_{\mathrm{corr}} = [\beta^{(0)}, \gamma^{(0)}, \delta^{(0)}, \dots, \beta^{(d-1)}, \gamma^{(d-1)}, \delta^{(d-1)}]$.

\textit{Step 6 --} Measure the output CJ fidelity $F_{\mathrm{CJ}}$ either deterministically or probabilistically. Deterministically, $F_{\mathrm{CJ}}$ is averaged over all measurement outcomes of the path DOF: 
\begin{equation}
F_{\mathrm{CJ, det}} = \sum_{j=0}^{d-1}P_{j}\langle\Phi^{+}|\rho^{\mathrm{out,corr}}_{\mathrm{ab}, j}|\Phi^{+}\rangle,
\label{eq:cj_fid_det}
\end{equation} 
where we have used Eq.~\eqref{eq:fid_simple}. 
Probabilistically, the measurement outcome with the largest $P_{j}$ is postselected:
\begin{equation}
F_{\mathrm{CJ, prob}} = \max\limits_{P_{j}} \langle\Phi^{+}|\rho^{\mathrm{out,corr}}_{\mathrm{ab}, j}|\Phi^{+}\rangle,
\label{eq:cj_fid_prob}
\end{equation} 
and we define the success probability
{\done
\begin{equation}
    p_{\mathrm{succ}} = \mathrm{max}(P_{j})
    \label{eq:succ_prob}
\end{equation}
such that $p_{\mathrm{succ}} \geq 1/d$.} Here, we postselect the \textit{most likely} outcome instead of the highest-fidelity outcome, since the latter may occur too infrequently to be practical. {\done In practice, both $F_{\mathrm{CJ}}$ and $p_{\mathrm{succ}}$ are estimated with a measurement budget of $N_{\mathrm{shots}}$ Bell pairs. If the path DOF outcome $j$ is measured $N_{j}$ times, $P_{j}$ is estimated as $N_{j}/N_{\mathrm{shots}}$, and we allocate $N_{j}/3$ measurements to each of the three operators $[\sigma_{i}]_{\mathrm{a}} \otimes [\sigma_{i}]_{\mathrm{b}}$ ($i=1,2,3$) (see Eq.~\eqref{eq:fid_estimate}). For simplicity, in this study we use a fixed measurement budget for each iteration (see Step~7), but it is generally more efficient to allocate measurement budgets adaptively throughout the optimization process~\cite{Shlosberg2023adaptive}; we expect the budget can be further reduced by Bayesian optimization which builds on prior knowledge~\cite{iannelli2022noisy,nicoli2023physics}.

In Sec.~\ref{sec:numerics_2_node}, we compare the deterministic and probabilistic schemes in the limit of infinite measurements $N_{\mathrm{shots}}\to \infty$, and give an example of the effect of a finite measurement budget. We stress that a large $N_{\mathrm{shots}}$ may be necessary if one wishes to optimize the post-measurement states of the less likely outcomes in the deterministic method.} {\done In addition, a truly deterministic scheme may be impractical due to attenuation and imperfect detection.} 

\textit{Step 7 --} Repeat Steps 1--6 iteratively to find the optimal parameters $\{\boldsymbol\theta_{\mathrm{c}1}, \boldsymbol\theta_{\mathrm{c}2}, \boldsymbol\theta_{\mathrm{corr}}\}_{\mathrm{opt}}$ that maximize $F_{\mathrm{CJ}}$. The optimization is performed on a classical processor, thus completing the feedback loop.

Note that for the probabilistic variant of our protocol, optimizing $F_{\mathrm{CJ}}$ alone may be insufficient if the success probability $p_{\mathrm{succ}}$ remains small; this is particularly problematic in resource-constrained settings.
{\done Instead, we optimize $F_{\mathrm{CJ}}$ and $p_{\mathrm{succ}}$ simultaneously by minimizing the cost function
\begin{equation}
\label{eq:cost_function}
    \nu(1-F_{\mathrm{CJ}})^2+(1-\nu)(1-p_{\mathrm{succ}})^2,
\end{equation}
where $0\leq\nu\leq1$ is an adjustable weight. A large $\nu \approx 1$ may not be enough to achieve a sufficiently large $p_{\mathrm{succ}}$, especially if the number of measurements per iteration $N_{\mathrm{shots}}$ used to estimate $F_{\mathrm{CJ}}$ and $p_{\mathrm{succ}}$ is finite. On the other hand, a very small $\nu$ will favor $p_{\mathrm{succ}}\approx 1$ at the expense of $F_{\mathrm{CJ}}$, potentially failing to produce an advantage. It is therefore important to choose a weight that balances fidelity and throughput. In the $N_{\mathrm{shots}}\to \infty$ limit, we find that for a wide range of $\nu$, the optimized $F_{\mathrm{CJ}}$ and $p_{\mathrm{succ}}$ are nearly independent of $\nu$ (see also Fig.~\ref{fig:multinode_complex}(c) and (e) for the effect of varying $\nu$).}

In practice, it is also convenient to employ a nested optimization scheme: for given path DOF unitary parameters $\boldsymbol{\theta}_{c1}, \boldsymbol{\theta}_{c2}$, we optimize the unitary correction parameters $\boldsymbol{\theta}_{\mathrm{corr}}$ for each path DOF measurement outcome in a very fast inner loop; $\boldsymbol{\theta}_{c1}, \boldsymbol{\theta}_{c2}$ are then optimized in the outer loop. This approach takes advantage of the fact that the cost function landscape is much simpler in the $\boldsymbol{\theta}_{\mathrm{corr}}$ parameter space. {\done For our noise model, owing to the absence of coherent noise, the inner loop is trivial in the probabilistic variant of the protocol~\cite{Rubino2021}.}

\textit{Step 8 --} Finally, we characterize the performance of the VQO by comparing the optimal coherent (superposed-path) fidelity $\mathrm{max}(F_{\mathrm{CJ}})$ against the incoherent (single-path) fidelity $F_{\mathrm{CJ}}^{0}$. For non-identical channels, $F_{\mathrm{CJ}}^{0}$ is regarded as the least noisy (i.e., highest-fidelity) path. From these, we compute an infidelity ratio
\begin{equation}
\mathcal{R} = \frac{1 -F_{\mathrm{CJ}}^{0}}{1 - \mathrm{max}(F_{\mathrm{CJ}})},
\label{eq:R_ratio}
\end{equation}
as a relative measure of improvement over the incoherent case. Any $\mathcal{R} > 1$ corresponds to noise mitigation: the larger $\mathcal{R}$ is, the greater the degree of mitigation.

\begin{table}
{\LinesNumberedHidden
    \begin{algorithm}[H]
        \SetKwInOut{Input}{Input}
        \SetKwInOut{Output}{Output}
        \SetAlgorithmName{Protocol}{}
    \noident
    \justifying \begin{enumerate}[leftmargin=*,itemindent=0pt]
      \setcounter{enumi}{-1}
          \item (I) Select $d$ noisy paths directly connecting Alice (`a') to Bob (`b') to superpose.

           \item (A) Prepare the Bell state $\rho_{\mathrm{ab}}^{\mathrm{in}} = \ket{\Phi^{+}}_{\mathrm{ab}}\bra{\Phi^{+}}_{\mathrm{ab}}$ in path $0$.

           \item (A) Apply the path DOF unitary $U_{\mathrm{c1}}$ to generate a superposition of the $d$ paths, encoded in the state $\rho^{\mathrm{in}}_{\mathrm{c, ab}}$ where system `c' is the path DOF. 

            \item (TR) Transmit system `b' of $\rho^{\mathrm{in}}_{\mathrm{c, ab}}$ through the $d$ superposed paths, with noise described by $s$ Kraus operators $\{K_{s_{i}}^{(i)}\}$ acting on each path $i$. 

            \item (B) Apply the path unitary $U_{\mathrm{c}2}$ to the transmitted state $\rho^{\mathrm{out}}_{\mathrm{c, ab}}$, then measure the path DOF in the $\ket{j}_{\mathrm{c}}$ basis ($j = 0, 1, \dots, d-1$).

            \item (B) Apply single-qubit correcting unitaries $V^{(j)}_{\mathrm{b}}$ to each of the $d$ postmeasured states $\rho^{\mathrm{out}}_{\mathrm{ab}, j}$.

            \item (O) Measure the fidelity $F_{\mathrm{CJ}}$ either deterministically (average on all outcomes) or probabilistically (on the most likely outcome only).

            \item (F) Repeat Steps 1--6 iteratively, updating all unitary parameters until $F_{\mathrm{CJ}}$ is maximized. 

            \item (O) Compute the infidelity ratio $\mathcal{R}$ (Eq.~\eqref{eq:R_ratio}), based on $\mathrm{max}(F_{\mathrm{CJ}})$ and the single-path (incoherent) fidelity $F^{0}_{\mathrm{CJ}}$.
        \end{enumerate}
    
   \justifying \textit{Legend}: I--protocol input, A--step at Alice, TR--transmission step, B--step at Bob, O--protocol output, F--feedback loop 
    
\caption{Self-configuring protocol, two nodes (Fig.~\ref{fig:vqo_2_node})} \label{table:two_node_protocol_summary}
\end{algorithm}}
\end{table}

\subsection{Performance analysis}
\label{sec:numerics_2_node}
We now investigate the performance of our two-node protocol using a combination of analytical calculations (see Apps.~\ref{app:analytical_ex} and \ref{app:noise_path_dof}) and numerical simulations with the QuTiP library~\cite{johansson2012qutip}.
Motivated by our discussion in Sec.~\ref{sec:superpos_traj}, we begin with a general analysis of vacuum interference in Sec.~\ref{sec:identical_chs_vac_coh}. Thereafter, we always adopt $F_{\mathrm{vio}}$ of the simple microscopic model in App.~\ref{app:microscopic_special} for concreteness. {\done While this choice is grounded in physically meaningful assumptions (see App.~\ref{app:microscopic_general}), the working principles of the underlying protocol are independent of the specific microscopic model.} 

For simplicity, {\done unless otherwise noted, we employ a minimal implementation of the $d$-path unitary $U(\boldsymbol{\theta})$}, which comprises a product of two-mode beamsplitter unitaries between adjacent paths:
\begin{equation}
    U(\boldsymbol{\theta})= T_{d-1,d}(\theta_{d-1})\cdots T_{2,3}(\theta_2)T_{1,2}(\theta_1),
    \label{eq:min_unitary}
\end{equation}
where the $d \times d$ unitary $T_{m,n}(\theta)$ is assumed to be a $Y$ rotation $e^{-i(\theta/2) Y}$ embedded in the two-dimensional subspace spanned by paths $m$ and $n$. The simplified parameterization Eq.~\eqref{eq:min_unitary} is sufficient for achieving maximum noise mitigation when the errors in the network are fully described by Eqs.~\eqref{eq:dephasingnoise}--\eqref{eq:amplitudedampingnoise} and have no coherent component.

{\done In most of this section, we also assume an infinite measurement budget $N_{\mathrm{shots}}\to \infty$, such that the figures of merit Eq.~\eqref{eq:cj_fid_det}--\eqref{eq:succ_prob} are evaluated without statistical noise. However, we demonstrate in Sec.~\ref{subsec:non_identical_chs_3} that the probabilistic variant of the two-node protocol is robust against the statistical noise due to a finite $N_{\mathrm{shots}}$.}

\subsubsection{Identical channels: Vacuum coherence analysis}
\label{sec:identical_chs_vac_coh}

We first examine the two-node scenario of Alice and Bob connected by $d$ identical paths with completely identical noisy channels. Assuming no path DOF noise, the optimal solution is simply the equal superposition of all paths. Nevertheless, this situation is analytically tractable, and provides qualitative insights on the typical performance of our path superposition protocol that can be generalized to more realistic situations. We focus on the interpretation of the results, and relegate the mathematical details to App.~\ref{app:analytical_ex}.

The vacuum amplitudes $\alpha_s$ and vacuum interference operators $F_{\mathrm{vio}}$ are essential in describing the action of noise on a path superposition. Indeed, we will demonstrate that vacuum interference as quantified by $\alpha_s$ and $F_{\mathrm{vio}}$ plays a fundamental role in the success of our protocol. To this end, we choose a specific form of $F_{\mathrm{vio}}$, $F_{\mathrm{vio}}=\alpha^*_0 K_0$, where the zeroth (no-jump) Kraus operator $K_0$ is given in Eqs.~\eqref{eq:dephasingnoise}--\eqref{eq:amplitudedampingnoise} for the three noise types we consider here, and $\alpha_0$ is the corresponding effective vacuum amplitude, assumed to be real and nonnegative for simplicity and satisfying $0\leq \alpha_0 \leq 1$. (Recall that, as shown in App.~\ref{app:Proof}, the effective vacuum amplitude vector is not necessarily normalized.) This choice of $F_{\mathrm{vio}}$ is inspired by the microscopic model in App.~\ref{app:microscopic_special} where averaging the environment over many realizations results in $F_{\mathrm{vio}}\propto K_0$. The effective vacuum amplitude $\alpha_0$ can be said to quantify the ``vacuum coherence'' of the channel: according to Eq.~\eqref{eq:vacuumamplitudes}, it describes how much the environment in the vacuum state remains unchanged after interacting with an information carrier passing through the channel.

\begin{figure*}
    \centering
    \includegraphics[width=0.9\textwidth]{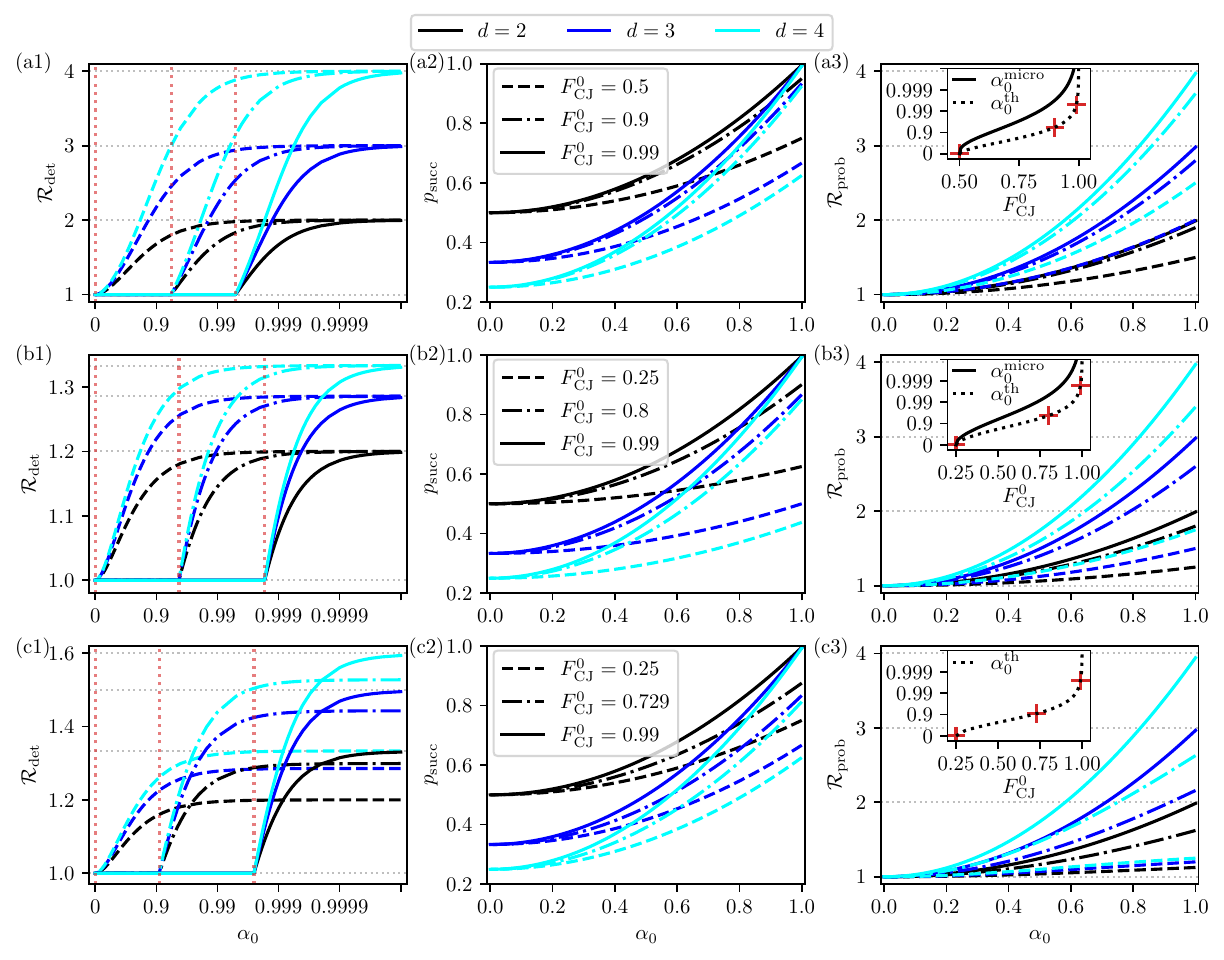}
    \caption{Role of vacuum coherence in the two-node protocol for $d$ identical paths. Assuming no path DOF noise, {\done $N_{\mathrm{shots}}\to\infty$,} and the vacuum interference operator is proportional to the zeroth Kraus operator, $F_{\mathrm{vio}}=\alpha^*_0 K_0$, we plot the optimal infidelity ratio $\mathcal{R}_{\mathrm{det}}$ for the deterministic variant of the protocol, together with the optimal postselection success probability $p_{\mathrm{succ}}$ and infidelity ratio $\mathcal{R}_{\mathrm{prob}}$ for the probabilistic variant of the protocol, as functions of the vacuum amplitude $\alpha_0$ in the case of (a) dephasing, (b) depolarizing or (c) amplitude damping noise for various incoherent (single-path) fidelities $F^{0}_{\mathrm{CJ}}$. The insets of (a3), (b3) and (c3) plot the thresholds $\alpha^{\mathrm{th}}_0$, below which the  protocol is no longer advantageous deterministically, as functions of $F^{0}_{\mathrm{CJ}}$ (dotted lines). In the dephasing and depolarizing cases, we also show $\alpha_0$ given by a simple microscopic model (solid lines), see Eqs.~\eqref{eq:dephasing_micro_vio} and \eqref{eq:depolarizing_micro_vio}. (In the amplitude damping case, the microscopic model always yields $\alpha_{0} = 1$; see Eq.~\eqref{eq:vio_micro_full_ad}.) Furthermore, in panels (a1), (b1) and (c1), the $\alpha^{\mathrm{th}}_0$ values corresponding to the chosen $F^{0}_{\mathrm{CJ}}$ are highlighted by vertical red lines.
    \label{fig:identical_detprob_vacamp}}
\end{figure*}

\begin{figure*}
    \centering
    \includegraphics[width=0.9\textwidth]{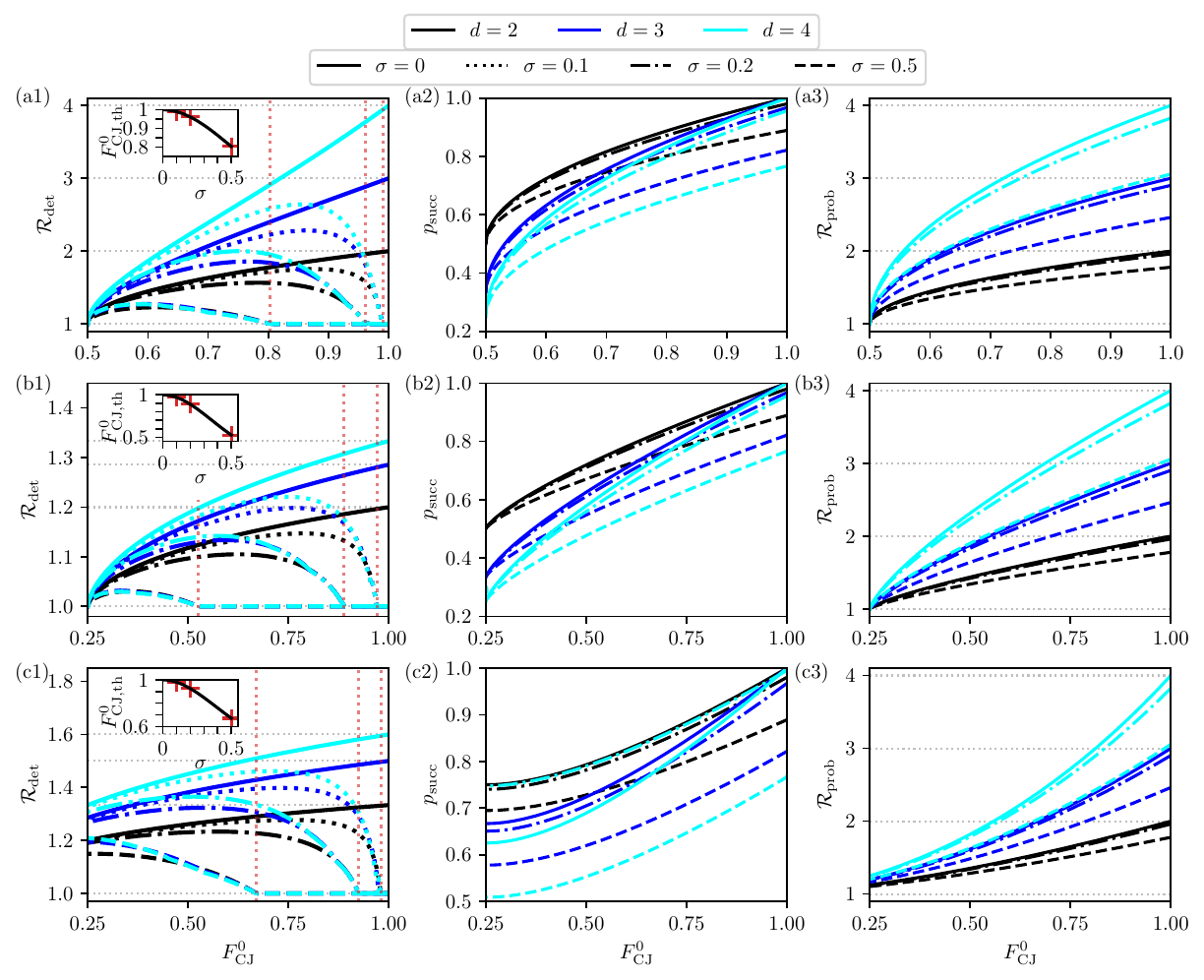}
    \caption{Two-node protocol for $d$ identical paths. The optimal $\mathcal{R}_{\mathrm{det}}$ for the deterministic variant and the optimal $p_{\mathrm{succ}}$ and $\mathcal{R}_{\mathrm{prob}}$ for the probabilistic variant are plotted as functions of $F^{0}_{\mathrm{CJ}}$ in the case of (a) dephasing, (b) depolarizing or (c) amplitude damping noise, where the vacuum interference operator $F_{\mathrm{vio}}$ is given by the microscopic model in App.~\ref{app:microscopic_special}, {\done and $N_{\mathrm{shots}}\to\infty$}. We also show the effect of varying the path DOF noise strength $\sigma$; note that $\sigma=0.1$ is not shown for the probabilistic variant as it visually overlaps with $\sigma=0$. In particular, the threshold values $F^{0}_{\mathrm{CJ,th}}$ above which the protocol is no longer advantageous deterministically are plotted as functions of $\sigma$ in the insets of (a1), (b1) and (c1), with $\sigma=0.1$, $0.2$ and $0.5$ highlighted by vertical red lines in each panel and red markers in its inset.}  
    \label{fig:identical_detprob_micro}
\end{figure*}

Figure~\ref{fig:identical_detprob_vacamp} demonstrates the relation between vacuum coherence and the protocol performance for dephasing, depolarizing and amplitude damping channels. We plot the optimal infidelity ratio of the deterministic variant of the protocol $\mathcal{R}_{\mathrm{det}}$, as well as the optimal postselection success probability $p_{\mathrm{succ}}$ and the optimal infidelity ratio $\mathcal{R}_{\mathrm{prob}}$ of the probabilistic variant of the protocol, as functions of $\alpha_0$ for varying incoherent (single-path) fidelity $F^{0}_{\mathrm{CJ}}$. For greater transparency, we temporarily neglect the path DOF noise.

For all noise types and any given $F^{0}_{\mathrm{CJ}}$, the figures of merit $\mathcal{R}_{\mathrm{det}}$, $p_{\mathrm{succ}}$ and $\mathcal{R}_{\mathrm{prob}}$ are all increasing functions of $\alpha_0$. In particular, they reach their maximum values for $\alpha_0=1$, and $\mathcal{R}_{\mathrm{det}}>1$ and $\mathcal{R}_{\mathrm{prob}}>1$ only when $\alpha_0\neq 0$. In other words, as expected, our protocol requires the presence of vacuum coherence to be advantageous, and stronger vacuum coherence results in a greater advantage. Furthermore, given the same $\alpha_0$ and $F^{0}_{\mathrm{CJ}}$, when the number of identical paths $d$ increases, $\mathcal{R}_{\mathrm{det}}$ and $\mathcal{R}_{\mathrm{prob}}$ also increase, whereas $p_{\mathrm{succ}}$ decreases. This indicates that having more paths in an equal coherent superposition can improve the best fidelity and the average fidelity of the outcome, but reduces the probability to obtain the best fidelity by postselection (from having a greater number of outcomes).

It is also important to distinguish between the behaviors of deterministic and probabilistic variants. In general, the probabilistic variant is more flexible and less demanding: it can tolerate a wider range of noise types and strengths, and requires less vacuum coherence compared to the deterministic variant. However, there is a possibility of failure which must be weighted against the gain in fidelity (recall the weighted cost function Eq.~\eqref{eq:cost_function}). In contrast, the deterministic variant necessarily succeeds whenever it is employed, but it imposes stricter requirements on both the noise characteristics and the amount of vacuum coherence needed for an appreciable advantage.

For the deterministic variant, in the strong vacuum coherence $\alpha_0\to 1$ and weak noise limit $F^{0}_{\mathrm{CJ}}\to 1$, the behavior of $\mathcal{R}_{\mathrm{det}}$ depends strongly on the type of noise: $\mathcal{R}_{\mathrm{det}}\to d$ for the dephasing noise, $\mathcal{R}_{\mathrm{det}}\to 3d/(2d+1)$ for the depolarizing noise, and $\mathcal{R}_{\mathrm{det}}\to 2d/(d+1)$ for the amplitude damping noise (dotted horizontal lines in Fig.~\ref{fig:identical_detprob_vacamp}(a1), (b1) and (c1)). The dephasing noise is special in that it does \emph{not} affect the state populations, such that in the $\alpha_0\to 1$ limit all $d-1$ post-measurement states that do not correspond to the maximally likely outcome can be completely corrected by local unitaries; therefore, these less likely measurement outcomes do not reduce $\mathcal{R}_{\mathrm{det}}$ from the probabilistic limiting value $d$ in the dephasing case. Interestingly, in the $\alpha_0\to 1$ limit, $\mathcal{R}_{\mathrm{det}}$ is independent of $F^{0}_{\mathrm{CJ}}$ for dephasing and depolarizing noise -- a consequence of the simple form of the Kraus operators.

As another important observation, for a given noise type and $F^{0}_{\mathrm{CJ}}$, there exists a threshold vacuum amplitude $\alpha^{\mathrm{th}}_0$, such that for any $\alpha_0<\alpha^{\mathrm{th}}_0$ and any $d$, $\mathcal{R}_{\mathrm{det}}=1$ (i.e., there is no deterministic advantage). Intuitively, this is because the interference between different paths becomes weaker for smaller $\alpha_0$, and at some point the post-measurement states corresponding to the $d-1$ less likely measurement outcomes can no longer be sufficiently corrected. For all three types of noise, $\alpha^{\mathrm{th}}_0$ approaches $1$ as $F^{0}_{\mathrm{CJ}} \to 1$; the full $F^{0}_{\mathrm{CJ}}$ dependence (see Eqs.~\eqref{eq:alpha_0_th_depha}, \eqref{eq:alpha_0_th_depol} and \eqref{eq:alpha_0_th_ad}) is plotted in the insets of Fig.~\ref{fig:identical_detprob_vacamp}(a3), (b3) and (c3), and $\alpha^{\mathrm{th}}_0$ corresponding to the chosen $F^{0}_{\mathrm{CJ}}$ values are also highlighted as vertical red lines in Fig.~\ref{fig:identical_detprob_vacamp}(a1), (b1) and (c1). The simple microscopic model in App.~\ref{app:microscopic_special} always yield an effective $\alpha_0$ above the threshold: For the dephasing and the depolarizing noise, we show the $\alpha_0$ given by the microscopic model (see Eqs.~\eqref{eq:dephasing_micro_vio} and \eqref{eq:depolarizing_micro_vio}) in the insets, while in the amplitude damping case, the microscopic model simply yields the maximum possible value $\alpha_0=1$ (see Eq.~\eqref{eq:vio_micro_full_ad}). The full analytical expressions of $\mathcal{R}_{\mathrm{det}}$ are given by Eqs.~\eqref{eq:rdet_depha_a2}, \eqref{eq:rdet_depol_a2} and \eqref{eq:rdet_ad_a2} for the three types of noise.

In contrast to the deterministic variant, for the probabilistic variant, $p_{\mathrm{succ}}\to 1$ and $\mathcal{R}_{\mathrm{prob}} \to d$ always apply in the $\alpha_0\to 1$ and $F^{0}_{\mathrm{CJ}}\to 1$ limit regardless of the noise type. The reason why $\mathcal{R}_{\mathrm{prob}} \to d$ is as follows: the equal superposition of $d$ paths places the same weight on each term in the path DOF density matrix ($d$ diagonal terms and $d(d-1)$ off-diagonal terms) such that, with $\alpha_0\to 1$, postselecting the most likely path DOF measurement outcome increases the relative likelihood of no error occurring by a factor of $d$. In fact, this argument applies to the more general case where the $d$ channels are not necessarily identical but have the same CJ fidelity, see App.~\ref{app:a1_prob_equal} for details. We also note that $\mathcal{R}_{\mathrm{prob}}$ and $p_{\mathrm{succ}}$ both smoothly depend on $\alpha_0$, such that the probabilistic variant can give an advantage ($\mathcal{R}_{\mathrm{prob}} > 1)$ for any $F^{0}_{\mathrm{CJ}}$ and any nonzero $\alpha_0$. The analytical expressions of $p_{\mathrm{succ}}$ and $\mathcal{R}_{\mathrm{prob}}$ for all three noise types are given in Eqs.~\eqref{eq:p0_depha_a2}, \eqref{eq:rprob_depha_a2}, \eqref{eq:p0_depol_a2}, \eqref{eq:rprob_depol_a2}, \eqref{eq:p0_ad_a2} and \eqref{eq:rprob_ad_a2}.


Finally, while vacuum coherence is crucial to the protocol performance, we stress that $\alpha_0$ and $F_{\mathrm{vio}}$ are inherent to the environment of the network and thus generally \emph{not} tunable. In practice, $F_{\mathrm{vio}}$ can be accessed experimentally by benchmarking procedures~\cite{paperbenchmark}, although neither the knowledge of $F_{\mathrm{vio}}$ nor any particular microscopic model is necessary for our protocol to succeed. Henceforth, we assume for concreteness that the $F_{\mathrm{vio}}$ of our noisy quantum channels are given by the microscopic model in App.~\ref{app:microscopic_special}, namely Eqs.~\eqref{eq:dephasing_micro_vio}, \eqref{eq:depolarizing_micro_vio} and \eqref{eq:vio_micro_full_ad}.

\subsubsection{Identical channels with path DOF noise}
\label{sec:identical_chs_micro}

We continue the study of the two-node setup with $d$ identical paths, but turn our attention to the microscopic models with varying $F^{0}_{\mathrm{CJ}}$, specifically in the presence of the path DOF noise.

To simulate imperfections in generating the path superposition, we consider an experimentally relevant DOF noise model: fluctuations in the parameters $\boldsymbol{\theta}=(\theta_1, \theta_2, \dots, \theta_{N})$ that control the path DOF unitary $U(\boldsymbol{\theta})\in\{U_{\mathrm{c}1} , U_{\mathrm{c}2} \}$~\cite{chen2006quantum}. The noisy action of $U(\boldsymbol{\theta})$ on the path DOF is an effective quantum channel $\mathcal{C}_{\boldsymbol{\theta}}$, represented by an ensemble average over individual noisy realizations $U(\boldsymbol{\tilde{\theta}})$ sampled from a probability density function $p(\boldsymbol{\tilde{\theta}, \boldsymbol{\theta}})$: 

\begin{equation}
\mathcal{C}_{\boldsymbol{\theta}}(\rho_{\mathrm{c}}) = \int_{\mathbb{R}^{N}}d\boldsymbol{\tilde{\theta}}p(\boldsymbol{\tilde{\theta}, \boldsymbol{\theta}})U(\boldsymbol{\tilde{\theta}})\rho_{\mathrm{c}} U^{\dagger}(\boldsymbol{\tilde{\theta}}).
\label{eq:control_noise}
\end{equation}
More concretely, we employ
\begin{equation}
p(\boldsymbol{\tilde{\theta}, \boldsymbol{\theta}}) = \frac{1}{(2\pi \sigma^2)^{N/2}} \exp\left(-\frac{|\boldsymbol{\tilde{\theta}}-\boldsymbol{\theta}|^2}{2\sigma^2} \right),
\label{eq:cnoise_norm_dist}
\end{equation}
an isotropic Gaussian distribution with variance $\sigma^{2}$ and mean $\boldsymbol{\theta}$. In the presence of such noise, we therefore have
\begin{equation}
\rho^{\mathrm{in}}_{\mathrm{c, ab}} = \mathcal{C}_{\boldsymbol{\theta}_{\mathrm{c1}}}(\ket{0}_{\mathrm{c}}\bra{0}_{\mathrm{c}}) \otimes \rho_{\mathrm{ab}}^{\mathrm{in}}
\end{equation}
and 
\begin{equation}
U_{\mathrm{c2}}\rho_{\mathrm{c, ab}}^{\mathrm{out}}U_{\mathrm{c2}}^\dagger \mapsto \mathcal{C}_{\boldsymbol{\theta}_{\mathrm{c2}}}(\rho_{\mathrm{c, ab}}^{\mathrm{out}})
\end{equation}
in Steps 2 and 4 of Protocol~\ref{table:two_node_protocol_summary}, respectively.

Apart from the probability distribution function, the noise model in Eq.~\eqref{eq:control_noise} also depends on the parameterization of $U(\boldsymbol{\theta})$. For our minimal implementation Eq.~\eqref{eq:min_unitary}, when $\sigma\ll 1$, the quantum channel described by Eq.~\eqref{eq:control_noise} has an error probability which is proportional to $\sigma^2$ and grows linearly with $d$; the analytical form of this channel can be found explicitly for $d=2$ (see App.~\ref{app:path_noise_channel}). For arbitrary $\sigma$, it is also straightforward to numerically implement Eq.~\eqref{eq:control_noise} as a superoperator.



We now discuss the performance of our two-node protocol, with $d$ identical paths described by the microscopic model in App.~\ref{app:microscopic_special} and the path DOF noise modeled by Eqs.~\eqref{eq:control_noise} and \eqref{eq:cnoise_norm_dist}. The results are reported in Fig.~\ref{fig:identical_detprob_micro} for noise strength $\sigma\leq 0.5$ and varying $F^{0}_{\mathrm{CJ}}$, using both the deterministic and the probabilistic variants of the protocol.

We begin from the $\sigma=0$ case. Once again, for all three types of noise with a fixed $F^{0}_{\mathrm{CJ}}$, having a larger $d$ increases both $\mathcal{R}_{\mathrm{det}}$ and $\mathcal{R}_{\mathrm{prob}}$ but reduces $p_{\mathrm{succ}}$. In the weak noise limit $F^{0}_{\mathrm{CJ}}\to 1$, our microscopic model recovers $\alpha_0 \to 1$ for all noise types (see App.~\ref{app:microscopic_general}). Therefore, the $F^{0}_{\mathrm{CJ}} \to 1$ limit in Fig.~\ref{fig:identical_detprob_micro} is consistent with the $F^{0}_{\mathrm{CJ}} \to 1, \alpha_0 \to 1$ limit in Fig.~\ref{fig:identical_detprob_vacamp} for both the deterministic and the probabilistic variants. On the other hand, in the strong noise limit $F^{0}_{\mathrm{CJ}}\to 1/2$ (for dephasing noise) and $F^{0}_{\mathrm{CJ}}\to 1/4$ (for depolarizing and amplitude damping noise), our protocol offers no advantage for the dephasing and depolarizing channels, but retains some advantage for the amplitude damping channel. This is again consistent with Fig.~\ref{fig:identical_detprob_vacamp}, since only the amplitude damping channel retains any vacuum coherence in the strong noise limit ($\alpha_{0} = 1$; see Eq.~\eqref{eq:vio_micro_full_ad}). $\mathcal{R}_{\mathrm{det}}$, $p_{\mathrm{succ}}$ and $\mathcal{R}_{\mathrm{prob}}$ all vary smoothly as functions of $F^{0}_{\mathrm{CJ}}$ between the weak and strong noise limits.

For $\sigma \neq 0$, deterministic and probabilistic variants are affected rather differently; we discuss the results qualitatively here, and give a quantitative analysis of the $d=2$ case to $O(\sigma^2)$ and $O(1-F^{0}_{\mathrm{CJ}})$ in App.~\ref{app:path_noise_effect}.

As shown in Fig.~\ref{fig:identical_detprob_vacamp}(a1), (b1) and (c1), the deterministic variant gives no advantage, $\mathcal{R}_{\mathrm{det}}=1$, when $F^{0}_{\mathrm{CJ}}$ exceeds a threshold value $F^{0}_{\mathrm{CJ,th}}$ that is dependent on $\sigma$ and the noise type but independent of $d$ (insets). An intuitive explanation is that the path DOF noise has an impact analogous to that of imperfect vacuum coherence $|\alpha_0|\neq 1$: in both cases, correcting the less likely measurement outcomes becomes less profitable. To $O(\sigma^2)$, we can show that the infidelity of each individual path at this threshold value is also $O(\sigma^2)$, comparable to the effective noise from $\sigma$ (see Eqs.~\eqref{eq:rdet_thres_deph_d}, \eqref{eq:rdet_thres_depo_d} and \eqref{eq:rdet_thres_ad_d}). Once $F^{0}_{\mathrm{CJ}}$ falls below the threshold, $\mathcal{R}_{\mathrm{det}}$ starts to grow, eventually reaching a maximum before approaching the $\sigma=0$ values as $F^{0}_{\mathrm{CJ}}$ further decreases. This behavior is reminiscent of the effect of the CSWAP noise in the related superposed quantum error mitigation protocol introduced in Ref.~\cite{sqem1}, which also acts on the DOF that controls the coherent superposition.

Although the threshold value for $F^{0}_{\mathrm{CJ}}$ is independent of $d$, for given $\sigma$ and $F^{0}_{\mathrm{CJ}}$ near the threshold, $\mathcal{R}_{\mathrm{det}}$ exhibits a saturating behavior as a function of $d$. This highlights a key tradeoff of the deterministic variant of the protocol: while larger values of $d$ generally offer greater advantages, they also exhibit increased sensitivity to the path DOF noise which usually grows with $d$.

The threshold behavior in the deterministic variant of the protocol is in stark contrast with the probabilistic variant, which is much less sensitive to the path DOF noise. As shown in Fig.~\ref{fig:identical_detprob_micro}(a2--3), (b2--3) and (c2--3), $p_{\mathrm{succ}}$ and $\mathcal{R}_{\mathrm{prob}}$ depend so slowly on $\sigma$ that their $\sigma=0$ and $\sigma=0.1$ values visually overlap, and we begin to observe a significant reduction from the $\sigma=0$ values only for $\sigma\gtrsim 0.2$.

\begin{figure}
    \centering
    \includegraphics[width=\linewidth]{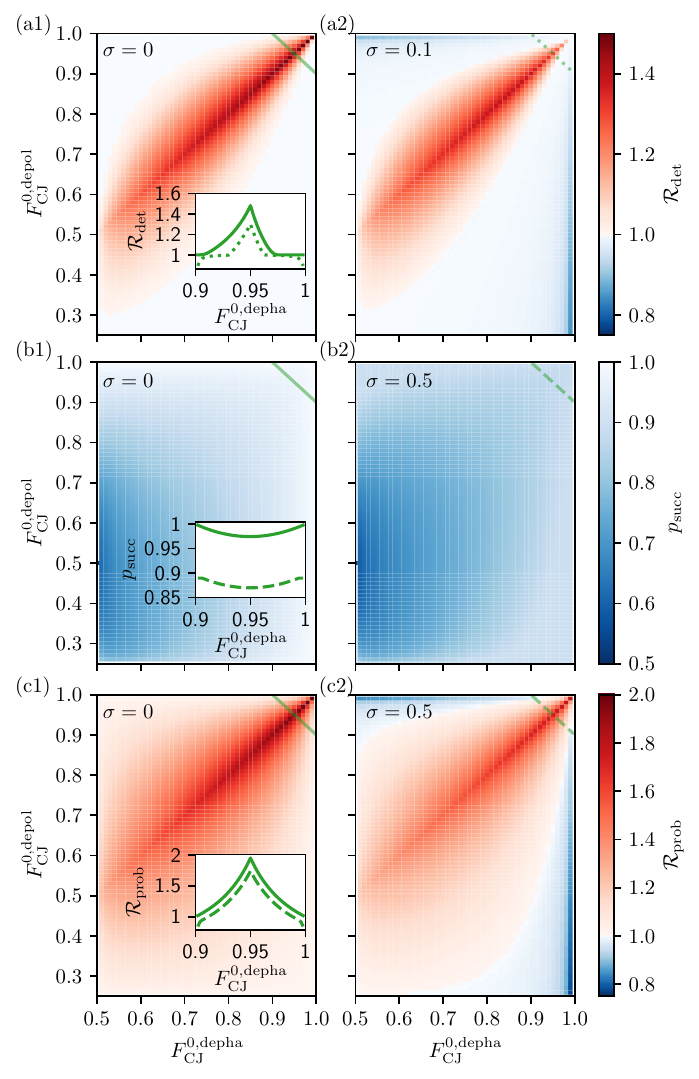}
    \caption{Two-node protocol for $d=2$ paths with dephasing and depolarizing noises, respectively. We plot the optimal values of (a) $\mathcal{R}_{\mathrm{det}}$ for the deterministic variant, (b) $p_{\mathrm{succ}}$ and (c) $\mathcal{R}_{\mathrm{prob}}$ for the probabilistic variant as functions of $F^{0}_{\mathrm{CJ}}$ of both channels, where the vacuum interference operators $F_{\mathrm{vio}}$ are given by the microscopic model in App.~\ref{app:microscopic_special}, {\done and $N_{\mathrm{shots}}\to\infty$}. The left column corresponds to results without the path DOF noise, and the right column corresponds to $\sigma=0.1$ (deterministic) and $\sigma=0.5$ (probabilistic). In each row, the inset of the left column compares the $\sigma=0$ (solid) and $\sigma\neq 0$ (dotted/dashed) cross sections along the line $F^{0,\mathrm{depha}}_{\mathrm{CJ}}+ F^{0,\mathrm{depol}}_{\mathrm{CJ}}=1.9$ (shown in the upper right corner of every panel).}
    \label{fig:two_ineq_channels}
\end{figure}

\subsubsection{Two non-identical channels\label{subsec:non_identical_chs_2}}
\label{sec:non_identical_chs}

We now turn to scenarios involving paths with different noise types and strengths, where the utility of our protocol begins to emerge. In Fig.~\ref{fig:two_ineq_channels}, we consider $d=2$ paths with dephasing and depolarizing channels, respectively, allowing the CJ fidelity of both paths, $F^{0,\mathrm{depha}}_{\mathrm{CJ}}$ and $F^{0,\mathrm{depol}}_{\mathrm{CJ}}$, to vary independently. For both the deterministic and the probabilistic variants, we observe that $\mathcal{R}_{\mathrm{det}}$ and $\mathcal{R}_{\mathrm{prob}}$ are maximal along the diagonal ridge where $F^{0,\mathrm{depha}}_{\mathrm{CJ}} = F^{0,\mathrm{depol}}_{\mathrm{CJ}}$, corresponding to symmetric noise strengths. The maximal $\mathcal{R}_{\mathrm{det}}$ in the absence of the path DOF noise is $3/2$ (see Eq.~\eqref{eq:rdet_dephadepol_a2}), which lies between the maximal $\mathcal{R}_{\mathrm{det}}$ values for two dephasing channels ($2$) and for two depolarizing channels ($6/5$), whereas the maximal $\mathcal{R}_{\mathrm{prob}}$ remains $2$, as discussed in App.~\ref{app:a1_prob_equal}.

Away from the diagonal, when one of the paths is much noisier than the other, the less noisy path dominates in the VQO, and the noisier path does little to improve the overall $F_{\mathrm{CJ}}$. This is reflected by $\mathcal{R}_{\mathrm{det}}$, $\mathcal{R}_{\mathrm{prob}}$ and $p_{\mathrm{succ}}$ all going to $1$. We also observe that the probabilistic variant maintains an advantage ($\mathcal{R}_{\mathrm{prob}}>1$) for a larger range of noise parameters than the deterministic variant does ($\mathcal{R}_{\mathrm{det}}>1$); in other words, the probabilistic variant is more robust against noise asymmetries.

Interestingly, when we focus on a linecut along the antidiagonal direction in the weak noise regime (i.e., with a constant $F^{0,\mathrm{depha}}_{\mathrm{CJ}}+ F^{0,\mathrm{depol}}_{\mathrm{CJ}} \approx 2$, see insets of Fig.~\ref{fig:two_ineq_channels}(a1), (b1) and (c1)), $\mathcal{R}_{\mathrm{det}}$ decreases more slowly away from the diagonal on the $F^{0,\mathrm{depha}}_{\mathrm{CJ}} < F^{0,\mathrm{depol}}_{\mathrm{CJ}}$ side than the opposite side. In other words, it is easier to improve a depolarizing channel deterministically by using a noisier dephasing channel compared with the opposite case. On the contrary, $p_{\mathrm{succ}}$ and $\mathcal{R}_{\mathrm{prob}}$ for the probabilistic variant are almost symmetric with respect to the diagonal.

Let us now examine the effect of the path DOF noise. As we have already noted for identical paths in Fig.~\ref{fig:identical_detprob_micro}, a finite $\sigma$ reduces $\mathcal{R}_{\mathrm{det}}$, $p_{\mathrm{succ}}$ and $\mathcal{R}_{\mathrm{prob}}$ everywhere. The reduction is particularly strong for the deterministic variant in the regime where both paths are less noisy than the path DOF unitaries themselves (upper right corner of Fig.~\ref{fig:two_ineq_channels}(a2)). Meanwhile, the path DOF noise affects the probabilistic variant much less drastically: for comparable reduction in $\mathcal{R}_{\mathrm{det}}$ and $\mathcal{R}_{\mathrm{prob}}$, a much larger $\sigma$ is needed in the probabilistic case (compare Fig.~\ref{fig:two_ineq_channels}(a2) and (c2)). It is worth mentioning that both $\mathcal{R}_{\mathrm{det}}$ and $\mathcal{R}_{\mathrm{prob}}$ can be less than unity when one of the channels is much noisier than the other, since we have to use the less noisy channel in the presence of the path DOF noise (which is in this case independent of the superposition amplitudes and phases).

To conclude this section, we mention that the above observations are also qualitatively applicable to other combinations of noise types, such as dephasing + dephasing, amplitude damping + amplitude damping, and depolarizing + amplitude damping. The most notable quantitative difference between these scenarios is that the maximum $\mathcal{R}_{\mathrm{det}}$ attainable in the deterministic variant can have different numerical values, which is already obvious from the case of identical channels (Fig.~\ref{fig:identical_detprob_micro}).

\subsubsection{Three non-identical channels\label{subsec:non_identical_chs_3}}

We now consider an example involving three non-identical paths with all three different noise types in Eqs.~\eqref{eq:dephasingnoise}--\eqref{eq:amplitudedampingnoise}. We denote the CJ fidelities of the dephasing, depolarizing and amplitude-damping channels as $F^{0,\mathrm{depha}}_{\mathrm{CJ}}$, $F^{0,\mathrm{depol}}_{\mathrm{CJ}}$, and $F^{0,\mathrm{ad}}_{\mathrm{CJ}}$, respectively. For simplicity, we restrict ourselves to a particular cross section of the parameter space, where $F^{0,\mathrm{depha}}_{\mathrm{CJ}}+ F^{0,\mathrm{depol}}_{\mathrm{CJ}}+F^{0,\mathrm{ad}}_{\mathrm{CJ}}=2.7$. We also assume no path DOF noise ($\sigma =0$); it can be verified that $\sigma < 0.1$ does not qualitatively impact the results reported below on this cross section.

\begin{figure}
    \centering
    \includegraphics[width=1\linewidth]{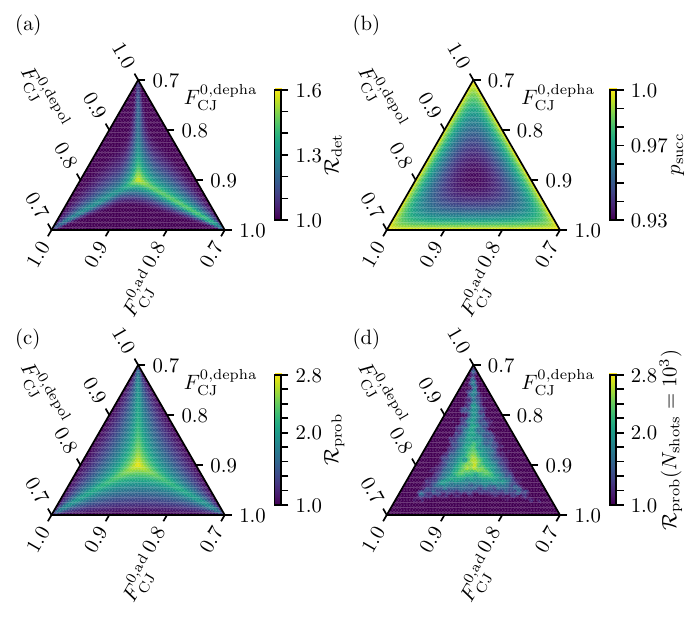}
    \caption{Two-node protocol for $d=3$ paths with dephasing, depolarizing and amplitude damping noises, respectively. The vacuum interference operators $F_{\mathrm{vio}}$ of all three channels are given by the microscopic model in App.~\ref{app:microscopic_special}. The sum of the CJ fidelities of all channels is $2.7$, and there is no path DOF noise ($\sigma = 0$). For the deterministic variant, panel (a) shows the optimal $\mathcal{R}_{\mathrm{det}}$ as a function of the three individual CJ fidelities $F^{0,\mathrm{depha}}_{\mathrm{CJ}}$, $F^{0,\mathrm{depol}}_{\mathrm{CJ}}$ and $F^{0,\mathrm{ad}}_{\mathrm{CJ}}$. For the probabilistic variant, the optimal $p_{\mathrm{succ}}$ and $\mathcal{R}_{\mathrm{prob}}$ are shown in panels (b) and (c) {\done for the $N_{\mathrm{shots}}\to \infty$ case; we further show $\mathcal{R}_{\mathrm{prob}}$ for a finite number of measurements per iteration, $N_{\mathrm{shots}}=10^3$.}}
    \label{fig:three_ineq_channels}
\end{figure}

The ternary plots in Fig.~\ref{fig:three_ineq_channels} show the optimal (a) $\mathcal{R}_{\mathrm{det}}$, (b) $p_{\mathrm{succ}}$ and (c) $\mathcal{R}_{\mathrm{prob}}$ within the triangular cross section~\cite{mpltern}. Remarkably, in the middle of the ternary plots where all three individual CJ fidelities are equal $F^{0,\mathrm{depha}}_{\mathrm{CJ}}= F^{0,\mathrm{depol}}_{\mathrm{CJ}}=F^{0,\mathrm{ad}}_{\mathrm{CJ}}=0.9$, all plotted quantities reach their extremum values: $\mathcal{R}_{\mathrm{det}}$ and $\mathcal{R}_{\mathrm{prob}}$ achieve their maxima and $p_{\mathrm{succ}}$ finds its minimum. This mirrors the $d=2$ results in Fig.~\ref{fig:two_ineq_channels}, and strongly hints at the general result that our protocol is most effective when all paths joining the superposition have comparable noise strengths. Away from the center, $\mathcal{R}_{\mathrm{det}}$ and $\mathcal{R}_{\mathrm{prob}}$ also show distinct ridge features where one of the three paths is very noisy and the other two have equal noise strengths; inspecting the optimal path DOF unitaries on these ridges close to the vertices of the triangular cross section, we find the VQO favors an equal superposition of the two less noisy paths, while avoiding the single noisiest path. Finally, on the edges of the triangle where one of the three paths is almost noiseless, the optimal path superposition is predictably dominated by this noiseless path, such that $\mathcal{R}_{\mathrm{det}}$, $p_{\mathrm{succ}}$ and $\mathcal{R}_{\mathrm{prob}}$ all tend to $1$, similar to Fig.~\ref{fig:two_ineq_channels}.

Comparing the deterministic and the probabilistic variants, we again find a maximum $\mathcal{R}_{\mathrm{det}}$ that is lower than the maximum $\mathcal{R}_{\mathrm{prob}}$. In the chosen cross section, the maximum $\mathcal{R}_{\mathrm{det}}$ and the maximum $\mathcal{R}_{\mathrm{prob}}$ are $1.589$ and $2.795$, respectively; in the weak noise limit $F^{0,\mathrm{depha}}_{\mathrm{CJ}}+ F^{0,\mathrm{depol}}_{\mathrm{CJ}}+F^{0,\mathrm{ad}}_{\mathrm{CJ}}\to 3$, these values become $1.636$ and $3$. The latter value further confirms that $\mathcal{R}_{\mathrm{prob}}\to d$ in the weak noise limit when all $d$ paths have equal $F^{0}_{\mathrm{CJ}}$ and almost perfect vacuum coherence (see App.~\ref{app:a1_prob_equal}). In addition, $\mathcal{R}_{\mathrm{prob}}$ has a rather symmetric dependence on the individual $F^{0}_{\mathrm{CJ}}$ as in the $d=2$ case. In contrast, $\mathcal{R}_{\mathrm{det}}$ is much less symmetric in comparison; as shown by the bottom third of Fig.~\ref{fig:three_ineq_channels}(a), it is difficult to deterministically improve a less noisy dephasing channel using noisier depolarizing and amplitude damping channels.

{\done Finally, in Fig.~\ref{fig:three_ineq_channels}(d), we examine the performance of the probabilistic variant of our protocol with a finite number of measurements $N_{\mathrm{shots}}=10^3$ per iteration. A finite $N_{\mathrm{shots}}$ results in statistical noise in the estimation of the cost function. Here we choose $\nu=0.909$ in the weighted cost function Eq.~\eqref{eq:cost_function}, the COBYLA method~\cite{COBYLA,COBYLA_modern,2020SciPy-NMeth} for the optimizer, and an equal superposition of all three paths as initial guess. This is a reasonable starting point since we are interested in utilizing as many paths in coherent superposition as possible. For each channel configuration in the triangular cross section, we perform one single optimization run using $N_{\mathrm{shots}}=10^3$ per iteration, then calculate the $\mathcal{R}_{\mathrm{prob}}$ and $p_{\mathrm{succ}}$ corresponding to the optimized variational parameters in the $N_{\mathrm{shots}}\to \infty$ limit to characterize the protocol's performance. Only $\mathcal{R}_{\mathrm{prob}}$ is plotted, since the behavior of $p_{\mathrm{succ}}$ is qualitatively similar to Fig.~\ref{fig:three_ineq_channels}(b).

For all channel configurations, the optimization always converges within $50$ iterations, such that the total measurement budget never exceeds $5\times 10^4$ Bell pairs. The advantage of the probabilistic scheme observed in the $N_{\mathrm{shots}}\to\infty$ limit largely persists for $N_{\mathrm{shots}}=10^3$ at the center of the triangular cross section, where the noise strengths in all paths are comparable. This is expected since our initial guess is an equal superposition. Interestingly, at locations very close to the vertices of the triangle, the optimizer does not always succeed in finding the optimal solution of the equal superposition of the two less noisy paths, and may resort to using only one of these two paths. This behavior stems from our minimal implementation of the path unitary, Eq.~\eqref{eq:min_unitary}, where the first path is treated differently from the other two paths. 

In Fig.~\ref{fig:three_ineq_channels}(d), we have adopted a path unitary implementation such that the first path is associated with the dephasing noise; correspondingly, the optimizer is more successful in the vicinity of the upper vertex $(F^{0,\mathrm{depha}}_{\mathrm{CJ}}, F^{0,\mathrm{depol}}_{\mathrm{CJ}}, F^{0,\mathrm{ad}}_{\mathrm{CJ}})\approx(0.7, 1, 1)$ than the other two vertices, with the infidelity ratio along the line $F^{0,\mathrm{depol}}_{\mathrm{CJ}}=F^{0,\mathrm{ad}}_{\mathrm{CJ}}$ approaching the $N_{\mathrm{shots}}\to\infty$ value of $2$. We also note the parameterization $U(\boldsymbol{\theta})= T_{1,3}(\theta_3) T_{2,3}(\theta_2)T_{1,2}(\theta_1)$ with one more variational parameter $\theta_{3}$ for each of $\boldsymbol{\theta}_{\mathrm{c}1}$ and $\boldsymbol{\theta}_{\mathrm{c}2}$ allows us to explore the immediate vicinity of all three vertices on equal footing, although the price is a larger total measurement budget (the maximum number of iterations required is $75$ rather than $50$). Overall, Fig.~\ref{fig:three_ineq_channels}(d) demonstrates the ability of our protocol to gain an advantage in the simple two-node scenario, even in the presence of statistical noise.} 


The results presented in Figs.~\ref{fig:two_ineq_channels} and \ref{fig:three_ineq_channels} support our observation that the probabilistic variant is more robust than the deterministic one for asymmetric noise types and strengths, and {\done further highlight the probabilistic variant’s resilience to both path DOF noise and statistical noise}.

\begin{figure}
{\includegraphics[width=\columnwidth]{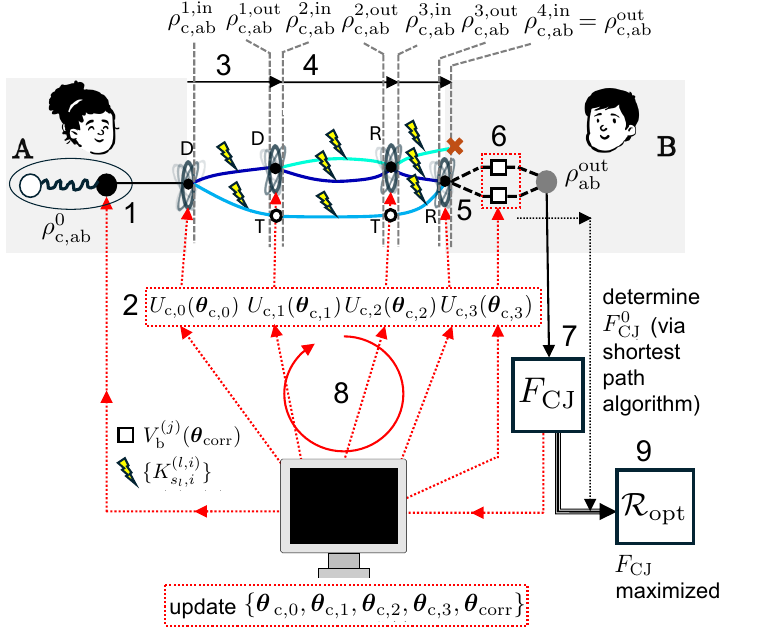}}
\caption{Schematic of the multi-node protocol (Protocol~\ref{table:multi_node_protocol_summary}). D, R, and T label dividing, recombining, and transit nodes respectively. All red lines/boxes involve the VQO algorithm. Orange dotted lines delineate the end of a stage $l$ and start of a new stage $l+1$ corresponding to intermediate states $\rho^{l,\mathrm{out}}_{\mathrm{c, ab}}$ and $\rho^{l+1, \mathrm{in}}_{\mathrm{c, ab}}$ respectively. Note that $\rho^{\mathrm{0, \mathrm{in}}}_{\mathrm{c, ab}} = \rho_{\mathrm{ab}}^{\mathrm{in}}$ and $\rho^{\mathrm{0, \mathrm{out}}}_{\mathrm{c, ab}} = \rho^{\mathrm{0}}_{\mathrm{c, ab}}$. `X' denotes paths that are disconnected from Bob and thus not postselected.  
\label{fig:vqo_multi_node}}
\end{figure}

\section{Multi-node network protocol}
\label{sec:multi_node_protocol}
To demonstrate the full capabilities and intent of our protocol, we move on to more complicated \emph{multi-node} scenarios that involve intermediate nodes and permit nested path superpositions. These larger-scale scenarios demonstrate the VQO’s flexibility in adapting to extended topologies, as well as its ability to maintain advantages even when each path is subjected to independently randomized noise.
\subsection{Protocol description}
\label{sec:multi_node_protocol_desc}
\label{subsec:multi_node}The multi-node protocol (shown schematically in Fig.~\ref{fig:vqo_multi_node}) proceeds similarly to the two-node protocol in Sec.~\ref{subsec:two_node}, with some key differences. It is summarized as Protocol~\ref{table:multi_node_protocol_summary} and described below. 

\textit{Step 0} -- Identify multiple non-interacting paths from Alice to Bob. These paths may now divide and recombine at intermediate nodes (not only at Alice and Bob). At dividing nodes, a path splits and interferes with vacuum; at recombining nodes, multiple paths interfere to form new ones (note some do not reach Bob and are discarded). The network is partitioned into $S+1$ stages ($l=0,1,2,\dots,S$), separated by dividing or recombining nodes. In stage $l$, the system contains a superposition of $d_{l}$ path segments, with $d_{0} = 1$ and $d_{l} \neq d_{l+1}$ in general. 


 When transitioning between stages, we add virtual ``transit'' nodes on unaffected paths. These nodes do not change the physics of the paths they lie on, but they enable bookkeeping of the path state in Steps 2--3.

\textit{Step 1} -- Prepare a input Bell state $\rho^{\mathrm{in}}_{\mathrm{ab}}=\ket{\Phi^{+}}_{\mathrm{ab}}\bra{\Phi^{+}}_{\mathrm{ab}}$ at Alice, giving initial state $\rho^{0}_{\mathrm{c,ab}}=\ket{0}_{\mathrm{c}}\bra{0}_{\mathrm{c}} \otimes \rho^{\mathrm{in}}_{\mathrm{ab}}$.

\textit{Step 2} -- At each dividing/recombining node in stage $l$, apply a parameterized path DOF unitary $U^{(l)}_{\mathrm{c}}$. It plays the roles of $U_{\mathrm{c}1}$ in Step 2 and $U_{\mathrm{c}2}$ in Step 4 of the two-node protocol. After projecting out any paths disconnected from Bob, the resulting state corresponds to the start of stage $l+1$:
\begin{equation}
    \label{eq:uc_l}
    \rho^{l+1,\mathrm{in}}_{\mathrm{c,ab}}=\Pi^{(l)}_{\mathrm{c}}U^{(l)}_{\mathrm{c}}\rho^{l,\mathrm{out}}_{\mathrm{c,ab}}U^{(l)\dagger}_{\mathrm{c}}\Pi^{(l)}_{\mathrm{c}}
\end{equation}
where $\rho^{l,\mathrm{in}}_{\mathrm{c,ab}}$ ($\rho^{l,\mathrm{out}}_{\mathrm{c,ab}}$) denotes the beginning (end) of stage $l$ and $\rho^{0,\mathrm{out}}_{\mathrm{c,ab}}=\rho^{0}_{\mathrm{c,ab}}$. 

The projector $\Pi^{(l)}_{\mathrm{c}}$ removes disconnected paths at each stage transition, and equals $\id$ if all paths remain connected. Note that this occurs at $l = 0$ (since $d_{1} > d_{0}$) and at Bob's node itself, making it distinct from the later path DOF measurement at Bob in Step 5. 
This projection is unique to the multi-node protocol, since it is not guaranteed that all paths directly connect Alice to Bob. It reflects the fact that any disconnected path is permanently lost to Bob and prevents us from postselecting the corresponding measurement outcome (see Step 7 below). Note that we are not required to discard any paths at an intermediate node; this decision is part of the underlying network topology, which ultimately depends on the available hardware.

One way to build $U^{(l)}_{\mathrm{c}}$ is to embed general parameterized $\mathrm{U}(2)$ matrices into a $\max\{d_l, d_{l+1}\}$-dimensional identity matrix, akin to two-mode optical beamsplitter decompositions~\cite{Reck1994, Clements2016},

\begin{equation}
\label{eq:2_mode_unry}
U(\alpha,\beta,\gamma,\theta) = 
\begin{pmatrix}
e^{\mathrm{i}(\alpha+\gamma)}\cos \theta & e^{\mathrm{i}(\beta+\gamma)}\sin \theta \\
-e^{\mathrm{i}(-\beta+\gamma)}\sin \theta & e^{\mathrm{i}(-\alpha+\gamma)}\cos \theta
\end{pmatrix}.
\end{equation}
Specifically, $U$ is embedded in each $2\times 2$ subspace affected by the dividing and/or recombining nodes separating stages $l$ and $(l+1)$, where $\cos^{2}\theta$ and $\sin^{2} \theta$ are the transmitted and reflected fractions for each incident beam respectively, $\alpha$ and $\beta$ denote the corresponding relative phases, and $\gamma$ is a phase relative to the remaining paths \footnote{A superposition of more than two paths can be realized through successive pairwise splitting in a tree-like structure, as in Fig.~\ref{fig:vqo_multi_node}. Alternatively, we can directly embed higher-dimensional parameterized unitary matrices in relevant subspaces.}. If $U^{(l)}_{\mathrm{c}}$ is affected by noise acting directly on the path DOF, $U^{(l)}_{\mathrm{c}}\rho^{l,\mathrm{out}}_{\mathrm{c,ab}}U^{(l)\dagger}_{\mathrm{c}} \mapsto \mathcal{C}^{(l)} (\rho^{l, \mathrm{out}}_{\mathrm{c, ab}})$, as given by e.g., Eq.~\eqref{eq:control_noise}.

\textit{Step 3} -- Transmit the latter half $\rho^{\mathrm{in}}_{\mathrm{ab}}$ through the noisy path segments until the next stage is encountered. For each path segment in stage $l$, diagonal blocks evolve as  
\begin{equation}
\rho^{l,\mathrm{out}(ii)}_{\mathrm{c,ab}}=\sum_{s_{l,i} }K^{(l,i)}_{s_{l,i}} \rho^{l,\mathrm{in}(ii)}_{\mathrm{c,ab}}K_{s_{l,i}}^{(l,i)\dagger},
\label{eq:incoh_part_multi}
\end{equation}
while off-diagonal blocks evolve as
\begin{equation}
\rho^{l,\mathrm{out}(ij)}_{\mathrm{c,ab}}=F^{(l,i)}_{\mathrm{vio}}\rho^{l,\mathrm{in}(ij)}_{\mathrm{c,ab}}F^{(l,j)\dagger}_{\mathrm{vio}} \ \ (i\neq j),
\label{eq:coh_part_multi}
\end{equation}
where $i,j\in \{0,1,\dots,d_l-1\}$, by direct analogy with Eqs.~\eqref{eq:supp_incoh_p_coh}--\eqref{eq:vio}. The $i$-th path segment in stage $l$ is characterized by the Kraus operators $K^{(l,i)}_{s_{l,i}}$ and the vacuum amplitudes $\alpha^{(l,i)}_{s_{l,i}}$. Correspondingly, the vacuum interference operators are:
\begin{equation}
F^{(l,i)}_{\mathrm{vio}}=\sum_{s_{l,i}}\alpha^{(l,i)*}_{s_{l,i}} K^{(l,i)}_{s_{l,i}}.
\label{eq:vio_multi} 
\end{equation}

\textit{Step 4} -- Repeat Steps 2--3 until the final output $\rho^{\mathrm{out}}_{\mathrm{c,ab}}=\rho^{S,\mathrm{in}}_{\mathrm{c,ab}}$ is obtained, representing the transmitted state at Bob. We denote the collection of all path DOF unitary parameters ($\alpha,\beta,\gamma$ and $\theta$ in Eq.~\eqref{eq:2_mode_unry}) acting at the end of stage $l$ by $\boldsymbol{\theta}_{\mathrm{c},l}$.

\textit{Step 5} -- Measure the path DOF in the $\ket{j}_{\mathrm{c}}$ basis ($j=0,1,\dots,d_S-1$). Analogous to Eq.~\eqref{eq:postmeas_st}, the post-measurement state for outcome $j$ is
\begin{equation}
\rho^{\mathrm{out}}_{\mathrm{ab}, j} = \frac{1}{P_j}\operatorname{tr}_{\mathrm{c}}\big(\ket{j}_{\mathrm{c}}\bra{j}_{\mathrm{c}}\rho_{\mathrm{c, ab}}^{\mathrm{out}}\big),
\label{eq:postmeas_st_multi}
\end{equation}
where $P_{j} = \operatorname{tr}\big(\ket{j}_{\mathrm{c}}\bra{j}_{\mathrm{c}}\rho_{\mathrm{c, ab}}^{\mathrm{out}}\big)$. The sum $\sum_j P_j =\operatorname{tr} \rho^{\mathrm{out}}_{\mathrm{c,ab}} \leq 1$ reflects transmission losses from discarded paths (if any). 

\textit{Step 6} -- Apply single-qubit unitary corrections $V_{\mathrm{b}}^{(j)}$ (parameterized by $\boldsymbol{\theta}_{\mathrm{corr}}$) to each output state $\rho^{\mathrm{out}}_{\mathrm{ab}, j}$, yielding $\rho^{\mathrm{out,corr}}_{\mathrm{ab}, j}$ as given by Eq.~\eqref{eq:u3_corr}.

\textit{Step 7} -- Measure the CJ fidelity of the output $F_{\mathrm{CJ}}$, either quasi-deterministically via Eq.~\eqref{eq:cj_fid_det}, or probabilistically via  Eq.~\eqref{eq:cj_fid_prob}. While Eq.~\eqref{eq:cj_fid_det} is deterministic in the two-node protocol, the potential losses at intermediate nodes mean that the multi-node protocol has a success probability $p_{\mathrm{succ}} \leq \operatorname{tr} \rho^{\mathrm{out}}_{\mathrm{c,ab}}$ where equality corresponds to keeping all paths connected to Bob.

\textit{Step 8} -- Repeat Steps 1--7 iteratively, optimizing all path and correction unitary parameters to find $\{\boldsymbol\theta_{\mathrm{c, 0}}, \boldsymbol\theta_{\mathrm{c, 1}},\dots, \boldsymbol\theta_{\mathrm{c}, S-1}, \boldsymbol\theta_{\mathrm{corr}}\}_{\mathrm{opt}}$ that maximize $F_{\mathrm{CJ}}$ {\done or minimize the weighted cost function in Eq.~\eqref{eq:cost_function}}. 

\textit{Step 9} -- As in the two-node protocol we characterize the protocol performance in our numerical simulations by calculating the infidelity ratio $\mathcal{R}$ in Eq.~\eqref{eq:R_ratio}; this requires a baseline CJ fidelity $F^{0}_{\mathrm{CJ}}$ without path superposition. For multiple nodes, $F^{0}_{\mathrm{CJ}}$ is again chosen to be the highest CJ fidelity of any path between Alice and Bob,
\begin{equation}
F_{\mathrm{CJ}}^{0} = \max_{p} F_{\mathrm{CJ}}^{(p)},
\label{eq:incoh_fid_multi}
\end{equation}
where $F_{\mathrm{CJ}}^{(p)}$ is the CJ fidelity of path $p$, and we consider two paths to be distinct as long as they do not overlap completely. In the weak noise limit, we may approximate $1-F_{\mathrm{CJ}}^{(p)}$ as:
\begin{equation}
1-F_{\mathrm{CJ}}^{(p)} \approx \prod_{s = 1}^{S-1} (1-F_{\mathrm{CJ}}^{(s, p)}) \hspace{0.1cm} \mathrm{for} \hspace{0.1cm} 1-F_{\mathrm{CJ}}^{(p)} \ll 1
\label{eq:incoh_infid_path_multi}
\end{equation}
for all segments $s$ along path $p$. In numerical simulations, since the density matrix of the system can be computed sequentially along a given path according to Eq.~\eqref{eq:channel}, we can efficiently maximize $F_{\mathrm{CJ}}^{(p)}$ via a modified Dijkstra algorithm~\cite{dijkstra} that proceeds through the network stage by stage.

\begin{table}
{\LinesNumberedHidden
    \begin{algorithm}[H]
        \SetKwInOut{Input}{Input}
        \SetKwInOut{Output}{Output}
        \SetAlgorithmName{Protocol}{}
    \noident
       \justifying \begin{enumerate}[leftmargin=*,itemindent=0pt]
        \setcounter{enumi}{-1}
          \item (I) Identify multiple non-interacting noisy paths containing dividing and recombining nodes from Alice (`a') to Bob (`b'). Divide the network into $S+1$ stages ($l=0,1,2,\dots,S$) delimited by these nodes. 

           \item (A) At stage $l=0$, prepare the Bell state $\rho_{\mathrm{ab}}^{\mathrm{in}} = \ket{\Phi^{+}}_{\mathrm{ab}}\bra{\Phi^{+}}_{\mathrm{ab}}$ in path 0.

           \item (A/TR) Evolve the state from stage $l$ to $l+1$ by applying the unitary $U^{l}_{\mathrm{c}}$ to the path DOF (system `c') followed by a projector $\Pi_{\mathrm{c}}^{(l)}$ which removes any paths that no longer reaches Bob after stage $l$. If no such path exists (e.g., for $l=0$), $\Pi_{\mathrm{c}}^{(l)}=\id$.  
        
            \item (TR) Within stage $l$, transmit system `b' of $\rho_{\mathrm{ab}}^{\mathrm{in}}$ through the $d_{l}$ superposed path segments, with noise described by $s$ Kraus operators $\{K_{s_{l,i}}^{(l, i)}$\} acting on each path segment $(l, i)$.

            \item (TR) Repeat Steps 3-4 until the final output $\rho^{\mathrm{out}}_{\mathrm{c, ab}}$ is obtained at $l = S$. 
            
            \item (B) Measure the path DOF in the $\ket{j}_{\mathrm{c}}$ basis ($j = 0, 1, \dots, d_S-1$).

            \item (B) Apply single-qubit correcting unitaries $V^{(j)}_{\mathrm{b}}$ to each of the $d$ postmeasured states $\rho^{\mathrm{out}}_{\mathrm{ab}, j}$.

            \item (O) Measure the fidelity $F_{\mathrm{CJ}}$ either quasi-deterministically (average on all outcomes accessible to Bob) or probabilistically (on the most likely outcome).

            \item (F) Repeat Steps 1--7 iteratively, updating all unitary parameters until $F_{\mathrm{CJ}}$ is maximized. 

            \item (O) Compute the infidelity ratio $\mathcal{R}$ (Eq.~\eqref{eq:R_ratio}), based on $\mathrm{max}(F_{\mathrm{CJ}})$ and the single-path (incoherent) fidelity $F^{0}_{\mathrm{CJ}}$. 
        \end{enumerate}
    
   \justifying \textit{Legend}: I--protocol input, A--step at Alice, TR--transmission step (through intermediate nodes), B--step at Bob, O--protocol output, F--feedback loop 
    
\caption{Self-configuring protocol, multi-node network (Fig.~\ref{fig:vqo_multi_node})} \label{table:multi_node_protocol_summary}
\end{algorithm}}
\end{table}

\subsection{Performance analysis}
\label{sec:numerics_multinode}

We now simulate the multi-node protocol in two networks involving intermediate nodes and nested superpositions. Specifically, we first explain the basics of the protocol in a simple network, which can be reduced to the two-node case in certain limits. We then showcase the performance of our protocol in a much larger network, where both the type and the strength of the noise in each path segment are randomly sampled. Our approach aims to demonstrate that VQO reliably achieves noise mitigation for a wide variety of experimental constraints. For simplicity, we assume no path DOF noise in this section ($\sigma=0$). This assumption is often justified in realistic quantum communication setups using photons as information carriers~\cite{micuda14}.

\begin{figure}
\centering
{\includegraphics[width=\linewidth]{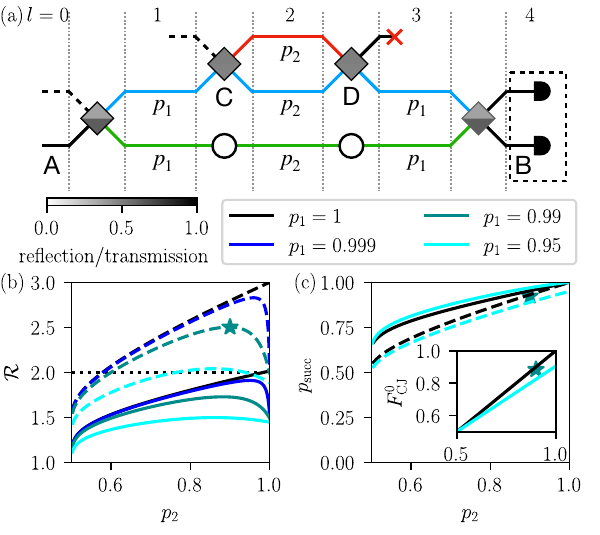}}
\caption{
\label{fig:multinode_simple} 
Multi-node protocol for a simple $4$-node network (with two transit nodes) featuring a nested superposition. (a) Sketch of the network, with its five stages ($l=0,1,\dots,4$) separated by dotted lines. Diamonds indicate dividing nodes (Alice and Charlie) and recombination nodes (Bob and David), and circles indicate transit nodes. Thick dashed lines indicate vacuum inputs, solid black lines indicate input and output paths, and red, blue and green lines indicate path segments with amplitude damping, dephasing and depolarizing noise, respectively. The output path at David is always discarded, whereas we can choose to retain one or two output paths at Bob, depending on whether the protocol is probabilistic or quasi-deterministic. The incoherent CJ fidelity of each path segment (assumed to be either $p_1$ or $p_2$) is given below the segment. Also shown inside the diamonds are the optimal reflection (upper half) and transmission (lower half) probabilities of the path DOF unitary at each node, valid for the probabilistic variant with $p_1=0.99,p_2=0.9$ (star symbol in (b) and (c)). (b)--(c) $\mathcal{R}$ and $p_{\mathrm{succ}}$ of the quasi-deterministic variant (solid lines) and the probabilistic variant (dashed lines) as functions of $p_2$ for varying $p_1$ {\done in the $N_{\mathrm{shots}}\to \infty$ limit}. The inset of (c) shows the incoherent CJ fidelity of the entire network $F^{0}_{\mathrm{CJ}}$. $p_{\mathrm{succ}}$ and $F^{0}_{\mathrm{CJ}}$ for $p_1=0.999$ and $p_1=0.99$ are not explicitly shown since they visually overlap with those for $p_1=1$.}

\end{figure}

\subsubsection{Basic connectivity}

In this section, we consider the simple network configuration shown in Fig.~\ref{fig:multinode_simple}(a), which contains a single nested superposition. It consists of $5$ stages ($S=4$, with $d_1=d_3=d_4=2$, $d_2=3$), with $2$ dividing nodes (Alice and Charlie) and $2$ recombination nodes (David and Bob); we have also added two virtual transit nodes for a convenient mathematical description. One of the two path segments (channels) emerging from David eventually reaches Bob, while the other is immediately discarded. The path segments directly connecting Alice with Bob are subject to depolarizing noise (green), one segment connecting Charlie with David is an amplitude damping channel (red), and the remaining segments are dephasing channels (blue). For simplicity, we assume that all $4$ path segments in stages $1$ and $3$ have CJ fidelity $p_1$, and all $3$ path segments in stage 2 are characterized by the CJ fidelity $p_2$. In the limit of $p_2= 1$, we can always choose the path unitaries at Charlie and David to be identity, thus recovering a simple $d=2$ two-node network with one dephasing channel and one depolarizing channel. In the limit of $p_1= 1$, we instead find a $d=3$ two-node network subject to the constraint of a discarded path.

In the current topology, since one path is discarded at David before reaching Bob, as explained in Sec.~\ref{subsec:multi_node}, we do not have a fully deterministic variant for this network. Instead, we can keep either both of the paths at Bob, or only the path with the higher $p_{\mathrm{succ}}$. We refer to these {\done two versions of the protocol} as quasi-deterministic and probabilistic, respectively.

Fig.~\ref{fig:multinode_simple}(b) and (c) show how the optimization results $\mathcal{R}$ and $p_{\mathrm{succ}}$ depend on the chosen protocol and the segment fidelities $p_1$ and $p_2$ {\done in the $N_{\mathrm{shots}}\to \infty$ limit}. Following the general trend in the two-node examples, the probabilistic variant has higher $\mathcal{R}$ and lower $p_{\mathrm{succ}}$ compared to the quasi-deterministic variant. We have also chosen a point in the parameter space $p_1=0.99$, $p_2=0.9$ (star symbol in panels (b) and (c)), and shaded each node in panel (a) by its corresponding reflection (top) and transmission (bottom) probabilities of the optimal solution found by the VQO. For this solution, Charlie and David function approximately as 50/50 beamsplitters, while the path DOF unitaries at Alice and Bob have a higher weight on the Charlie/David side, in order to exploit the nested path superposition between Charlie and David. For completeness, we have also plotted in the inset of Fig.~\ref{fig:multinode_simple}(c) the best single-path (incoherent baseline) CJ fidelity $F^{0}_{\mathrm{CJ}}$ of the network as a function of $p_2$; in this case, the single-path CJ fidelity is identical for all three paths from Alice to Bob.

We first discuss the limit of $p_1= 1$, where $\mathcal{R}$ increases monotonically with $p_2$. In particular, for $p_2 \to 1$, $\mathcal{R}_{\mathrm{prob}}\to 3$ in the probabilistic variant, corresponding to having three paths with identical CJ fidelity, while $\mathcal{R}_{\mathrm{det}}\to 2.015$ in the quasi-deterministic variant. This quasi-deterministic limiting value is still higher than the limiting value $\mathcal{R}_{\mathrm{det}}\to 1.636$ of a fully deterministic variant (see Sec.~\ref{subsec:non_identical_chs_3}), which would correspond to a modified network where David's currently discarded path is retained and rerouted to Bob instead.

Curiously, even if $p_1$ is reduced slightly from $1$, both $\mathcal{R}_{\mathrm{det}}$ and $\mathcal{R}_{\mathrm{prob}}$ are drastically reduced for $p_2$ close to $1$. This behavior is reminiscent of the effect of the path DOF noise discussed in Sec.~\ref{sec:identical_chs_micro}. Indeed, when $p_1$ is close to $1$, we can think of stage $1$ ($3$) as part of a noisy $d=3$ path DOF unitary spanning Alice and Charlie (David and Bob). However, this analogy is not perfect because, unlike the path DOF noise $\sigma$, the deviation of $p_1$ from $1$ is taken into account in the incoherent fidelity $F^{0}_{\mathrm{CJ}}$ of the network and thus also in the $\mathcal{R}_{\mathrm{det}}$ and $\mathcal{R}_{\mathrm{prob}}$. In the present case, if we take $p_2=1$, the quasi-deterministic variant becomes fully deterministic and yields $\mathcal{R}_{\mathrm{det}}\to 1.5$ (see Sec.~\ref{subsec:non_identical_chs_2}) while the probabilistic variant again gives $\mathcal{R}_{\mathrm{det}}\to 2$ when $p_1\to 1$. Therefore, for either protocol, with a fixed $p_1\neq 1$, $\mathcal{R}$ attains its maximum for some $p_2<1$ as demonstrated in Fig.~\ref{fig:multinode_simple}(b). That the limits $p_1\to 1$ and $p_2\to 1$ do not commute for the infidelity ratio $\mathcal{R}$ is a result of the nontrivial topology of the network. In comparison, $p_{\mathrm{succ}}$ depends slowly on both $p_1$ and $p_2$, as is evident from Fig.~\ref{fig:multinode_simple}(c); this is similar to the behavior of $p_{\mathrm{succ}}$ in the two-node examples.

Let us also briefly comment on the behavior at a fixed value of $p_2$. For either the deterministic or the probabilistic variant, $\mathcal{R}$ increases as $p_1$ approaches $1$, again in agreement with the picture of the effective noisy $d=3$ path DOF unitaries. Interestingly, as $p_1$ increases, $p_{\mathrm{succ}}$ decreases in the quasi-deterministic variant, but increases in the probabilistic variant.

\subsubsection{Complex connectivity}
\label{sec:multinode_complex}

\begin{figure*}[t]
{\includegraphics[width=\textwidth]{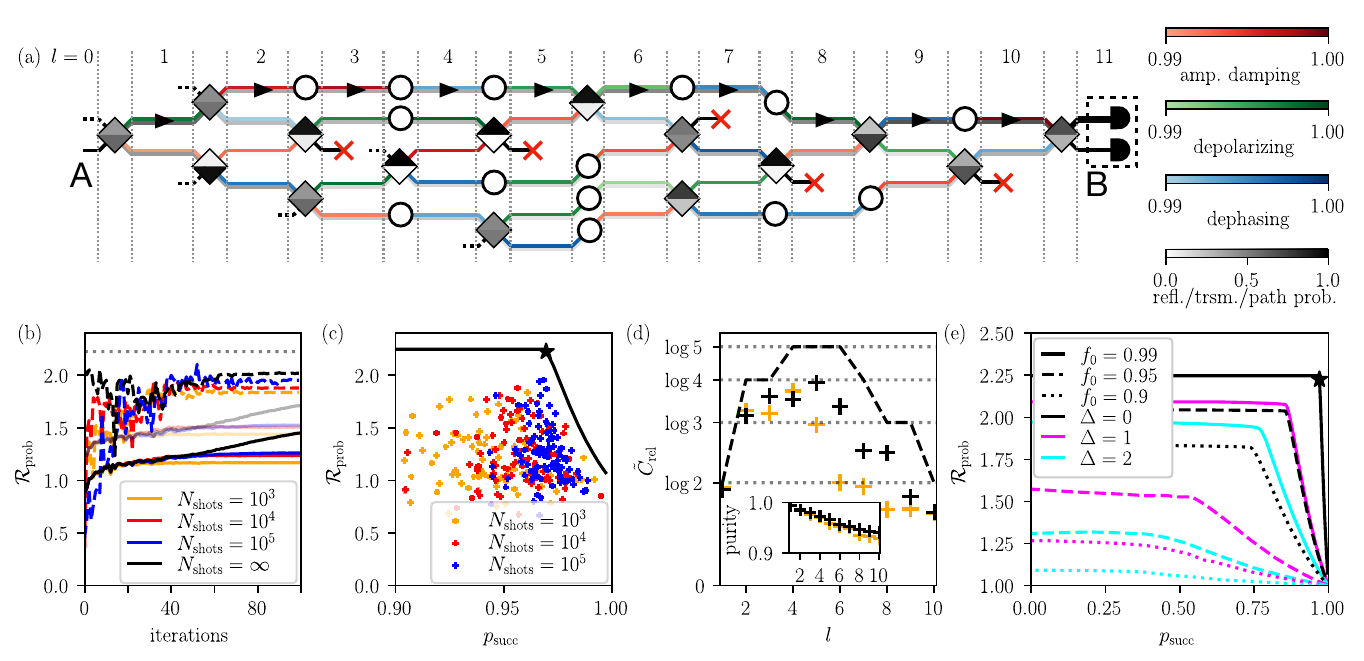}}
\caption{
\label{fig:multinode_complex}
Multi-node protocol for a $12$-stage network with $6$ dividing nodes, $9$ recombination nodes and $14$ transit nodes. (a) Sketch of the network. Components in the figure are similar to those in Fig.~\ref{fig:multinode_simple}; however, we now color-code each path segment according to both the noise type (amplitude damping, dephasing or depolarizing) and the incoherent CJ fidelity (the shade of color), which is for all noise types sampled from a uniform distribution between $0.99$ and $1$. The least noisy single path is marked by arrows. {\done We additionally show the optimal VQO solution for the probabilistic variant in the $N_{\mathrm{shots}}\to \infty$ limit, including the reflection and transmission probabilities at the parameterized nodes, and also the probability of each segment being traversed within that stage (grayscale). (b) Average probabilistic variant infidelity ratio $\bar{\mathcal{R}}_{\mathrm{prob}}$ (solid lines) and 1-sigma-above-average level $\bar{\mathcal{R}}_{\mathrm{prob}}+\sigma_{\mathcal{R}}$ (opaque lines) of $100$ optimization runs with distinct random initial guesses, as functions of the number of iterations for different values of $N_{\mathrm{shots}}$. Dashed lines are optimization runs which yield the highest $\mathcal{R}_{\mathrm{prob}}$ upon convergence or (in the $N_{\mathrm{shots}} \to \infty$ case) after $200$ iterations. The dotted line marks the maximum attainable $\mathcal{R}_{\mathrm{prob}}=2.226$ for $N_{\mathrm{shots}} \to \infty$. We assume one parameter per dividing/recombining node ($\alpha=\beta=\gamma=0$ in Eq.~\eqref{eq:2_mode_unry}), and set $\nu=0.909$ in the cost function Eq.~\eqref{eq:cost_function}. (c) Distributions of the final values of $p_{\mathrm{succ}}$ and $\mathcal{R}_{\mathrm{prob}}$ for the same $100$ optimization runs for each value of $N_{\mathrm{shots}}$. In addition, we show the trajectory of the optimal $p_{\mathrm{succ}}$ and $\mathcal{R}_{\mathrm{prob}}$ (i.e., the Pareto front) in the $N_{\mathrm{shots}} \to \infty$ limit obtained by smoothly varying $\nu$, with the $\nu=0.909$ values $p_{\mathrm{succ}}=0.970$ and $\mathcal{R}_{\mathrm{prob}}=2.226$ marked by a star symbol. (d) Relative entropy of coherence, $\tilde{C}_{\mathrm{rel}}$, and purity of reduced density matrix (inset) as functions of the stage index $l$ in the optimal solutions for $N_{\mathrm{shots}} \to \infty$ (black) and $N_{\mathrm{shots}}=10^3$ (orange). We also show $\log d_l $ (dashed), which is the upper bound of $\tilde{C}_{\mathrm{rel}}$. (e) Pareto fronts of $p_{\mathrm{succ}}$ and $\mathcal{R}_{\mathrm{prob}}$ for different noise strengths and vacuum coherence levels. The noise strength is measured by $f_0$, where we rescale the infidelities of all path segments in (a) by $(1-f_0)/0.01$. The vacuum coherence is quantified by the tuning parameter $\Delta \geq 0$ in the generic microscopic model in App.~\ref{app:microscopic_general}, with larger $\Delta$ corresponding to decreasing vacuum coherence. The $f_0=0.99$ (i.e., no rescaling), $\Delta=0$ Pareto front and the star symbol at $\nu=0.909$ are the same as in (b).}}
\end{figure*}

As another demonstration of the adaptability of the multi-node protocol, we study the complicated network depicted in Fig.~\ref{fig:multinode_complex}(a). It allows multiple nested superpositions, with $12$ stages ($S=11$), $6$ dividing nodes, $9$ recombination nodes and $14$ transit nodes~\footnote{Here, the transit nodes are physically meaningful and no longer ``virtual'' when separating different noise types.}. Each path segment is again subject to a dephasing, depolarizing or amplitude damping channel modeled by the microscopic model in {\done App.~\ref{app:microscopic}.}

We begin with the weak noise regime, where noise mitigation becomes highly desirable for quantum communication tasks such as teleportation or entanglement distribution. We randomly sample CJ fidelity values for each segment from a uniform distribution between a baseline fidelity $f_0 = 0.99$ and $1$, and color-code each segment by its noise type and $F^{0}_{\mathrm{CJ}}$ itself (via the shade of color). In the resulting network, the best single-path fidelity $F^{0}_{\mathrm{CJ}}=0.964$ corresponds to the topmost path from Alice to Bob.


Despite the complexity of the network, the VQO is still able to find a solution with nontrivial path superposition that improves on the best single-path fidelity $F^{0}_{\mathrm{CJ}}$. {\done Here we focus on the minimal parameterization scheme, where one parameter $\theta$ is allocated to each dividing/recombining node (we fix $\alpha=\beta=\gamma=0$ in Eq.~\eqref{eq:2_mode_unry}) for a total of $15$ parameters. Again, the minimal parameterization produces the same optimized $\mathcal{R}$ and $p_{\mathrm{succ}}$ as the full parameterization scheme with four variational parameters for each dividing/recombining node, since the noise described by Eqs.~\eqref{eq:dephasingnoise}--\eqref{eq:amplitudedampingnoise} has no coherent component.

Taking $N_{\mathrm{shots}} \in \{10^3$, $10^4$, $10^5$, $\infty\}$ per iteration and a cost function weight $\nu=0.909$, we execute $100$ optimization runs of the probabilistic variant of the protocol, where exactly one path arriving at Bob is postselected. We continue to employ the COBYLA optimizer, with random initial guesses of $\boldsymbol{\theta}_{\mathrm{c}}$ sampled uniformly between $0$ and $2\pi$; the same $100$ sets of initial guesses are used for all values of $N_{\mathrm{shots}}$. In Fig.~\ref{fig:multinode_complex}(b), for each value of $N_{\mathrm{shots}}$, we plot the resulting average $\bar{\mathcal{R}}_{\mathrm{prob}}$ and sample standard deviation $\sigma_{\mathcal{R}}$ of the advantage $\mathcal{R}_{\mathrm{prob}}$ as functions of the number of iterations, together with the ``best'' optimization run of $\mathcal{R}_{\mathrm{prob}}$. The latter refers to the run yielding the highest $\mathcal{R}_{\mathrm{prob}}$ upon convergence (or, in the $N_{\mathrm{shots}} \to \infty$ case, after $200$ iterations). All of the above are upper-bounded by the maximum attainable advantage in the $N_{\mathrm{shots}} \to \infty$ limit, $\mathcal{R}_{\mathrm{prob}}=2.226$, corresponding to $p_{\mathrm{succ}}=0.970$ for $\nu=0.909$ (star symbol in Fig.~\ref{fig:multinode_complex}(c) and (e)). While we do not explicitly show $p_{\mathrm{succ}}$ in Fig.~\ref{fig:multinode_complex}(b), we remark that it almost always converges faster than $\mathcal{R}_{\mathrm{prob}}$.

The effect of statistical noise on the optimization is apparent from Fig.~\ref{fig:multinode_complex}(b). For the finite values of $N_{\mathrm{shots}}$ we have examined, $\mathcal{R}_{\mathrm{prob}}$ generally stops improving after $ < 50$ iterations, and increasing $N_{\mathrm{shots}}$ from $10^3$ to $10^5$ offers little improvement on average, despite $\bar{\mathcal{R}}_{\mathrm{prob}}$ in the $N_{\mathrm{shots}} \to \infty$ limit increasing steadily within the first $100$ iterations. However, for each finite $N_{\mathrm{shots}}$ (e.g. $N_{\mathrm{shots}}=10^4$), both the best-case value and the 1-sigma mark of $\mathcal{R}_{\mathrm{prob}}$ ($1.88$ and $1.50$ respectively) are considerably higher than the average ($1.25$).

To shed more light on the convergence behavior, in Fig.~\ref{fig:multinode_complex}(c), we plot $\mathcal{R}_{\mathrm{prob}}$ versus $p_{\mathrm{succ}}$ at the end of all $100$ optimization runs for each value of $N_{\mathrm{shots}}\in \{10^3$, $10^4$, $10^5\}$. We also plot, in the $N_{\mathrm{shots}} \to \infty$ limit, the trajectory of the optimal values as the weight $\nu$ changes, which creates a Pareto front for the multiobjective optimization of $\mathcal{R}_{\mathrm{prob}}$ and $p_{\mathrm{succ}}$~\cite{myerson2013game}. As already mentioned in Sec.~\ref{subsec:two_node_protocol_desc}, $\mathcal{R}_{\mathrm{prob}}\to 1$ and $p_{\mathrm{succ}}\to 1$ for small values of $\nu$, while $\mathcal{R}_{\mathrm{prob}}$ saturates when $\nu$ is close to $1$. The $(p_{\mathrm{succ}},\mathcal{R}_{\mathrm{prob}})$ values for $\nu=0.909$ are located at a cusp of the Pareto front, approximately maximizing $\mathcal{R}_{\mathrm{prob}}$ without excessively penalizing $p_{\mathrm{succ}}$. As expected, the distribution of $(p_{\mathrm{succ}},\mathcal{R}_{\mathrm{prob}})$ approaches the Pareto front when $N_{\mathrm{shots}}$ increases. However, while $p_{\mathrm{succ}}$ improves significantly, the improvement of $\mathcal{R}_{\mathrm{prob}}$ is marginal, as already suggested by Fig.~\ref{fig:multinode_complex}(b).

Fig.~\ref{fig:multinode_complex}(b) and (c) together indicate the importance of choosing suitable initial guesses for the optimization in the presence of statistical noise, a topic that we discuss further in Sec.~\ref{sec:conclusionss}.} Note that in practice it is often sufficient to have a suboptimal solution with reasonably high $\mathcal{R}$ and $p_{\mathrm{succ}}$, and it is not always necessary to find the true globally optimal solution, which may incur an excessive computational cost.

{\done We can take a more in-depth look at the properties of the optimal VQO solution. For the optimal solution in the $N_{\mathrm{shots}}\to\infty$ limit,} Fig.~\ref{fig:multinode_complex}(a) shows in grayscale the reflection and transmission probabilities at each parameterized node, alongside the probability to find the information carrier in each path segment relative to other segments of the same stage. Six of the dividing and recombination nodes exhibit nearly completely reflecting or perfectly transmitting behavior, whereas the remaining nine parameterized nodes have nonzero reflection and transmission probabilities. In addition, $34$ out of $37$ path segments noticeably participate in the optimal solution, and only three segments (one in each of the stages $l=2,4,7$) see negligible probabilities of being traversed. These suggest that the optimal solution is indeed distributed across multiple paths, as we may expect from the infidelity ratio {\done $\mathcal{R}_{\mathrm{prob}}=2.226$} which is between $2$ and $5$.

To further quantify the coherent superposition across different paths, we evaluate the relative entropy of coherence~\cite{Baumgratz2014} for the path DOF, with the system in the output state of each stage $l$, i.e., $\rho^{l,\mathrm{out}}_{\mathrm{c,ab}}$. Specifically, we compute
\begin{equation}
    \tilde{C}_{\mathrm{rel}}(l) = \mathcal{S}(\rho^{l,\mathrm{out}}_{\mathrm{c,diag}}) - \mathcal{S}(\rho^{l,\mathrm{out}}_{\mathrm{c}}),
    \label{eq:crel}
\end{equation}
where $\rho^{l,\mathrm{out}}_{\mathrm{c}}=\operatorname{tr}_{\mathrm{ab}}\rho^{l,\mathrm{out}}_{\mathrm{c,ab}}$ is the reduced density matrix in the path DOF, $\rho^{l,\mathrm{out}}_{\mathrm{c,diag}}$ is obtained by removing all off-diagonal elements from $\rho^{l,\mathrm{out}}_{\mathrm{c}}$, and $\mathcal{S}(\rho) = -\operatorname{tr}(\rho \log \rho)$ is the von Neumann entropy. Obviously, $\tilde{C}_{\mathrm{rel}}=0$ for any classical mixture of paths; on the other hand, in a stage with $d$ parallel path segments, $\tilde{C}_{\mathrm{rel}}$ reaches its maximum value of $\log d$ for equal-weight superposition states. If $n$ paths form an equal-weight coherent superposition while the remaining $d-n$ paths have no weight, we have $\tilde{C}_{\mathrm{rel}}=\log n$. Therefore, $\tilde{C}_{\mathrm{rel}}$ accurately captures the extent to which the path DOF is in a superposition state in the path basis $\ket{j}_{\mathrm{c}}$.

{\done Figure~\ref{fig:multinode_complex}(d) depicts $\tilde{C}_{\mathrm{rel}}$ and $\log d_l$ as functions of the stage index $l$, both for the optimal $N_{\mathrm{shots}}\to\infty$ solution and the best $N_{\mathrm{shots}}=10^3$ solution in Figs.~\ref{fig:multinode_complex}(b) and (c). We see that $0<\tilde{C}_{\mathrm{rel}}<\log d_l$ throughout the network; in fact, $\tilde{C}_{\mathrm{rel}}>\log 2$ for stages $2\leq l \leq 8$ in the $N_{\mathrm{shots}}\to\infty$ solution and for $2\leq l \leq 6$ in the $N_{\mathrm{shots}}=10^3$ solution, indicating significant path coherence across the network. For stages $l=4,7$ in the $N_{\mathrm{shots}}\to\infty$ solution, $\tilde{C}_{\mathrm{rel}}<\log (d_l-1)$: this is consistent with our observation in Fig.~\ref{fig:multinode_complex}(a) that one path segment is almost completely untraversed. Note that $\tilde{C}_{\mathrm{rel}}$ directly characterizes the coherence of the path DOF, but not its decoherence due to the noise in the network. For the latter, a more appropriate quantifier is the purity of the reduced density matrix, $\operatorname{tr}(\rho^{l,\mathrm{out}}_{\mathrm{c}})^2$. This is plotted in the inset of Fig.~\ref{fig:multinode_complex}(d), and decreases monotonically across the network for both solutions, as expected.

Finally, we extend our analysis to the regime of stronger noise, and revisit the role of vacuum coherence discussed in Sec.~\ref{sec:identical_chs_vac_coh}. To modify the noise strength over the entire network, we rescale the infidelity of every segment sampled in Fig.~\ref{fig:multinode_complex}(a) instead of resampling the CJ fidelities of path segments each time. Here we rescale by a factor of $(1-f_0)/(1-0.99)$, such that $f_0=0.99$ means no rescaling, the noise vanishes as $f_0\to 1$, and a lower baseline fidelity $f_0$ corresponds to stronger noise; for instance, $f_0=0.95$ and $f_0=0.9$ correspond to the best single-path fidelities $F^{0}_{\mathrm{CJ}}=0.835$ and $F^{0}_{\mathrm{CJ}}=0.702$ respectively. To tune the vacuum coherence within the framework of the microscopic model, we introduce a phenomenological parameter $\Delta \geq 0$, which quantifies how the vacuum interference operator depends on the noise strength in the weak noise limit; see App.~\ref{app:microscopic_general} and in particular Eq.~\eqref{eq:vio_micro_tunable}. $\Delta = 0$ corresponds to the microscopic model in App.~\ref{app:microscopic_special} which we have employed thus far, and increasing $\Delta$ results in diminished vacuum coherence for a given noise strength.

In Fig.~\ref{fig:multinode_complex}(e), we plot the Pareto fronts on the $p_{\mathrm{succ}}$--$\mathcal{R}_{\mathrm{prob}}$ plane for different combinations of $f_0 \in \{0.99, 0.95, 0.9\}$ and $\Delta \in \{0, 1, 2\}$ in the $N_{\mathrm{shots}}\to\infty$ limit. We observe that the optimal $p_{\mathrm{succ}}$ and $\mathcal{R}_{\mathrm{prob}}$ located at the cusp of the Pareto front are both suppressed when we either increase the noise strength (by decreasing $f_0$) or decrease the vacuum coherence (by increasing $\Delta$). Provided the vacuum coherence is maintained, our protocol is able to produce a meaningful advantage even with a relatively high noise level $f_0=0.9$. On the other hand, increasing $\Delta$ has a much larger effect when the noise is strong. We attribute this to the power-law dependence of the vacuum interference operators on the channel fidelity in Eq.~\eqref{eq:vio_micro_tunable}, where the powers are linear functions of $\Delta$. Even though some advantage is retained in the probabilistic scheme in the $N_{\mathrm{shots}}\to\infty$ limit as long as $\Delta$ does not become infinite (recall Fig.~\ref{fig:identical_detprob_vacamp}(a3), (b3) and (c3)), such advantage is unlikely to survive the statistical noise associated with finite $N_{\mathrm{shots}}$ when $\Delta$ is too large. Therefore, a relatively low noise level is beneficial not only for the infidelity ratio and the success probability, but also in terms of the robustness against the loss of vacuum coherence~\footnote{Nevertheless, when the noise level is extremely low, the optimization loop cannot proceed without a very large $N_{\mathrm{shots}}$, even though path superposition may remain useful as a noise reduction mechanism.}.

}


{\done \subsection{Network topology considerations}
\label{sec:bottleneck}

Here, we briefly discuss how network topology affects the performance of our protocol.

The advantage of path superposition depends on the network topology. Regular lattice geometries (e.g., square and triangular) are favorable for our protocol, since they offer multiple path links with similar to equal lengths between pairs of nodes (Fig.~\ref{fig:topology}(a)). This can satisfy timing requirements for interference at beamsplitters, provide natural modularity for scaling up, and offer redundancy against links that fail or suffer from strong noise which our protocol automatically avoids \cite{kar2023routing, li2021effective, kiktenko2024routing}. For these reasons, we foresee that our protocol has a natural advantage in communication scenarios where lattice geometries are available, such as within a large data center or between city blocks.

In contrast, networks with single-link bottlenecks (e.g., linearly connected clusters of nodes) limits the performance of our protocol, as all paths must traverse the bottleneck (Fig.~\ref{fig:topology}(b)) \cite{simmons2024scalable, kar2023routing}. 
One possible example of a naturally occurring network bottleneck is a satellite link in long-distance quantum communication. Satellite links are among the most efficient long-distance quantum channels~\cite{bedington2017progress, shabani2025building}, with successful entanglement and quantum key distribution protocols demonstrated over distances of more than $1000$ km~\cite{yin2017satellite, liao2017satellite, liao2017entanglement, zhang2022realization}. However, low-Earth orbit satellites have short overhead pass durations (5--10 minutes)~\cite{roger2023realtime, bedington2017progress, liao2017satellite}; in addition, satellite-to-ground optical links can be blocked by weather conditions~\cite{peng2025security, abasifard2024satellite, jaouni2025predicting}, and are subject to rapidly changing noise environments in the atmosphere~\cite{Maharjan2022atmospheric}. These conditions imply that it may not always be efficient to optimize a path superposition involving a satellite link. Nevertheless, we expect our protocol to remain beneficial in the ground-based parts of hybrid satellite-fiber networks~\cite{shao2025hybrid}, and the bottleneck effect of satellite links can be mitigated by, e.g., encoding information in a larger Hilbert space to increase throughput~\cite{villoresi2015temporal, wang2021temporal} and creating a multi-satellite ``constellation'' to add redundancy~\cite{wang2016ground}.

\begin{figure}[t]
{\includegraphics[width=\columnwidth]{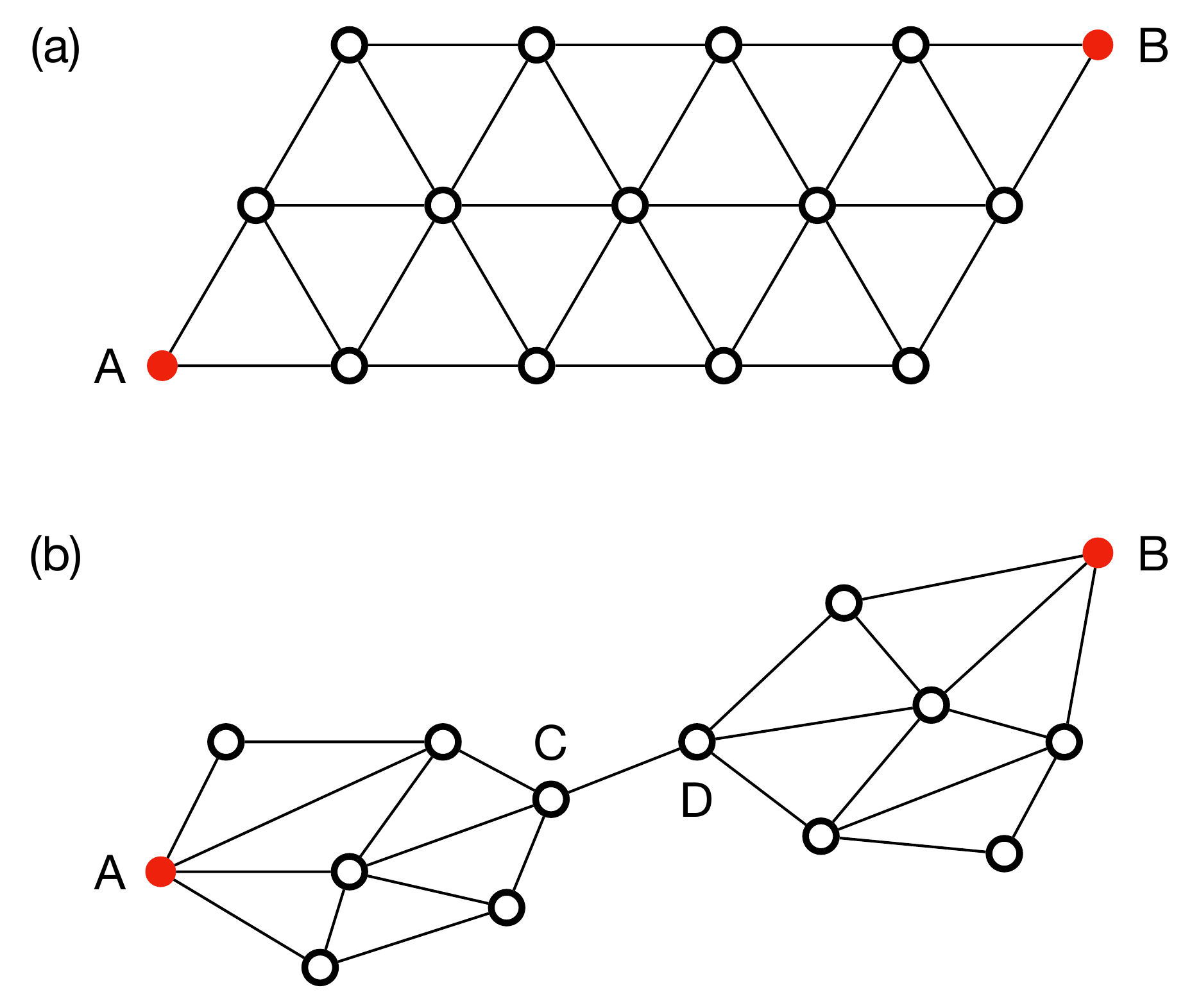}}
{\done \caption{\label{fig:topology} Qualitative effect of quantum network topology on protocol performance. (a) Regular lattice network geometries contain multiple equal-length paths with (typically) similar noise properties connecting Alice and Bob, enabling coherent superposition and interference at intermediate nodes. (b) Clusters connected by single links suffer from traversal bottlenecks, since all incoming paths must traverse single links that do not permit path superposition. Here, one can still fully exploit path superposition in the clusters connecting Alice to Charlie and David to Bob, but the overall advantage of the protocol remains limited by the noise properties of the Charlie--David bottleneck.
}}
\end{figure}

Our protocol has the algorithmic flexibility to be compatible with different network architectures and dynamic network conditions.
Nevertheless, a practical advantage can only be realized through careful consideration of the network. Therefore, it remains important to exploit redundant routes and reduce bottlenecks as much as possible.

}

{\done \section{Prospects for experimental implementation}
\label{sec:exp_impl}
Before concluding, we discuss prospects for near-term experimental realization of our self-configuring protocol.

\textit{Proof-of-principle feasibility \& scalability} -- A possible experimental realization of our protocol operates entirely within the photonic domain: path superpositions are generated via beamsplitters (or interferometers), transmission occurs via optical fibers, and Bob performs projective measurements and possibly unitary corrections on the information-carrying DOF (e.g., polarization). These require only standard photonic components alongside classical optimization hardware. As both represent mature technologies, this suggests that proof-of-principle protocol demonstrations are within reach~\cite{azuma2015all, ikuta2019first, li2019experimental, takeuchi2016photonic}.


Looking forward, several technological developments can enhance the scalability of the protocol within the transmission distance limit imposed by photon loss. These encompass fast reconfigurable photonic circuits \cite{Wang2018multi, llewellyn2020chip, ding2017high} that can implement path degree-of-freedom unitaries with low latency and allow for fast variational optimization loops \cite{liu2016fully, zhang2025reconfigurable, perez-lopez2025large}, and integrated circuits that combine sources, interferometers, and detectors on chip to minimize losses and stabilize relative phases at the network nodes.  

\textit{Experimental challenges} -- We anticipate two primary challenges in the practical implementation of our protocol, namely, achieving sufficient vacuum coherence and ensuring interference between different paths.

We have previously discussed vacuum coherence as an inherent property of the noisy quantum channel, which in general cannot be tuned. Therefore, it is always possible that the existing degree of vacuum coherence in a given setup is not enough for our protocol to identify an advantageous path superposition within a specified measurement budget. Nevertheless, perfect vacuum coherence is not a requirement for the success of our protocol, especially for the probabilistic variant which maintains an advantage in the presence of any nonzero vacuum coherence (see Fig.~\ref{fig:identical_detprob_vacamp}). Furthermore, the variational nature of our protocol ensures that paths with the strongest vacuum coherence are automatically prioritized in a path superposition. We also note that vacuum coherence is generally strong in the weak-noise limit, regardless of the concrete noise mechanism (see App.~\ref{app:microscopic_general}). For these reasons, we do not expect vacuum coherence to pose a fundamental obstacle to the experimental verification of our protocol, particularly in the low-noise regime, as we have demonstrated in Sec.~\ref{sec:multinode_complex} for $f_0$ close to unity.

On the experimental side, it is crucial that information carriers traversing different paths interfere at Bob's and any other intermediate recombining nodes. Similar path synchronization tasks have been accomplished in other contexts, for example by inserting optical delay lines as a passive, memory-free approach~\cite{Mendoza2016,quan2016demonstration,kim2022quantum,lal2024synchronized}. The interference between paths with large length differences can in principle be fully guaranteed using quantum memories~\cite{rancic2018coherence, ma2021hour, liu2025millisecond}, such that earlier-arriving states are stored until they can be recombined with later-arriving states. The use of quantum memories may even allow superposing sequentially transmitted states in the same quantum channel, further expanding the applicability of our protocol. Overall, we expect that the timing and path-length matching requirements of our protocol will be straightforward to satisfy in a future quantum network with memory-based quantum repeaters, given sufficient coherence times and photon conversion efficiencies. 
}

\section{Conclusions and outlook} \label{sec:conclusionss}

In this work, we introduced a self-configuring protocol for a quantum network augmented by coherent path superposition, which finds the configuration with the most effective noise mitigation and the best transmission fidelity using a quantum-classical feedback loop. The protocol works by sending half of a Bell pair from Alice to Bob. At Bob’s node, the Choi–Jamiołkowski fidelity and the postselection success probability are optimized. These quantities are functions of the path DOF unitaries applied at the network nodes and the unitary corrections performed by Bob. We described the protocol both in the two-node case and in the multi-node case, and showcased its performance under different conditions, specifically in the presence of a form of path DOF noise. We found that the protocol depends crucially on the vacuum coherence of quantum channels within the network, having demonstrated the vacuum coherence of three typical quantum channels using a microscopic model. {\done We showed that the probabilistic variant, while exhibiting a nonzero probability of failure, is generally more effective in improving the fidelity, more robust against the path DOF noise, and has less stringent vacuum coherence requirements. Finally, we numerically demonstrated the adaptability of our protocol in a complicated network configuration with multiple nodes and path segments, as well as the effect of statistical noise from a finite measurement budget.}

Our protocol can be readily generalized to high-dimensional quantum communication~\cite{Cozzolino2019}, where the information carriers are qudits (with local Hilbert space dimension $d>2$) rather than qubits~\cite{Wang2020}. In this regard, an important practical issue is the efficient estimation of $F_{\mathrm{CJ}}$ for qudits. {\done While fidelity estimation is much less expensive than state tomography, which quickly becomes prohibitively expensive for large $d$~\cite{thew02}, it remains a relevant question whether methods that are more efficient and scalable than full tomography, such as partial tomography~\cite{martinez25,garberoglio25,kakihara25}, SWAP tests~\cite{garciaescartin13}, and entanglement witnesses~\cite{kadlec25}, can be adapted to further enhance the efficiency of fidelity estimation.} In addition, we can potentially further reduce the resource cost of our protocol by sharing nonmaximally entangled qudit pairs between Alice and Bob~\cite{kraft18}.

{\done A key feature of our protocol is that it does not require prior knowledge of the noise types and strengths in the network.} Nevertheless, we can avoid barren plateaus and greatly accelerate convergence by ``warm starts''~\cite{Huggins2019,Mele2022,Dborin2022,Grimsley2023,Rudolph2023,Chan2024,Miao2024,Puig2025}, in which the variational parameters are initialized close to the optimal solution (or at least an advantageous solution) in a problem-informed fashion. Intuitively, if we know in advance that a certain path segment is noisier and/or has lower vacuum coherence than the segments parallel to it, we should choose our initial path unitary parameters $\boldsymbol{\theta}_{\mathrm{c}}$ such that the path unitaries place a lower weight on this segment. As another example, when we embed a smaller network into a larger one by turning the initial and final nodes into intermediate nodes (e.g., Charlie and David in Fig.~\ref{fig:multinode_simple}), it is reasonable to expect that the optimal $\boldsymbol{\theta}_{\mathrm{c}}$ in the smaller network remain close to optimal for the larger network. Therefore, to achieve faster convergence in a large network, we could first estimate $\boldsymbol{\theta}_{\mathrm{c}}$ by benchmarking the fidelity and vacuum coherence of individual path segments~\cite{paperbenchmark}, then run our protocol on smaller networks consisting of subsets of nodes before using the resulting optimized $\boldsymbol{\theta}_{\mathrm{c}}$ as the initial guesses of the large network. {\done We leave vacuum coherence benchmarking, warm starts, as well as other experimental prospects (Sec.~\ref{sec:exp_impl}) for future work.}

Another potential research direction concerns the use of quantum-controlled operations in conjunction with the superposition of paths~\cite{Guerin2019,Rubino2021}. In our current protocol, we have limited ourselves to applying postprocessing unitaries to the information carrier after measuring the path DOF at Bob. In practice, we can apply unitary operations to any path before or after the measurement, and even at any intermediate node in the multi-node protocol. It is known that the error mitigation performance using quantum-controlled operations exceeds or equals that of path superposition; nevertheless, further investigations are warranted to characterize the interaction between quantum-controlled operations and path superposition, especially in the context of VQO.

{\done Finally, we remark that our protocol does not rely on quantum memories, nor does it make use of entanglement generated at intermediate nodes. This fundamentally distinguishes our protocol from the paradigm of quantum repeaters~\cite{Dur99, briegel1998quantum, hensen2015loophole, Azuma2023}, which are experimentally far more demanding. Nevertheless, in a network equipped with quantum repeaters, we expect our protocol to remain applicable when multiple transmission paths exist between pairs of nodes (see also discussions on employing quantum memories for path synchronization in Sec.~\ref{sec:exp_impl}). In particular, for \textit{all-photonic} quantum repeaters~\cite{azuma2015all, niu2023all, benchasattabuse2024engineering}, the probabilistic variant of our protocol may be used to distribute repeater graph states between adjacent nodes with increased fidelity.}

\section*{Acknowledgments}

This work was supported by the Austrian Science Fund (FWF) through projects No. P36009-N and No. P36010-N. Finanziert von der Europäischen Union - NextGenerationEU. Furthermore, we acknowledge support from the Natural Sciences and Engineering Research
Council of Canada (NSERC), the Canada First Research Excellence Fund (CFREF, Transformative Quantum Technologies), New Frontiers in Research Fund (NFRF), Ontario Early Researcher Award, and the Canadian Institute for Advanced Research (CIFAR). LD acknowledges the EPSRC Quantum Career Development grant EP/W028301/1 and Standard Research grant EP/Z534250/1.

\bibliography{BCPbiblio}

\clearpage
\appendix
\renewcommand{\theequation}{\Alph{section}.\arabic{equation}}
\renewcommand{\thefigure}{\thesection.\arabic{figure}}
\section{Analytical examples}
\label{app:analytical_ex}
\setcounter{equation}{0} 
\setcounter{figure}{0}

In this appendix, we focus on the two-node setup and analytically study the strong vacuum coherence limit $|\alpha_0| \to 1$.

\subsection{Probabilistic variant, channels with equal CJ fidelity}
\label{app:a1_prob_equal}

We first consider applying the probabilistic variant to $d$ distinct channels labeled by $i=0,1,\dots,d-1$ which have the same CJ fidelity $F^{0}_{\mathrm{CJ}}$ but generally different Kraus operators. For channel $i$, a Kraus operator $K^{(i)}_{s_i}$ can be expanded in the orthogonal basis $\{\id, X, Y, Z\}$ as follows:
\begin{equation}
    K^{(i)}_{s_i}=\sum_{j=0}^{3} c^{(i)}_{s_i, j}\sigma_j,
    \label{eq:pauli_kraus_expand}
\end{equation}
where again $\sigma_j = \id, X, Y, Z$ for $j=0, 1, 2, 3$, respectively.

For concreteness, we assume the noise represented by each channel has no coherent component, and a unique no-jump Kraus operator $K^{(i)}_{0}$ exists for each channel, whose effective vacuum amplitude $|\alpha^{(i)}_0| \to 1$. Here ``no-jump'' refers to a Kraus operator with a nonzero identity component; in other words, $c^{(i)}_{s_i, 0}=0$ unless $s_i=0$. If we further assume $c^{(i)}_{0, 0}>0$ without loss of generality (the phase of $c^{(i)}_{0, 0}$ can be absorbed into the phase of the vacuum amplitude vector), then ``having no coherent component'' means that $c^{(i)}_{0, j}$ are real for all $j\neq 0$, since any imaginary part of $c^{(i)}_{0, j}$ could be absorbed into a unitary -- a coherent noise component. We note that these assumptions are satisfied by all three noise types considered in this work, Eqs.~\eqref{eq:dephasingnoise}--\eqref{eq:amplitudedampingnoise}.

Let us examine the properties of the expansion coefficients $c^{(i)}_{s_i, j}$. Inserting Eq.~\eqref{eq:pauli_kraus_expand} into Eqs.~\eqref{eq:rho_out_fcj} and \eqref{eq:fid_simple}, and noting the Bell state $\ket{\Phi^{+}}_{\mathrm{ab}}$ satisfies the following relation,
\begin{equation}
    \bra{\Phi^{+}}[\id_{\mathrm{a}} \otimes(\sigma_j)_{\mathrm{b}}]\ket{\Phi^{+}}_{\mathrm{ab}}=0\ (j\neq 0),
    \label{eq:bell_trace}
\end{equation}
we easily express $c^{(i)}_{0, 0}$ (already assumed to be positive) in terms of the CJ fidelity $F^{0}_{\mathrm{CJ}}$ of each channel,
\begin{equation}
    c^{(i)}_{0, 0} = \sqrt{F^{0}_{\mathrm{CJ}}},\ c^{(i)}_{s_i, 0}=0\ (s_i \neq 0).
    \label{eq:c00_a1}
\end{equation}
We can glean more insight from the normalization condition $\sum_{s_i}K^{(i)\dagger}_{s_i}K^{(i)}_{s_i}=\id$. Comparing the coefficients of the $\id$ term, we readily obtain
\begin{equation}
    \sum_{s_i} \sum_{j=0}^{3} |c^{(i)}_{s_i, j}|^2=1.
    \label{eq:norm_a1}
\end{equation}
Furthermore, comparing the coefficients of the $Z$ term, we have
\begin{align}
    0=&\sum_{s_i} \operatorname{Re}(c^{(i)*}_{s_i, 0} c^{(i)}_{s_i, 3}+\mathrm{i}c^{(i)*}_{s_i, 1} c^{(i)}_{s_i, 2})\nonumber\\
    = &\sqrt{F^{0}_{\mathrm{CJ}}} c^{(i)}_{0, 3}-\sum_{s_i \neq 0} \operatorname{Im}(c^{(i)*}_{s_i, 1} c^{(i)}_{s_i, 2}),
    \label{eq:z_expand_a1}
\end{align}
recalling Eq.~\eqref{eq:c00_a1} as well as our assumption of $c^{(i)}_{0, j}$ being real for all $j\neq 0$. In the $F^{0}_{\mathrm{CJ}}\to 1$ limit, Eqs.~\eqref{eq:norm_a1} and \eqref{eq:z_expand_a1} together indicate that
\begin{equation}
    c^{(i)}_{0, j} = O(1-F^{0}_{\mathrm{CJ}}) \ (j\neq 0),
    \label{eq:c03_a1}
\end{equation}
where the $j=1,2$ cases are similar to the $j=3$ case.

We now calculate the infidelity ratio $\mathcal{R}_{\mathrm{prob}}$. We write $\alpha^{(i)}_0=|\alpha^{(i)}_0|e^{\mathrm{i}\zeta_{i}}$, and consider the action of all channels on the following equal superposition of paths:
\begin{equation}
    \ket{\psi}^\mathrm{in}=\sum_{i=0}^{d-1}\frac{e^{\mathrm{i}\zeta_{i}}}{\sqrt{d}}\ket{i}_{\mathrm{c}}\otimes\ket{\Phi^+}_{\mathrm{ab}}.
\end{equation}
Eq.~\eqref{eq:supp_incoh_p_coh} is now evaluated in a straightforward fashion, with
\begin{subequations}
    \begin{align}
        \rho^{\mathrm{out}(ii)}
        &
        = 
        \frac{1}{d}\sum_{s_i }\sum_{jj'}c^{(i)}_{s_i,j}c^{(i)*}_{s_i,j'}\ket{i}\bra{i}_{\mathrm{c}}\otimes \varrho^{jj'}_{\mathrm{ab}}
        ,
        \\
        \rho^{\mathrm{out}(ii')}
        & 
        = 
        \frac{1}{d}|\alpha^{(i)}_0 \alpha^{(i')}_0|\sum_{jj'}c^{(i)}_{0,j}c^{(i')*}_{0,j'}\ket{i}\bra{i'}_{\mathrm{c}}\otimes \varrho^{jj'}_{\mathrm{ab}}\\
        & \ (i\neq i'),\nonumber
    \end{align}
    \label{eq:output_a1_block}
\end{subequations}
where we have used the shorthand
\begin{equation}
    \varrho^{jj'}_{\mathrm{ab}}=[\id_{\mathrm{a}} \otimes (\sigma_j)_{\mathrm{b}} ]\proj{\Phi^+}_{\mathrm{ab}}[\id_{\mathrm{a}} \otimes (\sigma_{j'})_{\mathrm{b}} ].
    \label{eq:varrho_ab}
\end{equation}
Further applying the path DOF unitary $U_{\mathrm{c}2}$ and projecting to path $k$, we find the post-measurement state Eq.~\eqref{eq:postmeas_st} to be
\begin{align}
&P_k\rho^{\mathrm{out}}_{\mathrm{ab}, k} \nonumber\\
 = &\frac{1}{d}\sum_{jj'}\varrho^{jj'}_{\mathrm{ab}}\Big[
\sum_{i}\sum_{s_i}(U_{\mathrm{c}2})_{ki}(U_{\mathrm{c}2})^*_{ki} c^{(i)}_{s_i,j}c^{(i)*}_{s_i,j'}\nonumber\\
&+\sum_{i\neq i' }(U_{\mathrm{c}2})_{ki}(U_{\mathrm{c}2})^*_{ki'}|\alpha^{(i)}_0 \alpha^{(i')}_0|c^{(i)}_{0,j}c^{(i')*}_{0,j'}\Big].
\label{eq:postmeas_st_gen_a1}
\end{align}
We now focus on $k=0$ and choose a unitary that creates an equal superposition in path $0$, $(U_{\mathrm{c}2})_{0i}=1/\sqrt{d}$ for every $i$. This gives
\begin{align}
P_0\rho^{\mathrm{out}}_{\mathrm{ab}, 0} 
 = &\frac{1}{d^2}\sum_{jj'}\varrho^{jj'}_{\mathrm{ab}}\Big[
\sum_{i}\sum_{s_i} c^{(i)}_{s_i,j}c^{(i)*}_{s_i,j'}\nonumber\\
&+\sum_{i\neq i' }|\alpha^{(i)}_0 \alpha^{(i')}_0|c^{(i)}_{0,j}c^{(i')*}_{0,j'}\Big].
\label{eq:postmeas_st_0_a1}
\end{align}
It then follows that $P_0$, the probability of measuring the path DOF in $\ket{0}_{\mathrm{c}}$, has the form
\begin{align}
    & P_0= \operatorname{tr}(P_0\rho^{\mathrm{out}}_{\mathrm{ab}, 0})\nonumber\\
    = &\frac{1}{d^2}\big[d+\sum_{i\neq i' }|\alpha^{(i)}_0 \alpha^{(i')}_0|F^{0}_{\mathrm{CJ}}\nonumber\\
    &+\sum_{j\neq 0}\sum_{i\neq i' }|\alpha^{(i)}_0 \alpha^{(i')}_0|c^{(i)}_{0,j}c^{(i')*}_{0,j}\big].
    \label{eq:p0_a1}
\end{align}
We have used Eqs.~\eqref{eq:bell_trace}, \eqref{eq:c00_a1} and \eqref{eq:norm_a1}; in particular, Eq.~\eqref{eq:bell_trace} ensures that only the $j=j'$ terms contribute to $P_0$. In the $F^{0}_{\mathrm{CJ}}\to 1$ and $|\alpha_0^{(i)}|\to 1$ limit, inserting Eq.~\eqref{eq:c03_a1} into Eq.~\eqref{eq:p0_a1}, we obtain
\begin{equation}
    P_0 = 1-(1-\frac{1}{d})(1-F^{0}_{\mathrm{CJ}})+O((1-F^{0}_{\mathrm{CJ}})^2).
    \label{eq:p0_limit_a1}
\end{equation}
As an aside, if $c^{(i)}_{0,j}=0$ for any $i$ and for any $j\neq 0$, which is the case when every channel is either dephasing or depolarizing, the double sum in Eq.~\eqref{eq:p0_a1} simply vanishes regardless of the value of $F^{0}_{\mathrm{CJ}}$.

To maximize the CJ fidelity of the output density matrix, we generally need to apply a unitary correction, which amounts to rotating the 4-vector $c^{(i)}_{s_i,j}$ ($j=0,1,2,3$) for all $i$ and $s_i$ simultaneously. Under our current assumptions, no unitary correction is necessary for the path DOF measurement outcome $0$. In Eq.~\eqref{eq:postmeas_st_0_a1}, only the $j=j'=0$ term contributes to the CJ fidelity:
\begin{equation}
    \bra{\Phi^+}P_0\rho^{\mathrm{out}}_{\mathrm{ab}, 0}\ket{\Phi^+}_{\mathrm{ab}} = \frac{1}{d^2}(d+\sum_{i \neq i'} |\alpha_0^{(i)}\alpha_0^{(i')}|)F^{0}_{\mathrm{CJ}}.
    \label{eq:P0_rho_out0_tr}
\end{equation}
Combining Eqs.~\eqref{eq:p0_limit_a1} and \eqref{eq:P0_rho_out0_tr}, in the $F^{0}_{\mathrm{CJ}}\to 1$ and $|\alpha_0^{(i)}|\to 1$ limit, we find the following behavior for the infidelity ratio,
\begin{equation}
    \mathcal{R}_{\mathrm{prob}}
    =\frac{1-F^{0}_{\mathrm{CJ}}}{1-\frac{1}{P_0}\bra{\Phi^+}P_0\rho^{\mathrm{out}}_{\mathrm{ab}, 0}\ket{\Phi^+}_{\mathrm{ab}}}\to d,
\end{equation}
independent of the details of the channels.

\subsection{Identical channels}
\label{app:a2_identical}

Another case where we can easily gain analytical insight is applying either the probabilistic or the deterministic variant to $d$ identical channels in equal superposition. In this case, we can write $\alpha^{(i)}_0=\alpha_0$ and $c^{(i)}_{s_i,j}=c_{s_i,j}$. Eq.~\eqref{eq:postmeas_st_0_a1} continues to hold for the path DOF measurement outcome $0$, while for any other measurement outcome $k=1,\dots,d-1$, the unitarity of $U_{\mathrm{c}2}$ results in
\begin{equation}
    \sum_i |(U_{\mathrm{c}2})_{ki}|^2=1,\ \sum_i (U_{\mathrm{c}2})_{ki} = 0\ (k\neq 0);
\end{equation}
therefore, Eqs.~\eqref{eq:postmeas_st_gen_a1} and \eqref{eq:postmeas_st_0_a1} become
\begin{align}
P_0\rho^{\mathrm{out}}_{\mathrm{ab}, 0} 
 = &\frac{1}{d}\sum_{jj'}\varrho^{jj'}_{\mathrm{ab}}\Big(
\sum_{s_i} c_{s_i,j}c^*_{s_i,j'}\nonumber\\
&+(d-1)|\alpha_0|^2 c_{0,j}c^*_{0,j'}\Big),
\label{eq:postmeas_st_0_a2}
\end{align}
and
\begin{align}
P_k\rho^{\mathrm{out}}_{\mathrm{ab}, k} = &\frac{1}{d}\sum_{jj'}\varrho^{jj'}_{\mathrm{ab}}\Big(
\sum_{s_i} c_{s_i,j}c^*_{s_i,j'}\nonumber\\
& -|\alpha_0|^2 c_{0,j}c^*_{0,j'}\Big)\ (k\neq 0).
\label{eq:postmeas_st_gen_a2}
\end{align}

We now evaluate Eqs.~\eqref{eq:postmeas_st_0_a2} and \eqref{eq:postmeas_st_gen_a2} for different types of noise.
\subsubsection{Dephasing channels}
For $d$ dephasing channels,
\begin{equation}
    P_0\rho^{\mathrm{out}}_{\mathrm{ab}, 0}= \frac{1}{d}\Big[(1-p_0)(1+(d-1)|\alpha_0|^2)\varrho^{00}_{\mathrm{ab}}+p_0 \varrho^{33}_{\mathrm{ab}}\Big],
\end{equation}
\begin{align}
    P_k\rho^{\mathrm{out}}_{\mathrm{ab}, k} &= \frac{1}{d}\Big[(1-p_0)(1-|\alpha_0|^2)\varrho^{00}_{\mathrm{ab}}+p_0 \varrho^{33}_{\mathrm{ab}}\Big]\nonumber\\ &(k\neq 0).
\end{align}
Thus, for the measurement outcome $0$,
\begin{subequations}
\begin{equation}
    P_0 = \frac{1}{d}\Big[1+(1-p_0)(d-1)|\alpha_0|^2\Big],
    \label{eq:p0_depha_a2}
\end{equation}
the correcting unitary $V_0=\id$ as discussed in Sec.~\ref{app:a1_prob_equal},
\begin{align}
    & \bra{\Phi^+}P_0\rho^{\mathrm{out}}_{\mathrm{ab}, 0}\ket{\Phi^+}_{\mathrm{ab}} \nonumber\\
    =&  \frac{1}{d}(1-p_0)[1+(d-1)|\alpha_0|^2];
\label{eq:f0_depha_a2}
\end{align}
for the measurement outcome $k\neq 0$,
\begin{equation}
    P_k = \frac{1}{d}\Big[1-(1-p_0)|\alpha_0|^2\Big],
    \label{eq:pk_depha_a2}
\end{equation}
and if $|\alpha_0|^2 > 1-p_0/(1-p_0)$, the correcting unitary $V^{(k)}=Z$,
\begin{equation}
    \bra{\Phi^+}P_k\rho^{\mathrm{out,corr}}_{\mathrm{ab}, k}\ket{\Phi^+}_{\mathrm{ab}} = \frac{p_0}{d},
\end{equation}
while if $|\alpha_0|^2 \leq 1-p_0/(1-p_0)$, $V^{(k)}=\id$,
\begin{equation}
    \bra{\Phi^+}P_k\rho^{\mathrm{out,corr}}_{\mathrm{ab}, k}\ket{\Phi^+}_{\mathrm{ab}} = \frac{1}{d}(1-p_0)(1-|\alpha_0|^2).
\end{equation}
\end{subequations}
Therefore, using $F^{0}_{\mathrm{CJ}}=1-p_0$, we find
\begin{align}
    &\mathcal{R}_{\mathrm{prob}}  =\frac{1-F^{0}_{\mathrm{CJ}}}{1-\bra{\Phi^+}\rho^{\mathrm{out,corr}}_{\mathrm{ab}, 0}\ket{\Phi^+}_{\mathrm{ab}}}\nonumber\\
    =& [1-(1-p_0)]/\Bigg[1-\frac{(1-p_0)(1+(d-1)|\alpha_0|^2)}{1+(1-p_0)(d-1)|\alpha_0|^2}\Bigg]\nonumber\\
    =& 1+(d-1)|\alpha_0|^2 F^{0}_{\mathrm{CJ}},
    \label{eq:rprob_depha_a2}
\end{align}
\begin{align}
    &\mathcal{R}_{\mathrm{det}}  =\frac{1-F^{0}_{\mathrm{CJ}}}{1-\sum_{k=0}^{d-1} \bra{\Phi^+}P_k\rho^{\mathrm{out,corr}}_{\mathrm{ab}, k}\ket{\Phi^+}_{\mathrm{ab}}}\nonumber\\
    =& [1-(1-p_0)]/\Bigg\{1-\Big[\frac{1}{d}(1-p_0)(1+(d-1)|\alpha_0|^2)\nonumber\\
    & +\frac{d-1}{d}\max\{p_0, (1-p_0)(1-|\alpha_0|^2)\}\Big]\Bigg\}\nonumber\\
    = &\bigg\{\begin{matrix}
    1 ,& \mathrm{if}\ |\alpha_0|^2 \leq 2-\frac{1}{F^{0}_{\mathrm{CJ}}}, \\
    \frac{d(1-F^{0}_{\mathrm{CJ}})}{(1-F^{0}_{\mathrm{CJ}})+(d-1)(1-|\alpha_0|^2)F^{0}_{\mathrm{CJ}}} ,& \mathrm{if}\ |\alpha_0|^2> 2-\frac{1}{F^{0}_{\mathrm{CJ}}}. \\
    \end{matrix}
    \label{eq:rdet_depha_a2}
\end{align}

Equations~\eqref{eq:p0_depha_a2}, \eqref{eq:rprob_depha_a2} and \eqref{eq:rdet_depha_a2} are plotted in Fig.~\ref{fig:identical_detprob_vacamp}(a1)--(a3). Note that these expressions are valid for arbitrary noise strength $p_0$ and vacuum amplitude $\alpha_0$; we have \emph{not} taken the weak noise limit $p_0 \to 0$ or the strong vacuum coherence limit $|\alpha_0|\to 1$. In particular, for given $F^{0}_{\mathrm{CJ}}$, Eq.~\eqref{eq:rdet_depha_a2} predicts a threshold vacuum amplitude
\begin{equation}
    \alpha^{\mathrm{th}}_0 = \sqrt{2-\frac{1}{F^{0}_{\mathrm{CJ}}}}\ (\text{dephasing}),
    \label{eq:alpha_0_th_depha}
\end{equation}
below which the deterministic variant yields no advantage; this is plotted in the inset of Fig.~\ref{fig:identical_detprob_vacamp}(a3). In the $|\alpha_0|\to 1$ limit, for arbitrary $F^{0}_{\mathrm{CJ}}$, Eqs.~\eqref{eq:rprob_depha_a2} and \eqref{eq:rdet_depha_a2} are reduced to
\begin{equation}
    \mathcal{R}_{\mathrm{prob}}\to d, \ \mathcal{R}_{\mathrm{det}}\to d\ (\text{dephasing}).
\end{equation}
It is important to mention that the limits $|\alpha_0|\to 1$ and $F^{0}_{\mathrm{CJ}}\to 1$ do not commute: $F^{0}_{\mathrm{CJ}}= 1$ automatically leads to $|\alpha_0|= 1$ (see App.~\ref{app:microscopic_general}), but the converse is not true.

\subsubsection{Depolarizing channels}
For $d$ depolarizing channels,
\begin{align}
    P_0\rho^{\mathrm{out}}_{\mathrm{ab}, 0} =& \frac{1}{d}\Big[(1-p_0)(1+(d-1)|\alpha_0|^2)\varrho^{00}_{\mathrm{ab}}\nonumber\\
    &+\frac{p_0}{3}(\varrho^{11}_{\mathrm{ab}}+\varrho^{22}_{\mathrm{ab}}+\varrho^{33}_{\mathrm{ab}})\Big],
\end{align}
\begin{align}
    P_k\rho^{\mathrm{out}}_{\mathrm{ab}, k} =& \frac{1}{d}\Big[(1-p_0)(1-|\alpha_0|^2)\varrho^{00}_{\mathrm{ab}}\nonumber\\&+\frac{p_0}{3}(\varrho^{11}_{\mathrm{ab}}+\varrho^{22}_{\mathrm{ab}}+\varrho^{33}_{\mathrm{ab}})\Big]\ (k\neq 0).
\end{align}
Thus, Eqs.~\eqref{eq:p0_depha_a2}, \eqref{eq:f0_depha_a2} and \eqref{eq:pk_depha_a2} continue to be valid; in particular,
\begin{subequations}
\begin{equation}
    P_0 = \frac{1}{d}\Big[1+(1-p_0)(d-1)|\alpha_0|^2\Big].
    \label{eq:p0_depol_a2}
\end{equation}
If $|\alpha_0|^2 > 1-p_0/(3-3p_0)$, $V^{(k)}$ can be any traceless $\mathrm{U}(2)$ unitary,
\begin{equation}
    \bra{\Phi^+}P_k\rho^{\mathrm{out,corr}}_{\mathrm{ab}, k}\ket{\Phi^+}_{\mathrm{ab}} = \frac{p_0}{3d},
\end{equation}
while if $|\alpha_0|^2 \leq 1-p_0/(3-3p_0)$, $V^{(k)}=\id$,
\begin{equation}
    \bra{\Phi^+}P_k\rho^{\mathrm{out,corr}}_{\mathrm{ab}, k}\ket{\Phi^+}_{\mathrm{ab}} = \frac{1}{d}(1-p_0)(1-|\alpha_0|^2).
\end{equation}
\end{subequations}
Therefore, using $F^{0}_{\mathrm{CJ}}=1-p_0$, we have
\begin{equation}
    \mathcal{R}_{\mathrm{prob}}=1+(d-1)|\alpha_0|^2 F^{0}_{\mathrm{CJ}},\label{eq:rprob_depol_a2}
\end{equation}
which is identical to Eq.~\eqref{eq:rprob_depha_a2}, and
\begin{align}
    &\mathcal{R}_{\mathrm{det}}\nonumber\\
    =& [1-(1-p_0)]/\Bigg\{1-\Big[\frac{1}{d}(1-p_0)\big[1+(d-1)|\alpha_0|^2\big]\nonumber\\
    & +\frac{d-1}{d}\max\{\frac{p_0}{3}, (1-p_0)(1-|\alpha_0|^2)\}\Big]\Bigg\}\nonumber\\
    = &\bigg\{\begin{matrix}
    1 ,& \mathrm{if}\ |\alpha_0|^2 \leq \frac{4}{3}-\frac{1}{3F^{0}_{\mathrm{CJ}}}, \\
    \frac{d(1-F^{0}_{\mathrm{CJ}})}{\frac{2d+1}{3}(1-F^{0}_{\mathrm{CJ}})+(d-1)(1-|\alpha_0|^2)F^{0}_{\mathrm{CJ}}} ,& \mathrm{if}\ |\alpha_0|^2> \frac{4}{3}-\frac{1}{3F^{0}_{\mathrm{CJ}}}. \\
    \end{matrix}
    \label{eq:rdet_depol_a2}
\end{align}
Equations~\eqref{eq:p0_depol_a2}, \eqref{eq:rprob_depol_a2} and \eqref{eq:rdet_depol_a2} are plotted in Fig.~\ref{fig:identical_detprob_vacamp}(b1)--(b3). Again, they are valid for arbitrary $p_0$ and $\alpha_0$. Equation~\eqref{eq:rdet_depol_a2} yields the threshold vacuum amplitude
\begin{equation}
    \alpha^{\mathrm{th}}_0 = \sqrt{\frac{4}{3}-\frac{1}{3F^{0}_{\mathrm{CJ}}}}\ (\text{depolarizing}),
    \label{eq:alpha_0_th_depol}
\end{equation}
which is plotted in the inset of Fig.~\ref{fig:identical_detprob_vacamp}(b3).
In the $|\alpha_0|\to 1$ limit, for arbitrary $F^{0}_{\mathrm{CJ}}$, Eqs.~\eqref{eq:rprob_depol_a2} and \eqref{eq:rdet_depol_a2} are reduced to
\begin{equation}
    \mathcal{R}_{\mathrm{prob}}\to d, \ \mathcal{R}_{\mathrm{det}}\to \frac{3d}{2d+1}\ (\text{depolarizing}).
\end{equation}

\subsubsection{Amplitude damping channels}
For $d$ amplitude damping channels,
\begin{align}
&P_0\rho^{\mathrm{out}}_{\mathrm{ab}, 0} \nonumber\\
 = &\frac{1}{4d}\Bigg\{
[1+(d-1)|\alpha_0|^2]\Big[(1+\sqrt{1-p_0})^2 \varrho^{00}_{\mathrm{ab}} \nonumber\\
&+(1-\sqrt{1-p_0})^2 \varrho^{33}_{\mathrm{ab}}+p_0(\varrho^{03}_{\mathrm{ab}}+\varrho^{30}_{\mathrm{ab}})\Big] \nonumber\\
&+p_0(\varrho^{11}_{\mathrm{ab}}-\mathrm{i}\varrho^{12}_{\mathrm{ab}}+\mathrm{i}\varrho^{21}_{\mathrm{ab}}+\varrho^{22}_{\mathrm{ab}})
\Bigg\},
\end{align}
and
\begin{align}
&P_k\rho^{\mathrm{out}}_{\mathrm{ab}, k} \nonumber\\
 = &\frac{1}{4d}\Bigg\{
(1-|\alpha_0|^2)\Big[(1+\sqrt{1-p_0})^2 \varrho^{00}_{\mathrm{ab}} \nonumber\\
&+(1-\sqrt{1-p_0})^2 \varrho^{33}_{\mathrm{ab}}+p_0(\varrho^{03}_{\mathrm{ab}}+\varrho^{30}_{\mathrm{ab}})\Big] \nonumber\\
&+p_0(\varrho^{11}_{\mathrm{ab}}-\mathrm{i}\varrho^{12}_{\mathrm{ab}}+\mathrm{i}\varrho^{21}_{\mathrm{ab}}+\varrho^{22}_{\mathrm{ab}})
\Bigg\}\ (k\neq 0).
\end{align}
Therefore,
\begin{subequations}
\begin{equation}
    P_0 = \frac{1}{2d}\Big[[1+(d-1)|\alpha_0|^2](2-p_0)+p_0 \Big],
    \label{eq:p0_ad_a2}
\end{equation}
$V_0=\id$, 
\begin{align}
    &\bra{\Phi^+}P_0\rho^{\mathrm{out}}_{\mathrm{ab}, 0}\ket{\Phi^+}_{\mathrm{ab}} \nonumber\\
    = & \frac{1}{4d}[1+(d-1)|\alpha_0|^2](1+\sqrt{1-p_0})^2;
    \label{eq:f0_ad_a2}
\end{align}
\begin{equation}
    P_k = \frac{1}{2d}\Big[(1-|\alpha_0|^2)(2-p_0)+p_0 \Big]\ (k\neq 0),
    \label{eq:pk_ad_a2}
\end{equation}
and if $|\alpha_0|^2 > 1-p_0/(1+\sqrt{1-p_0})^2$, $V^{(k)}$ can be any purely off-diagonal $\mathrm{U}(2)$ unitary,
\begin{equation}
    \bra{\Phi^+}P_k\rho^{\mathrm{out,corr}}_{\mathrm{ab}, k}\ket{\Phi^+}_{\mathrm{ab}} = \frac{p_0}{4d};
\end{equation}
if $|\alpha_0|^2 \leq 1-p_0/(1+\sqrt{1-p_0})^2$, $V^{(k)}=\id$,
\begin{align}
    &\bra{\Phi^+}P_k\rho^{\mathrm{out,corr}}_{\mathrm{ab}, k}\ket{\Phi^+}_{\mathrm{ab}}\nonumber\\
    = & \frac{1}{4d}(1-|\alpha_0|^2)(1+\sqrt{1-p_0})^2.
\end{align}
\end{subequations}
Using $F^{0}_{\mathrm{CJ}}=(1+\sqrt{1-p_0})^2/4$, we obtain
\begin{align}
    &\mathcal{R}_{\mathrm{prob}} \nonumber\\
    = & \Big[[1+(d-1)|\alpha_0|^2](1-\frac{p_0}{2})+\frac{p_0}{2} \Big] \nonumber\\
    & \times \frac{1-\frac{1}{4}(1+\sqrt{1-p_0})^2}{\frac{p_0}{2}+\frac{1}{4}(1-\sqrt{1-p_0})^2 [1+(d-1)|\alpha_0|^2]},
    \label{eq:rprob_ad_a2}
\end{align}
\begin{align}
    &\mathcal{R}_{\mathrm{det}} \nonumber\\
    =& [1-\frac{1}{4}(1+\sqrt{1-p_0})^2]/\Bigg\{1-\Big[\nonumber\\
    & \frac{1}{4d}(1+\sqrt{1-p_0})^2 [1+(d-1)|\alpha_0|^2]+\frac{d-1}{4d}\nonumber\\
    & \times\max\{p_0, (1+\sqrt{1-p_0})^2(1-|\alpha_0|^2)\}\Big]\Bigg\}.
    \label{eq:rdet_ad_a2}
\end{align}

Equations~\eqref{eq:p0_ad_a2}, \eqref{eq:rprob_ad_a2} and \eqref{eq:rdet_ad_a2} are plotted in Fig.~\ref{fig:identical_detprob_vacamp}(c1)--(c3). Once again, if $|\alpha_0| \leq \alpha^{\mathrm{th}}_0$ where 
\begin{equation}
    \alpha^{\mathrm{th}}_0 = \sqrt{2-\frac{1}{\sqrt{F^{0}_{\mathrm{CJ}}}}}\ (\text{amplitude damping}),
    \label{eq:alpha_0_th_ad}
\end{equation}
we simply recover $\mathcal{R}_{\mathrm{det}}=1$. In addition, in the $p_0\to 0$ and $|\alpha_0|= 1$ limit, Eqs.~\eqref{eq:rprob_ad_a2} and\eqref{eq:rdet_ad_a2} are reduced to
\begin{equation}
    \mathcal{R}_{\mathrm{prob}}\to d, \ \mathcal{R}_{\mathrm{det}} \to \frac{2d}{d+1} \ (\text{amplitude damping}).
\end{equation}

\subsubsection{Nonidentical channels: an example}
We mention that the availability of analytic solutions is not limited to identical channels. For instance, if we have one dephasing channel and one depolarizing channel with equal $p_0$, then in the $|\alpha_0|\to 1$ limit, the only other path DOF measurement outcome is $\ket{1}_{\mathrm{c}}$:
\begin{equation}
P_1\rho^{\mathrm{out}}_{\mathrm{ab}, 1}= \frac{p_0}{12}(\varrho^{11}_{\mathrm{ab}}+\varrho^{22}_{\mathrm{ab}}+4\varrho^{33}_{\mathrm{ab}});
\label{eq:postmeas_st_depha_depol_a2}
\end{equation}
in this case, $P_1=p_0/2$, $V^{(1)}=Z$,
\begin{equation}
    \bra{\Phi^+}P_1\rho^{\mathrm{out,corr}}_{\mathrm{ab}, 1}\ket{\Phi^+}_{\mathrm{ab}} = \frac{p_0}{3},
\end{equation}
and
\begin{align}
    \mathcal{R}_{\mathrm{det}} 
    & =\frac{1-(1-p_0)}{1-[(1-p_0)+\frac{p_0}{3}]}\nonumber\\
    & = \frac{3}{2} \ (\text{dephasing+depolarizing}),
    \label{eq:rdet_dephadepol_a2}
\end{align}
which is numerically verified in Fig.~\ref{fig:two_ineq_channels}(a1).

\section{General form of the vacuum interference operators\label{app:Proof}}
\setcounter{equation}{0} 
The form of $F_\mathrm{vio}$ appearing in Eq.~\eqref{eq:vio} requires that the vacuum amplitudes $\sum_{s} |\alpha_{s}|^{2}$ = 1. This is because $\alpha_{s}$ are the expansion coefficients of the environment state $\ket{\varepsilon}_{\mathrm{e}}$ in a certain complete orthonormal basis $\ket{s}_{\mathrm{e}}$. This basis corresponds to a potentially very large set of physical Kraus operators $K_{s}$, which are generally not orthogonal to each other. However, in practice we are usually interested in an effective representation of the quantum channel with orthogonal Kraus operators instead, e.g. Eqs.~\eqref{eq:dephasingnoise}--\eqref{eq:amplitudedampingnoise}. In this appendix, we will examine the properties of such an effective representation; in particular, we will show that the magnitude of the effective vacuum amplitude vector is no greater than $1$.

We denote the orthogonal effective Kraus operators by $\tilde{K}_{r}$ and the corresponding effective vacuum amplitudes by $\tilde{\alpha}_{r}$. $\tilde{K}_{r}$ satisfies the orthogonality condition
\begin{equation}
    \operatorname{tr}(\tilde{K}_{r}\tilde{K}^\dagger_{r'})=\operatorname{tr}(\tilde{K}_{r}\tilde{K}^\dagger_{r}) \delta_{rr'}.
    \label{eq:kraus_eff_orthogonal}
\end{equation}
Importantly, Eq.~\eqref{eq:kraus_eff_orthogonal} does not generally apply to the original physical Kraus operators $K_s$. It is useful to expand $K_s$ in the $\tilde{K}_{r}$ basis,
\begin{equation}
    K_{s}=\sum_{r}L_{sr}\tilde{K}_{r}.
    \label{eq:kraus_transform}
\end{equation}
Since $K_{s}$ and $\tilde{K}_{r}$ represent the same channel, we have for an arbitrary density matrix $\rho$
\begin{equation}
    \sum_{s}\sum_{kl}(K_{s})_{ik}\rho_{kl}(K^\dagger_{s})_{lj}=\sum_{r}\sum_{kl}(\tilde{K}_{r})_{ik}\rho_{kl}(\tilde{K}^\dagger_{r})_{lj},
\end{equation}
or equivalently
\begin{equation}
    \sum_{s}\sum_{rr'}L_{sr} L^*_{sr'}(\tilde{K}_{r})_{ik}(\tilde{K}_{r'})^*_{jl}=\sum_{r}(\tilde{K}_{r})_{ik}(\tilde{K}_{r})^*_{jl},
    \label{eq:kraus_orthonormal_0}
\end{equation}
where we have inserted Eq.~\eqref{eq:kraus_transform}. Multiplying Eq.~\eqref{eq:kraus_orthonormal_0} by $(\tilde{K}_{r_1})^*_{ik}(\tilde{K}_{r_2})_{jl}$, summing over all state indices $i,j,k,l$ and using Eq.~\eqref{eq:kraus_eff_orthogonal}, we obtain
\begin{equation}
    \sum_{s}L_{sr_1} L^*_{sr_2}=\delta_{r_1 r_2},\ L^\dagger L=\id.
    \label{eq:kraus_orthonormal}
\end{equation}
Therefore, if we perform a singular value decomposition
\begin{equation}
    L=U\Sigma V^\dagger
    \label{eq:kraus_transform_svd}
\end{equation}
where $U_{ss'}$ and $V_{rr'}$ are unitary matrices, and $\Sigma_{sr}$ is a rectangular diagonal matrix with real and nonnegative diagonal elements, then all singular values (diagonal elements of $\Sigma$) are $1$.

The vacuum interference operator $F_{\mathrm{vio}}$ should also be the same in both representations:
\begin{equation}
    F_{\mathrm{vio}}=\sum_r \tilde{\alpha}^*_{r} \tilde{K}_{r}=\sum_s \alpha^*_{s} K_{s} = \sum_{sr}\alpha^*_{s} L_{sr}\tilde{K}_{r}.
\end{equation}
Using Eq.~\eqref{eq:kraus_eff_orthogonal} again immediately gives the relation between the two vacuum amplitude vectors $\alpha$ and $\tilde\alpha$:
\begin{equation}
    \tilde{\alpha}^*_{r} = \sum_{s}\alpha^*_{s} L_{sr},\  \tilde{\alpha}=L^\dagger \alpha.
\end{equation}
Therefore, the effective vacuum amplitude vector $\tilde{\alpha}$ satisfies
\begin{align}
    \sum_{r}|\tilde{\alpha}_{r}|^2 & = \tilde{\alpha}^\dagger \tilde{\alpha}=\alpha ^\dagger (L L^\dagger) \alpha = (\alpha ^\dagger U) (\Sigma\Sigma^\dagger ) (U^\dagger\alpha) \nonumber\\
    & \leq (\alpha ^\dagger U)(U^\dagger\alpha)=1,
    \label{eq:vav_eff_bound}
\end{align}
where we have inserted the singular value decomposition Eq.~\eqref{eq:kraus_transform_svd}, and taken into account the fact that the diagonal elements of $\Sigma\Sigma^\dagger $ are either $1$ or $0$.

Eq.~\eqref{eq:vav_eff_bound} indicates that the effective vacuum amplitude vector cannot have a length greater than $1$. A simple example is a dephasing channel with two nonequivalent $Z$ Kraus operators, $K_0=(1/\sqrt{2})\id, K_1=K_2=(1/2)Z$ and $\alpha=(1/\sqrt{2},1/2,-1/2)$. To reproduce the vacuum interference operator $F_\mathrm{vio}=(1/2)\id$, we can choose the effective channel to be $\tilde{K}_{0}=(1/\sqrt{2})\id,\tilde{K}_1=(1/\sqrt{2})Z$ and $\tilde{\alpha}=(1/\sqrt{2},0)$; it is clear that $|\tilde{\alpha}|=1/\sqrt{2}<1=|\alpha|$. The microscopic model in App.~\ref{app:microscopic_special} is another example; see the insets of Fig.~\ref{fig:identical_detprob_vacamp}(a3) (dephasing) and (b3) (depolarizing), and Eq.~\eqref{eq:vio_micro_full_ad} (amplitude damping).

\section{Microscopic model of quantum channels with nontrivial vacuum interference}
\label{app:microscopic}
In this appendix, we discuss a microscopic model of the quantum channels Eqs.~\eqref{eq:dephasingnoise}--\eqref{eq:amplitudedampingnoise}, where a qubit interacts sequentially with many environment qubits~\cite{attal06,grimmer16}. We will show that this model naturally yields nontrivial vacuum interference operators which depend on the noise parameter; furthermore, it gives us a concrete example on which we can base our numerical demonstrations of the path superposition protocol.

\subsection{A concrete example}
\label{app:microscopic_special}

We consider an information carrier qubit $\mathbf{\sigma}$ and a collection of $N_{\mathrm{env}}$ environment qubits $\mathbf{s}_{n}$, $n=0,1,\dots,N_{\mathrm{env}}-1$. During the short time interval $t_n<t<t_{n+1}$, where $t_n=n\Delta t$, $\mathbf{\sigma}$ interacts with $\mathbf{s}_{n}$ through an (anisotropic) interaction
\begin{equation}
    H_n=\sum_{l=1}^{3} \hbar \lambda_l \sigma_l s_{n,l}.
    \label{eq:micro_ham}
\end{equation}
For simplicity we assume the interaction strength in the direction $l$, $\lambda_l$, is independent of $n$. We further assume that the environment qubit $\mathbf{s}_{n}$ is in the state $\rho_{s_n}$ at time $t_{n}$. 

To find the effective quantum channel for this microscopic model, we denote the density matrix at $t_{n}$ by $\rho_{n}$. We can express $\rho_{n+1}$ in terms of $\rho_{n}$ by calculating the joint unitary time evolution of the system and the environment and then tracing out the environment:
\begin{align}
    \rho_{n+1} &= \operatorname{tr}_{\mathbf{s}_n}\Big[e^{-\mathrm{i}\Delta t\sum_{l} \lambda_l \sigma_l s_{n,l}}\rho_{n}\otimes\rho_{s_n} \nonumber\\
    &\times e^{\mathrm{i}\Delta t\sum_{l'} \lambda_{l'} \sigma_{l'} s_{n,l'}}\Big].
    \label{eq:rho_diag_evo}
\end{align}

In the presence of path superposition, we can similarly find the time evolution of the full density matrix including the path DOF. Note that Eq.~\eqref{eq:micro_ham} represents the contribution of a path to the total Hamiltonian only when the information carrier qubit is traversing that path; it is reasonable to assume that there is no contribution when the information carrier is not traversing that path. Based on this assumption, we can write the time evolution of the off-diagonal blocks of the density matrix as
\begin{align}
    \rho^{(ij)}_{n+1}&=\operatorname{tr}_{\mathbf{s}^{(i)}_n,\mathbf{s}^{(j)}_n}\Big[e^{-\mathrm{i}\Delta t \sum_{l} \lambda^{(i)}_l \sigma_l s^{(i)}_{n,l}}\rho^{(ij)}_{n}\otimes\rho^{(i)}_{s_n} \nonumber\\
    &\otimes\rho^{(j)}_{s_n} e^{\mathrm{i}\Delta t \sum_{l'} \lambda^{(j)}_{l'} \sigma_{l'} s^{(j)}_{n,l'}}\Big],
    \label{eq:rho_offdiag_evo}
\end{align}
where we have added the path index $i$ and $j\neq i$ to the environment qubits $\mathbf{s}^{(i)}_n$, the system-environment couplings $\lambda_l^{(i)}$ and the environment states $\rho^{(i)}_{s_n}$. If the paths are independent, Eq.~\eqref{eq:rho_offdiag_evo} factorizes into the form of Eq.~\eqref{eq:coh_part}, $\rho^{(ij)}_{n+1}=F^{(i)}_{\mathrm{vio}}\rho^{(ij)}_{n}F^{(j)\dagger}_{\mathrm{vio}}$, allowing us to extract the vacuum interference operator for path $i$:
\begin{equation}
    F^{(i)}_{\mathrm{vio}}=\operatorname{tr}_{\mathbf{s}^{(i)}_n}\Big[e^{-\mathrm{i}\Delta t \sum_{l} \lambda^{(i)}_l \sigma_l s^{(i)}_{n,l}}\rho^{(i)}_{s_n}\Big].
    \label{eq:vio_micro_gen}
\end{equation}

To evaluate the partial traces in Eqs.~\eqref{eq:rho_diag_evo} and \eqref{eq:vio_micro_gen}, we need to know the distribution of the environment states $\rho^{(i)}_{s_n}$. We will see below that this distribution has a profound impact on the nature of the quantum channel.

\subsubsection{Dephasing channel and depolarizing channel}

We first consider the particularly simple case where the environment states are completely random, so that $\rho^{(i)}_{s_n}$ is uniformly distributed on the Bloch sphere and $\operatorname{tr}_{\mathbf{s}^{(i)}_n}\sigma_l\rho^{(i)}_{s_n}$ vanishes for $l=1,2,3$. To $O(\lambda^2)$, Eq.~\eqref{eq:rho_diag_evo} turns into the operator-sum form
\begin{align}
    \rho_{n+1} =& \sum_{l=0}^{3} K_l \rho_{n} K_l^\dagger,\\
    K_0=&\Big[1-\frac{\Delta t^2}{2}(\lambda_1^2+\lambda_2^2+\lambda_3^2)\Big]\id, K_l=\Delta t\lambda_l \sigma_l.
    \label{eq:channel_micro_infntsm_kr}
\end{align}
Meanwhile, to $O(\lambda^2)$, Eq.~\eqref{eq:vio_micro_gen} simply becomes
\begin{equation}
    F_{\mathrm{vio}}=\Big[1-\frac{\Delta t^2}{2}(\lambda_1^2+\lambda_2^2+\lambda_3^2)\Big]\id,
    \label{eq:vio_micro_infntsm}
\end{equation}
where we have suppressed the path index.

We are often concerned with the continuum limit, where the total time the information qubit interacts with the environment is fixed to $t$, but there are a large number of environment qubits $N_{\mathrm{env}}\to \infty$, such that $\Delta t=t/N_{\mathrm{env}}\to 0$. The resulting quantum channel amounts to iterating the channel in Eq.~\eqref{eq:channel_micro_infntsm_kr} by $N_{\mathrm{env}}$ times. This is most easily achieved in the superoperator formalism: once we define $\mu_l=\lambda_l^2 \Delta t$, the superoperator corresponding to Eq.~\eqref{eq:channel_micro_infntsm_kr} takes the form
\begin{widetext}
\begin{align}
    \Lambda 
    =
    \ \ \begin{blockarray}{cccc}
    \ket{0}\bra{0} & \ket{1}\bra{0} & \ket{0}\bra{1} & \ket{1}\bra{1} \\
    \begin{block}{(cccc)}
        1-\Delta t(\mu_1+\mu_2)& 0 & 0 &\Delta t(\mu_1+\mu_2) \\
        0&1-\Delta t(\mu_1+\mu_2+2\mu_3) & \Delta t(\mu_1-\mu_2)& 0 \\
        0& \Delta t(\mu_1-\mu_2)&1-\Delta t(\mu_1+\mu_2+2\mu_3) & 0 \\
        \Delta t(\mu_1+\mu_2)& 0 & 0 &1-\Delta t(\mu_1+\mu_2) \\
    \end{block}
    \end{blockarray}\ \, ,
\end{align}
\end{widetext}
and has eigenvalues $1$, $1-2\Delta t(\mu_1+\mu_2)$, $1-2\Delta t(\mu_2+\mu_3)$ and $1-2\Delta t(\mu_3+\mu_1)$. Iterating the infinitesimal channel by $N_{\mathrm{env}}=t/\Delta t$ times corresponds to the superoperator $\Lambda^{N_{\mathrm{env}}}$, which has eigenvalues $1$, $e^{-2t(\mu_1+\mu_2)}$, $e^{-2t(\mu_2+\mu_3)}$ and $e^{-2t(\mu_3+\mu_1)}$. Therefore, the Kraus operators of the continuum limit channel have the form
\begin{align}
    K_0= &\sqrt{\frac{1+e^{-2t(\mu_1+\mu_2)}+e^{-2t(\mu_2+\mu_3)}+e^{-2t(\mu_3+\mu_1)}}{4}}\id,\\
    K_3= &\sqrt{\frac{1+e^{-2t(\mu_1+\mu_2)}-e^{-2t(\mu_2+\mu_3)}-e^{-2t(\mu_3+\mu_1)}}{4}}\sigma_z,
    \label{eq:channel_micro_full_kr}
\end{align}
and $K_1$ and $K_2$ are obtained from $K_3$ by cyclic permutations of indices. On the other hand, the iteration of the infinitesimal vacuum interference operator Eq.~\eqref{eq:vio_micro_infntsm} is even more straightforward:
\begin{equation}
    F_{\mathrm{vio}}= e^{-\frac{1}{2}t(\mu_1+\mu_2+\mu_3)}\id.
    \label{eq:vio_micro_full}
\end{equation}

Eq.~\eqref{eq:channel_micro_full_kr} represents an anisotropic depolarizing channel. In the special cases of $\mu_3=\mu, \mu_1=\mu_2=0$ and $\mu_l=\mu$, Eq.~\eqref{eq:channel_micro_full_kr} is reduced to a dephasing channel and an isotropic depolarizing channel, respectively. In both cases, we can now easily relate the error probability to the vacuum interference operator of the channel. For the dephasing channel, comparing with Eq.~\eqref{eq:dephasingnoise}, we find
\begin{equation}
    p_0=\frac{1-e^{-2\mu t}}{2},\ F_{\mathrm{vio}}=(1-2p_0)^{\frac{1}{4}}\id;
    \label{eq:dephasing_micro_vio}
\end{equation}
for the isotropic depolarizing channel, comparing with Eq.~\eqref{eq:depolarizingnoise}, we find
\begin{equation}
    p_0=\frac{3(1-e^{-4\mu t})}{4},\ F_{\mathrm{vio}}=(1-\frac{4}{3}p_0)^{\frac{3}{8}}\id.
    \label{eq:depolarizing_micro_vio}
\end{equation}
It is clear from Eqs.~\eqref{eq:dephasing_micro_vio}, \eqref{eq:depolarizing_micro_vio} that our dephasing and depolarizing channels have strong vacuum coherence in the short-time limit $t\to 0$ and $p_0 \to 0$, and lose vacuum coherence completely in the long-time limit $t\to \infty$ and $p_0 \to 1/2$ or $p_0 \to 3/4$. Interestingly, Eqs.~\eqref{eq:dephasing_micro_vio} and \eqref{eq:depolarizing_micro_vio} are in agreement with an alternative microscopic model in Ref.~\cite{sqem2}, where the system qubit is coupled to an environment represented by a fluctuating magnetic field.

\subsubsection{Amplitude damping channel}

Another possibility for the environment state is that $\rho^{(i)}_{s_n}$ is always a pure state. This may happen if the environment DOFs tend to rapidly relax to a ground state after each use of the quantum channel. For simplicity, we choose $\rho^{(i)}_{s_n}=\proj{0}$, $\lambda_3=0$, and $\lambda_1=\lambda_2=\lambda$. Then, to $O(\lambda^2)$, Eqs.~\eqref{eq:rho_diag_evo} and \eqref{eq:vio_micro_gen} are simplified as
\begin{align}
    \rho_{n+1} =& \sum_{l=0}^{1} K_l \rho_{n} K_l^\dagger,\\
    K_0=&\Big[1-(\lambda\Delta t)^2\Big]\id+(\lambda\Delta t)^2 \sigma_3,\nonumber\\ K_1=&\lambda\Delta t( \sigma_1+i \sigma_2),
    \label{eq:channel_micro_infntsm_kr_ad}
\end{align}
and
\begin{equation}
    F_{\mathrm{vio}}=\Big[1-(\lambda\Delta t)^2\Big]\id+(\lambda\Delta t)^2 \sigma_3.
    \label{eq:vio_micro_infntsm_ad}
\end{equation}
Again, denoting $\mu=\lambda^2 \Delta t$, we find the superoperator corresponding to Eq.~\eqref{eq:channel_micro_infntsm_kr_ad} to be
\begin{equation}
    \Lambda = \begin{pmatrix}
        1& 0 & 0 &4\mu\Delta t \\
        0&1-2\mu\Delta t & 0& 0 \\
        0& 0&1-2\mu\Delta t & 0 \\
        0& 0 & 0 &1-4\mu\Delta t
    \end{pmatrix},
\end{equation}
with eigenvalues $1$, $1-2\mu\Delta t$, $1-2\mu\Delta t$ and $1-4\mu\Delta t$. Therefore, the continuum limit channel has eigenvalues $1$, $e^{-2\mu t}$, $e^{-2\mu t}$ and $e^{-4\mu t}$, and the continuum limit Kraus operators are
\begin{equation}
    K_0=\proj{0}+e^{-2\mu t}\proj{1},\
    K_1=\sqrt{1-e^{-4\mu t}}\ket{0}\bra{1}.
    \label{eq:channel_micro_full_kr_ad}
\end{equation}
The iteration of the infinitesimal vacuum interference operator Eq.~\eqref{eq:vio_micro_infntsm_ad} is again straightforward:
\begin{equation}
    F_{\mathrm{vio}}=\proj{0}+e^{-2\mu t}\proj{1}=K_0.
    \label{eq:vio_micro_full_ad}
\end{equation}
Therefore, for our microscopic model, the amplitude damping channel has $F_{\mathrm{vio}}=K_0$ independent of the noise parameter $p_0=1-e^{-4\mu t}$ (see Eq.~\eqref{eq:amplitudedampingnoise}).

\subsection{General considerations}
\label{app:microscopic_general}

The above example inspires us to discuss general properties of an arbitrary microscopic model, which describes one of the three channels we have considered in this work Eqs.~\eqref{eq:dephasingnoise}, \eqref{eq:depolarizingnoise} and \eqref{eq:amplitudedampingnoise}.

We first note the fact that when concatenating two channels of the same type, we obtain a new channel of the same type as the original channels. Assuming the error probabilities of the original channels are $p_0^{(1)}$ and $p_0^{(2)}$ and that of the concatenated channel is $p_0$, we easily show that
\begin{align}
    & \text{Dephasing: } (1-2p_0^{(1)})(1-2p_0^{(2)})=1-2p_0, \label{eq:dephasingconcat}\\
   & \text{Depolarizing: } (1-\frac{4}{3}p_0^{(1)})(1-\frac{4}{3}p_0^{(2)})=1-\frac{4}{3}p_0, 
   \label{eq:depolarizingconcat}\\
    & \text{Amplitude damping: } (1-p_0^{(1)})(1-p_0^{(2)})=1-p_0.\label{eq:amplitudedampingconcat}
\end{align}

To facilitate our discussion below, we also mention that the vacuum coherence is perfect when there is no interaction with the environment. This is easily justified by noting that the only Kraus operator is the identity and the environment Hilbert space is one-dimensional, which necessitates (up to an unimportant phase) a trivial vacuum amplitude     vector $\alpha_0=1$ and a perfect vacuum interference operator $F_{\mathrm{vio}}=\id$.

We now make an important conjecture: for a given microscopic model, the vacuum interference operator $F_{\mathrm{vio}}$ of a channel is a smooth function of its error probability $p_0$. This conjecture follows naturally if, for instance, the interactions with the environment DOFs are spatially (temporally) uniformly distributed along the path of the information (during the action of the channel), such that both $F_{\mathrm{vio}}$ and $p_0$ are smooth functions of the length (traversal time) of the channel. With the above conjecture, we consider again the concatenation of two channels with the same microscopic model but different error probabilities $p_0^{(1)}$ and $p_0^{(2)}$. If we place the concatenated channel in a path superposition, Eq.~\eqref{eq:coh_part} suggests that the concatenated vacuum interference operator should satisfy
\begin{equation}
    F_{\mathrm{vio}}(p_0)=F_{\mathrm{vio}}(p_0^{(2)})F_{\mathrm{vio}}(p_0^{(1)}).
    \label{eq:vio_concat}
\end{equation}

Equation~\eqref{eq:vio_concat} places a strong constraint on the form of $F_{\mathrm{vio}}$. For the dephasing channel, for instance, we can expand $F_{\mathrm{vio}}$ in the weak-noise limit:
\begin{equation}
    F_{\mathrm{vio}}(p_0)=(1+\sum_{n=1}^{\infty} f^{(0)}_{n}p_0^n)\id+(\sum_{n=1}^{\infty} f^{(z)}_{n}p_0^n)\sigma_3.
    \label{eq:vio_depha_expand}
\end{equation}
Inserting Eqs.~\eqref{eq:dephasingconcat} and \eqref{eq:vio_depha_expand} into Eq.~\eqref{eq:vio_concat}, we find the higher order coefficients are fully determined by the first-order coefficients $f^{(0)}_{1}$ and $f^{(z)}_{1}$:
\begin{align}
    f^{(0)}_{2}&=\frac{1}{2}\left[\left(2f^{(0)}_{1}+(f^{(0)}_{1})^2\right)+(f^{(z)}_{1})^2\right],\nonumber\\
    f^{(z)}_{2}&=f^{(z)}_{1}\left[1+f^{(0)}_{1}\right],\nonumber\\
    f^{(0)}_{3}&=\frac{1}{6}\Big[\left(8f^{(0)}_{1}+6(f^{(0)}_{1})^2+(f^{(0)}_{1})^3\right)\nonumber\\
    &+(f^{(z)}_{1})^2 \left(6+3f^{(0)}_{1}\right)\Big],\nonumber\\
    f^{(z)}_{3}&=\frac{1}{6}\left[f^{(z)}_{1}\left(8+12f^{(0)}_{1}+3(f^{(0)}_{1})^2\right)+(f^{(z)}_{1})^3 \right],\nonumber\\
    &\dots \label{eq:vio_micro_depha_coeff}
\end{align}
In other words, the vacuum coherence properties of any microscopic model are completely encoded in $f^{(0)}_{1}$ and $f^{(z)}_{1}$.

It is worth pointing out that $f^{(0)}_{1}$ and $f^{(z)}_{1}$ themselves must satisfy further constraints. We can upper-bound the norm of $F_{\mathrm{vio}}$ as follows:
\begin{align}
    &\frac{1}{2}\operatorname{tr}F^\dagger_{\mathrm{vio}}F_{\mathrm{vio}} =(1-p_0)|\tilde{\alpha}_{0}|^2 +p_0 |\tilde{\alpha}_{1}|^2 \nonumber\\
     \leq & (1-p_0)|\tilde{\alpha}_{0}|^2 +p_0 (1-|\tilde{\alpha}_{0}|^2)\leq 1-p_0,
     \label{eq:vio_norm_bound1}
\end{align}
where the effective vacuum amplitudes $\tilde{\alpha}_{0}$ and $\tilde{\alpha}_{1}$ correspond to the two effective Kraus operators in Eq.~\eqref{eq:dephasingnoise}, and we have used Eq.~\eqref{eq:vav_eff_bound} and the facts that $p_0 \leq 1/2$   and $|\tilde{\alpha}_{0}|^2 \leq 1$. On the other hand, to $O(p_0^3)$, Eqs.~\eqref{eq:vio_depha_expand} and \eqref{eq:vio_micro_depha_coeff} together lead to
\begin{align}
    &\frac{1}{2}\operatorname{tr}F^\dagger_{\mathrm{vio}}F_{\mathrm{vio}}
    \nonumber\\= & 1+(2\operatorname{Re}f^{(0)}_{1})p_0+(|f^{(0)}_{1}|^2+2\operatorname{Re}f^{(0)}_{2}+|f^{(z)}_{1}|^2)p_0^2\nonumber\\
    =&1+(2\operatorname{Re}f^{(0)}_{1})p_0+[2\operatorname{Re}f^{(0)}_{1}\nonumber\\
    +&2(\operatorname{Re}f^{(0)}_{1})^2+2(\operatorname{Re}f^{(z)}_{1})^2]p_0^2.
    \label{eq:vio_norm_bound2}
\end{align}
Comparing Eqs.~\eqref{eq:vio_norm_bound1} and \eqref{eq:vio_norm_bound2}, we find
\begin{equation}
    \operatorname{Re}f^{(0)}_{1} \leq -\frac{1}{2},
    \label{eq:re_f0_1}
\end{equation}
and, when $|\operatorname{Re}f^{(0)}_{1} +1/2|\ll 1$ so that the $O(p_0^2)$ term is as important as the $O(p_0)$ term,
\begin{equation}
    (\operatorname{Re}f^{(z)}_{1})^2 \leq -\operatorname{Re}f^{(0)}_{1}
    -(\operatorname{Re}f^{(0)}_{1})^2\approx \frac{1}{4}.
    \label{eq:re_fz_1}
\end{equation}
Equations~\eqref{eq:re_f0_1} and \eqref{eq:re_fz_1} are both constraints on the real parts rather than the imaginary parts, since at the lowest nontrivial order the imaginary parts simply correspond to $p_0$-dependent overall phases of $F_{\mathrm{vio}}$.

If we let $f^{(z)}_{1}=0$, it is possible to calculate the expansion coefficients in Eq.~\eqref{eq:vio_micro_depha_coeff} to all orders and obtain a closed-form solution
\begin{equation}
    F_{\mathrm{vio}}(p_0)=(1-2p_0)^{-\frac{1}{2}f^{(0)}_{1}}\id.
    \label{eq:vio_dephase_no_z}
\end{equation}
Equation~\eqref{eq:dephasing_micro_vio} is simply a special case of Eq.~\eqref{eq:vio_dephase_no_z} with $f^{(0)}_{1}=-1/2$.

It is straightforward to derive similar results for the depolarizing channel and the amplitude damping channel, whose vacuum coherence properties are likewise always encoded in a few parameters. {\done For the depolarizing channel, in the special case that $F_{\mathrm{vio}}\propto \id$,
Eq.~\eqref{eq:vio_concat} leads to
\begin{equation}
    F_{\mathrm{vio}}(p_0)=(1-\frac{4}{3}p_0)^{-\frac{3}{4}f^{(0)}_{1}}\id,
    \label{eq:vio_depol_no_xyz}
\end{equation}
while the constraint on $f^{(0)}_{1}$, Eq.~\eqref{eq:re_f0_1}, continues to apply.

For the amplitude damping channel, an ansatz consistent with Eq.~\eqref{eq:vio_concat} is
\begin{align}
    &F_{\mathrm{vio}}(p_0)=(1-p_0)^\Delta K_0\nonumber\\
    = & (1-p_0)^\Delta\proj{0}+(1-p_0)^{\Delta+\frac{1}{2}}\proj{1},
    \label{eq:vio_ad_arb}
\end{align}
where $\Delta$ is a model-dependent real number that is independent of $p_0$. Thus
\begin{align}
    &\frac{1}{2}\operatorname{tr}F^\dagger_{\mathrm{vio}}F_{\mathrm{vio}} =(1-p_0)^{2\Delta}(1-\frac{p_0}{2}) \nonumber\\
    =&(1-\frac{p_0}{2})|\tilde{\alpha}_{0}|^2 +\frac{p_0}{2} |\tilde{\alpha}_{1}|^2\nonumber\\
    \leq &(1-p_0)|\tilde{\alpha}_{0}|^2 +\frac{p_0}{2} \leq 1-\frac{p_0}{2},
\end{align}
resulting simply in $\Delta \geq 0$. Interestingly, in the weak-noise limit $p_0 \ll 1$, $F_{\mathrm{vio}}$ has an expansion similar to Eq.~\eqref{eq:vio_depha_expand}:
\begin{equation}
    F_{\mathrm{vio}}(p_0)=(1+ f^{(0)}_{1}\frac{p_0}{2})\id+f^{(z)}_{1}\frac{p_0}{2}\sigma_3+O(p_0^2),
    \label{eq:vio_ad_expand}
\end{equation}
where $f^{(0)}_{1}=-2\Delta-1/2$ and $f^{(z)}_{1}=1/2$ satisfy the constraints Eq.~\eqref{eq:re_f0_1} and \eqref{eq:re_fz_1}. Note that $p_0/2$ rather than $p_0$ appears in Eq.~\eqref{eq:vio_ad_expand} since $F^{0}_{\mathrm{CJ}}=1-p_0/2+O(p_0^2)$ for the amplitude damping channel.


The weak-noise expansions Eqs.~\eqref{eq:vio_depha_expand} and \eqref{eq:vio_ad_expand} imply that we can parameterize the vacuum interference of the microscopic models described by Eqs.~\eqref{eq:vio_dephase_no_z}, \eqref{eq:vio_depol_no_xyz} and \eqref{eq:vio_ad_arb} on equal footing via the coefficient $f^{(0)}_{1}$, or equivalently the more convenient parameter $\Delta=-f^{(0)}_{1}/2-1/4$. In terms of $\Delta$, Eqs.~\eqref{eq:vio_dephase_no_z}, \eqref{eq:vio_depol_no_xyz} and \eqref{eq:vio_ad_arb} read
\begin{align}
    & \text{Dephasing: } F_{\mathrm{vio}}(p_0)=(1-2p_0)^{\frac{1}{4}+\Delta}\id, \nonumber\\
    & \text{Depolarizing: } F_{\mathrm{vio}}(p_0)=(1-\frac{4}{3}p_0)^{\frac{3}{8}+\frac{3}{2}\Delta}\id, \nonumber\\
    & \text{Amplitude damping: } \nonumber\\
    &F_{\mathrm{vio}}(p_0)=(1-p_0)^\Delta\proj{0}+(1-p_0)^{\Delta+\frac{1}{2}}\proj{1}.
    \label{eq:vio_micro_tunable}
\end{align}
In our multi-node protocol demonstration (Fig.~\ref{fig:multinode_complex}), we adopt the microscopic models defined by Eq.~\eqref{eq:vio_micro_tunable}. Here the parameter $\Delta \geq 0$ depends on the underlying noise mechanism; $\Delta=0$ corresponds to the microscopic model discussed in App.~\ref{app:microscopic_special}, and larger $\Delta$ indicates weaker vacuum coherence for a given fidelity.}

\section{Noise on the path DOF}
\label{app:noise_path_dof}

In this appendix, we derive the effective quantum channel for our path DOF noise model, describe how it is implemented in numerical simulations, and quantitatively explain the effects of the path DOF noise on our protocol. For simplicity, we limit ourselves to our minimal implementation of the path DOF unitary $U(\boldsymbol{\theta})$ adopted in Sec.~\ref{sec:numerics_2_node}.

\subsection{Effective quantum channel}
\label{app:path_noise_channel}

In the $d=2$ case, only one parameter is needed, and $U$ has the particularly simple form
\begin{equation}
    U(\theta)=\exp(-i\frac{\theta}{2}Y)=\begin{pmatrix}
        \cos \frac{\theta}{2} & -\sin \frac{\theta}{2} \\
        \sin \frac{\theta}{2} & \cos \frac{\theta}{2}
    \end{pmatrix}.
\end{equation}
For finite $\sigma$, we can numerically implement the path DOF noise by writing down the corresponding superoperator $U(\cdot)U^\dagger$,
\begin{align}
     &U(\theta)(\cdot)[U(\theta)]^\dagger \nonumber\\
    =&
    \begin{pmatrix}
        \cos^2 \frac{\theta}{2} & -\frac{1}{2}\sin \theta & -\frac{1}{2}\sin \theta & \sin^2 \frac{\theta}{2} \\
        \frac{1}{2}\sin \theta & \cos^2 \frac{\theta}{2} & -\sin^2 \frac{\theta}{2} & -\frac{1}{2}\sin \theta \\
        \frac{1}{2}\sin \theta & -\sin^2 \frac{\theta}{2} & \cos^2 \frac{\theta}{2} & -\frac{1}{2}\sin \theta \\
        \sin^2 \frac{\theta}{2} & \frac{1}{2}\sin \theta & \frac{1}{2}\sin \theta & \cos^2 \frac{\theta}{2} \\
    \end{pmatrix},
    \label{eq:u_dof_simple_super}
\end{align}
and calculating the $\theta$ integral in Eq.~\eqref{eq:control_noise} to obtain $\mathcal{C}_{\theta}$ in the form of a noise-averaged superoperator. On the other hand, in the weak path DOF noise limit $\sigma \to 0$, we can analytically expand $U(\cdot)U^\dagger$ to $O((\theta-\bar{\theta})^2)$:
\begin{align}
    & U(\tilde\theta)\rho[U(\tilde\theta)]^\dagger \nonumber\\
    = & e^{-i\frac{Y}{2}\theta}e^{-i\frac{Y}{2}(\tilde\theta-\theta)}\rho e^{i\frac{Y}{2}(\tilde\theta-\theta)}e^{i\frac{Y}{2}\theta} \nonumber\\
    = & e^{-i\frac{Y}{2}\theta}[\id - i\frac{1}{2}(\tilde\theta-\theta)Y-\frac{1}{8}(\tilde\theta-\theta)^2 \id]\nonumber\\
    & \times\rho [\id + i\frac{1}{2}(\tilde\theta-\theta)Y-\frac{1}{8}(\tilde\theta-\theta)^2 \id]e^{i\frac{Y}{2}\theta}.
\end{align}
Integrating over $\tilde\theta$, we are left with only the constant and quadratic terms,
\begin{equation}
    [U(\theta)]^\dagger \mathcal{C}_{\theta}(\rho)U(\theta) = (1-\frac{\sigma^2}{4})\rho+\frac{\sigma^2}{4}Y\rho Y.
    \label{eq:c_theta_twopaths}
\end{equation}
Therefore, our implementation of the path DOF noise with strength $\sigma \ll 1$ has a particularly simple effect in the case of $d=2$: a channel with a $Y$ error that occurs with a probability of $\sigma^2/4$.

To conclude this section, we briefly comment on the implementation of arbitrary $d$. As in Eq.~\eqref{eq:u_dof_simple_super}, it is straightforward to numerically find the channel $\mathcal{C}_{\boldsymbol{\theta}}$ corresponding to Eq.~\eqref{eq:min_unitary} in the superoperator representation; indeed, we can factorize the $\boldsymbol{\tilde\theta}$ integral and consider the contribution of each two-path unitary $T_{m,n}$ separately. In addition, in the $\sigma\to 0$ limit, the $O(\sigma^2)$ scaling of the error probability continues to apply. However, for $d>2$, a complication arises in that different $T_{m,n}$ usually do not commute, such that the form of the effective channel is much less transparent than Eq.~\eqref{eq:c_theta_twopaths}.

\subsection{Role of path DOF noise in protocol}
\label{app:path_noise_effect}

Let us now study how the path DOF noise affects our two-node protocol for $d=2$ using Eq.~\eqref{eq:c_theta_twopaths}. We assume an equal superposition of two identical channels with noise parameter $p_0$, both described by the microscopic model in App.~\ref{app:microscopic_special}.

\subsubsection{Dephasing channels}

In the case of dephasing channels, after some algebra, we find the post-measurement states at $O(\sigma^2)$ and $O(p_0)$ to be
\begin{subequations}
\begin{equation}
    P_0\rho^{\mathrm{out}}_{\mathrm{ab}, 0} = (1-p_0-\frac{\sigma^2}{2})\varrho^{00}_{\mathrm{ab}}+\frac{p_0}{2}\varrho^{33}_{\mathrm{ab}},
\end{equation}
\begin{equation}
    P_1\rho^{\mathrm{out}}_{\mathrm{ab}, 1} = \frac{\sigma^2}{2}\varrho^{00}_{\mathrm{ab}}+\frac{p_0}{2}\varrho^{33}_{\mathrm{ab}},
\end{equation}
\end{subequations}
where $\varrho^{jj'}_{\mathrm{ab}}$ is defined in Eq.~\eqref{eq:varrho_ab}.
Here we have used $F_{\mathrm{vio}}=(1-p_0/2) \id+O(p_0^2)$ for the microscopic dephasing channel (see Eq.~\eqref{eq:dephasing_micro_vio}). As a result,
\begin{subequations}
\begin{align}
    &P_0 = 1-\frac{p_0}{2}-\frac{\sigma^2}{2}, V_0=\id,\nonumber\\
    &\bra{\Phi^+}P_0\rho^{\mathrm{out}}_{\mathrm{ab}, 0}\ket{\Phi^+}_{\mathrm{ab}} =  1-p_0-\frac{\sigma^2}{2};\nonumber\\
    &P_1 = \frac{p_0}{2}+\frac{\sigma^2}{2},
    \label{eq:meas_out_noisy_deph_d}
\end{align}
and if $p_0> \sigma^2$, $V_1=Z$,
\begin{equation}
    \bra{\Phi^+}P_1\rho^{\mathrm{out,corr}}_{\mathrm{ab}, 1}\ket{\Phi^+}_{\mathrm{ab}} =  \frac{p_0}{2};
\end{equation}
if $p_0\leq\sigma^2$, $V_1=\id$,
\begin{equation}
    \bra{\Phi^+}P_1\rho^{\mathrm{out,corr}}_{\mathrm{ab}, 1}\ket{\Phi^+}_{\mathrm{ab}} =  \frac{\sigma^2}{2}.
\end{equation}
\end{subequations}
Therefore, to $O(\sigma^2)$, the probabilistic advantage is minimally affected,
\begin{equation}
    \mathcal{R}_{\mathrm{prob}}=\frac{p_0}{1-\frac{1-p_0-\frac{\sigma^2}{2}}{1-\frac{p_0}{2}-\frac{\sigma^2}{2}}}\to 2,
    \label{eq:rprob_noisy_deph_d}
\end{equation}
whereas the effect on the deterministic variant is much more drastic:
\begin{align}
    \mathcal{R}_{\mathrm{det}}=&\frac{p_0}{1-(1-p_0-\frac{\sigma^2}{2}+\max\{\frac{p_0}{2},\frac{\sigma^2}{2}\})}\nonumber\\
    =&\bigg\{\begin{matrix}
    1 ,& \mathrm{if}\ p_0\leq \sigma^2, \\
    \frac{2p_0}{p_0+\sigma^2} ,& \mathrm{if}\ p_0> \sigma^2. \\
    \end{matrix}
    \label{eq:rdet_thres_deph_d}
\end{align}
Equation~\eqref{eq:rdet_thres_deph_d} indicates that a deterministic advantage appears only if the path DOF noise is weaker than the noise in each individual channel (up to a constant factor), $p_0 > \sigma^2$. This is indeed numerically observed in Fig.~\ref{fig:identical_detprob_micro}(a1), and corresponds to a threshold CJ fidelity of $F^{0}_{\mathrm{CJ,th}}=1-\sigma^2$.

\subsubsection{Depolarizing channels}

For the depolarizing noise,
\begin{subequations}
\begin{equation}
    P_0\rho^{\mathrm{out}}_{\mathrm{ab}, 0} = (1-p_0-\frac{\sigma^2}{2})\varrho^{00}_{\mathrm{ab}}+\frac{p_0}{6}(\varrho^{11}_{\mathrm{ab}}+\varrho^{22}_{\mathrm{ab}}+\varrho^{33}_{\mathrm{ab}}),
\end{equation}
\begin{equation}
    P_1\rho^{\mathrm{out}}_{\mathrm{ab}, 1} = \frac{\sigma^2}{2}\varrho^{00}_{\mathrm{ab}}+\frac{p_0}{6}(\varrho^{11}_{\mathrm{ab}}+\varrho^{22}_{\mathrm{ab}}+\varrho^{33}_{\mathrm{ab}}).
\end{equation}
\end{subequations}
Correspondingly, Eqs.~\eqref{eq:meas_out_noisy_deph_d} continue to be valid. If $p_0> 3\sigma^2$, $V_1$ can be any traceless $\mathrm{U}(2)$ unitary, and
\begin{subequations}
\begin{equation}
    \bra{\Phi^+}P_1 \rho^{\mathrm{out,corr}}_{\mathrm{ab}, 1}\ket{\Phi^+}_{\mathrm{ab}} =  \frac{p_0}{6};
\end{equation}
if $p_0\leq 3\sigma^2$, $V_1=\id$,
\begin{equation}
    \bra{\Phi^+}P_1\rho^{\mathrm{out,corr}}_{\mathrm{ab}, 1}\ket{\Phi^+}_{\mathrm{ab}} =  \frac{\sigma^2}{2}.
\end{equation}
\end{subequations}
Therefore, Eq.~\eqref{eq:rprob_noisy_deph_d} continues to be valid, and the deterministic infidelity ratio has the form
\begin{align}
    \mathcal{R}_{\mathrm{det}}=&\frac{p_0}{1-(1-p_0-\frac{\sigma^2}{2}+\max\{\frac{p_0}{6},\frac{\sigma^2}{2}\})}\nonumber\\
    =&\bigg\{\begin{matrix}
    1 ,& \mathrm{if}\ p_0\leq 3\sigma^2, \\
    \frac{6p_0}{5p_0+3\sigma^2} ,& \mathrm{if}\ p_0> 3\sigma^2, \\
    \end{matrix}
    \label{eq:rdet_thres_depo_d}
\end{align}
which predicts a threshold CJ fidelity of $F^{0}_{\mathrm{CJ,th}}=1-3\sigma^2$; see Fig.~\ref{fig:identical_detprob_micro}(b1).

\subsubsection{Amplitude damping channels}
For the amplitude damping noise,
\begin{subequations}
\begin{align}
    P_0\rho^{\mathrm{out}}_{\mathrm{ab}, 0} = &(1-\frac{p_0}{2}-\frac{\sigma^2}{2})\varrho^{00}_{\mathrm{ab}}+\frac{p_0}{4}(\varrho^{03}_{\mathrm{ab}}+\varrho^{30}_{\mathrm{ab}})\nonumber\\
    &+\frac{p_0}{8}(\varrho^{11}_{\mathrm{ab}}-\mathrm{i}\varrho^{12}_{\mathrm{ab}}+\mathrm{i}\varrho^{21}_{\mathrm{ab}}+\varrho^{22}_{\mathrm{ab}}),
\end{align}
\begin{align}
    P_1\rho^{\mathrm{out}}_{\mathrm{ab}, 1} = &\frac{\sigma^2}{2}\varrho^{00}_{\mathrm{ab}}\nonumber\\
    & +\frac{p_0}{8}(\varrho^{11}_{\mathrm{ab}}-\mathrm{i}\varrho^{12}_{\mathrm{ab}}+\mathrm{i}\varrho^{21}_{\mathrm{ab}}+\varrho^{22}_{\mathrm{ab}}),
\end{align}
\end{subequations}
correspondingly,
\begin{subequations}
\begin{align}
    &P_0 = 1-\frac{p_0}{4}-\frac{\sigma^2}{2}, V_0=\id,\nonumber\\
    &\bra{\Phi^+}P_0\rho^{\mathrm{out}}_{\mathrm{ab}, 0}\ket{\Phi^+}_{\mathrm{ab}} =  1-\frac{p_0}{2}-\frac{\sigma^2}{2};\nonumber\\
    &P_1 = \frac{p_0}{4}+\frac{\sigma^2}{2},
    \label{eq:meas_out_noisy_ad_d}
\end{align}
if $p_0> 4\sigma^2$, $V_1$ can be any purely off-diagonal $\mathrm{U}(2)$ unitary, and

\begin{equation}
    \bra{\Phi^+}P_1 \rho^{\mathrm{out,corr}}_{\mathrm{ab}, 1}\ket{\Phi^+}_{\mathrm{ab}} =  \frac{p_0}{8};
\end{equation}
if $p_0\leq 4\sigma^2$, $V_1=\id$,
\begin{equation}
    \bra{\Phi^+}P_1\rho^{\mathrm{out,corr}}_{\mathrm{ab}, 1}\ket{\Phi^+}_{\mathrm{ab}} =  \frac{\sigma^2}{2}.
\end{equation}
\end{subequations}
Therefore,
\begin{equation}
    \mathcal{R}_{\mathrm{prob}}=\frac{1-(1-\frac{p_0}{2})}{1-\frac{1-\frac{p_0}{2}-\frac{\sigma^2}{2}}{1-\frac{p_0}{4}-\frac{\sigma^2}{2}}}\to 2,
    \label{eq:rprob_noisy_ad_d}
\end{equation}
and
\begin{align}
    \mathcal{R}_{\mathrm{det}}=&\frac{1-(1-\frac{p_0}{2})}{1-(1-\frac{p_0}{2}-\frac{\sigma^2}{2}+\max\{\frac{p_0}{8},\frac{\sigma^2}{2}\})}\nonumber\\
    =&\bigg\{\begin{matrix}
    1 ,& \mathrm{if}\ p_0\leq 4\sigma^2, \\
    \frac{4p_0}{3p_0+4\sigma^2} ,& \mathrm{if}\ p_0> 4\sigma^2, \\
    \end{matrix}
    \label{eq:rdet_thres_ad_d}
\end{align}
where we have used $F^{0}_{\mathrm{CJ}}=1-p_0/2+O(p_0^2)$. This gives a threshold CJ fidelity of $F^{0}_{\mathrm{CJ,th}}=1-2\sigma^2$, as shown in Fig.~\ref{fig:identical_detprob_micro}(c1).

\end{document}